\documentclass{amsart}

\usepackage{tikz-cd}

\usepackage[utf8]{inputenc}
\usepackage{amsmath}
\usepackage{amsfonts}
\usepackage{amssymb}
\usepackage{xfrac,color,url}
\usepackage[normalem]{ulem}

\usepackage{tabularx}
\usepackage{algorithm}
\usepackage{algpseudocode}
\usepackage{longtable}

\usepackage{ltablex}
\usepackage{makecell}
\keepXColumns

\theoremstyle{definition}
	\newtheorem{theorem}{Theorem}[section]
	\newtheorem{definition}[theorem]{Definition}

	\theoremstyle{remark}
	\newtheorem*{remark}{Remark}

\newcommand{\vect}[1]{\ifmmode\mathbf{#1}\else\textbf{#1}\fi}
\DeclareMathOperator*{\argmax}{arg\,max}

\title[Respiration from Fingertip]{Fingertip Micro-Motion as a Source of Respiratory Information During Sleep Using Triaxial Accelerometers}

\author{Jeanne Lin}
\address{Courant Institute of Mathematical Sciences, New York University, New York, 10012, USA}

\author{Lily Liu}
\address{Courant Institute of Mathematical Sciences, New York University, New York, 10012, USA}

\author{Chih-Wei Hsu}
\address{1PranaQ Pte. Ltd., 331 North Bridge Road, \#12-03 Oden Towers, Singapore, 188720}

\author{Hau-Tieng Wu}
\address{Courant Institute of Mathematical Sciences, New York University, New York, 10012, USA}

\begin{document}
\maketitle

\begin{abstract}

Objective: Triaxial accelerometers (TAAs) are widely used in home care medicine. This study investigates whether TAA signals recorded at the fingertip encode respiratory information, particularly instantaneous respiratory rate (IRR) and respiratory effort, during sleep. 
Method: We propose an antiderivative-based nonlinear transformation to convert TAA signals into a respiratory surrogate, termed TAA-resp. To quantify the embedded respiration-induced oscillation, a modern time-frequency analysis tool is applied to derive an index, referred to as the {\em respiratory motion index} (RMI). The proposed TAA-resp and RMI are validated on two datasets comprising 39 Asians and 10 white or African Americans with full-night recordings from simultaneous polysomnography (PSG) and a fingertip TAA measurement. Criteria for labeling TAA-resp signal quality as good, moderate, or poor are established, and two experts' independent annotations with consensus are obtained.
Result: On average, TAA-resp encodes high-quality respiratory information in over 21.25\%$\pm$16.11\% of full-night recordings, reaching up to 58.33\% in some cases. TAA-resp shows stronger correlation with thoracic and abdominal motion than with airflow, indicating predominant capture of respiratory effort. High-quality TAA-resp segments offer an accurate IRR estimate with root mean square error $0.027\pm 0.022$ Hz. RMI is higher for high-quality segments and lower for poor-quality segments, and its distribution aligns with physiology, with higher values during REM, N2, and N3 sleep and in the absence of apnea or hypopnea events. In leave-one-subject-out cross-validation, RMI predicts quality labels with 0.84 sensitivity and 0.88 specificity.  
Conclusion: Fingertip-mounted TAAs encode meaningful respiratory information intermittently, and our method recycles underutilized respiratory information.
\end{abstract}

\section{Introduction}

Triaxial accelerometers (TAA) have been widely adopted across scientific disciplines, particularly in medicine, due to their low cost, portability, and ability to noninvasively and nonintrusively capture continuous motion signals. They are commonly used to quantify physical activity, including  movement disorder detection and  rehabilitation progress monitoring \cite{patel2012review},  gait, cadence, and balance analysis \cite{wu2023application},  sleep stage estimation \cite{chen2025validation}, and  respiratory information extraction, among other applications.
In the context of respiratory information extraction, TAAs have been deployed in a variety of anatomical locations, resulting in a diverse and extensive body of literature. Prior studies have investigated single-sensor configurations placed 
on the chest wall or abdomen to capture respiratory motion \cite{bates2010respiratory}, 
on waist-worn belts to track abdominal movement \cite{liu2011estimation}, 
and directly on the chest to reflect the anterior-posterior thoracic motion \cite{fekr2014tidal}.
Additional placements include the suprasternal notch, enabling measurement of upper airway airflow or tracheal motion \cite{bucklin2010inexpensive,dehkordi2011validation,hafezi2020sleep}, 
the xiphoid process to approximate diaphragmatic movement  \cite{jin2009performance}, 
and the intersection of the lower costal margin and the midclavicular line to estimate chest wall angular motion \cite{mann2011simultaneous}. 
In this work, we focus specifically on studies employing a single TAA sensor positioned on the fingertip pad, a location that, to our knowledge, has not been previously explored for respiratory monitoring; multi-sensor configurations and related approaches are discussed in the Discussion section. 

The objective of this study is to explore the extent to which fingertip TAA signals can detect respiratory information during sleep using an existing FDA-approved home sleep apnea test device that includes a TAA sensor. We demonstrate that such measurements can intermittently capture respiratory information, including instantaneous respiratory rate (IRR) and respiratory effort, with good accuracy. This capability enables new opportunities to recycle the underutilized TAA signals to obtain respiratory dynamics during sleep. While it is not the focus of this paper, we expect this recycled information could be further applied in the sensor-limited home-based sleep monitoring setting.

Our exploration is justified by physiological knowledge. During sleep, voluntary motor activity is markedly reduced. Despite this reduction in gross body movement, respiration persists continuously to maintain gas exchange. Breathing is driven by rhythmic activation of respiratory muscles, primarily the diaphragm and intercostal muscles, controlled by brainstem respiratory centers \cite{west2020west}. Their contraction produces cyclic expansion and recoil of the thoracic cavity, generating mechanical motion of the chest wall, abdomen, and torso that can propagate throughout the body. These mechanical movements propagate through the body and {\em may} be detectable by a TAA placed at peripheral locations such as the fingertip pad.
On this basis, we posit that the oscillations observed in fingertip TAA signals during sleep predominantly reflect mechanical oscillations arising from respiratory effort. This physiological rationale motivates the proposed algorithmic framework for extracting respiratory information from fingertip TAA signals.

In this paper, we propose a phenomenological model, called the {\em multivariable adaptive non-harmonic model} (mANHM), to characterize the {\em respiration-induced mechanical oscillation}, or respiration-induced oscillation for brevity, in TAA signals recorded from a fingertip-mounted TAA during sleep. A key novelty of this framework is its explicit formulation of what constitutes an oscillation, which is an essential consideration given that respiration-induced oscillation may not always be detectable at the fingertip-mounted TAA. Despite its fundamental importance in biomedical signal processing, to our knowledge, this question remains insufficiently quantified. Building on this model and focusing on information recycling, we propose a nonlinear transformation that maps the TAA signal to a derived signal, termed TAA-resp. The key novelty includes improving signal-to-noise ratio by taking the antiderivative and fusing three axes using physiological constraints, which jointly help extract weak respiration-induced oscillations from fingertip-mounted measurements.
This approach can be viewed as a generalization of existing methods for extracting respiratory signals from TAA recordings acquired on the torso \cite{jin2009performance,bates2010respiratory,bucklin2010inexpensive,liu2011estimation,mann2011simultaneous,dehkordi2011validation,fekr2014tidal,hafezi2020sleep}. The quality of TAA-resp is annotated by experts and quantified via a novel metric, the {\em respiratory motion index} (RMI). RMI is designed to quantify the strength of respiration-induced oscillation, where the phase-based unwrapping of TAA-resp using the synchrosqueezing transform (SST) \cite{DaLuWu2011} is applied to handle time-varying amplitude and frequency.
To assess the extent of respiratory information encoded in TAA-resp, we compare it against simultaneously recorded excursions of chest (THO) and abdomen (ABD) via respiratory inductance plethysmography (RIP) sensors, as well as airflow signals obtained from PSG. The results demonstrate that TAA-resp predominantly reflects respiratory effort and can provide reliable IRR information. Finally, we show that RMI reliably predicts TAA-resp quality, enabling automated identification of segments suitable for downstream analysis.

The paper is organized as follows. In Section \ref{section mANHM}, we present the details of the mANHM framework for the TAA signal. Section \ref{section algo} describes the proposed algorithms, including the construction of TAA-resp from the TAA signal and the methods used to compute RMI. The materials and statistical analyses are provided in Section \ref{section material}, with results presented in Section \ref{section result}. The paper concludes with a discussion in Section \ref{section discussion}. A brief review of SST is included in Section \ref{section app review sst} for the purpose of self-containedness.

\section{A Phenomenological Model for fingertip TAA}\label{section mANHM}

The recorded measurements by a TAA are three-dimensional time series, which we refer to as {\em TAA signals}. These signals contain contributions from body motion-induced acceleration, the gravitational component, and small sensor vibrations or measurement and thermal noise. In this section, we clarify the relationship between respiratory dynamics and respiratory effort, and provide a {\em phenomenological} model to quantify the TAA signal that encodes respiration-induced oscillation.

\subsection{Respiratory dynamics and effort}

We first clarify the relationship between {\em respiratory dynamics} and {\em respiratory effort} \cite{west2020west,kaminsky2023respiratory}. Formally, respiratory effort refers to the pressure-generating activity (i.e., neural drive and resulting mechanical work) of the respiratory muscles required to produce ventilation. It is not a single directly measurable quantity, but rather a latent construct reflecting how strongly the body is ``trying'' to breathe. In practice, respiratory effort is assessed via surrogate measurements, including
esophageal pressure \cite{yoshida2018esophageal}, transdiaphragmatic pressure \cite{laporta1985assessment}, airway pressure swings \cite{webster2009medical}, chest and abdominal excursions via piezoelectric or RIP sensors  \cite{costanzo2020respiratory}, diaphragm electromyogram \cite{luo2008diaphragm}, etc.
In contrast, respiratory dynamics refers more broadly to the time-varying behavior of the respiratory system, encompassing, but not limited to, the evolution and interaction of airflow, pressure, and volume during breathing. Respiratory effort constitutes one component of these dynamics.
Beyond the aforementioned measures, respiratory dynamics can be observed through a variety of sensing modalities. For example, airflow can be recorded via pressure transducer or thermistors  \cite{costanzo2020respiratory}; torso motion via TAA \cite{jin2009performance,bates2010respiratory,bucklin2010inexpensive,liu2011estimation,mann2011simultaneous,dehkordi2011validation,fekr2014tidal,hafezi2020sleep}; impedance-derived signals via electrocardiogram (ECG) electrodes (impedance pneumography) \cite{mcerlean2024unsupervised}; electrical activity of the diaphragm (EAdi) via specialized nasogastric electrodes \cite{beck2001electrical}; end-tidal CO2 via capnography \cite{webster2009medical}; and respiration-induced intensity variations in the photoplethysmogram (PPG) \cite{shelley2006use}. See \cite{costanzo2020respiratory,vanegas2020sensing} for reviews of many other techniques. Each modality captures a distinct projection of the underlying respiratory dynamics and differs based on sensor characteristics. For example, airflow and chest, abdominal, and torso motions collectively reflect the interplay between inspiratory muscle activation and expiratory recoil; during obstructive apnea, however, airflow is markedly attenuated or absent, whereas chest, abdominal, and torso motion signals may exhibit characteristic paradoxical patterns. These physiological considerations lead to the claim that oscillations observed in the TAA signal primarily encode mechanical vibrations associated with respiratory effort. Direct evidence supporting this claim is presented in the Results section.

\subsection{Empirical Characteristics of Respiratory Dynamics}

Respiratory dynamics are typically quasi-periodic with cycle-to-cycle variation known as breathing pattern variability, reflecting interactions among central respiratory pattern generators, autonomic regulation, and feedback from chemoreceptors and mechanoreceptors \cite{benchetrit2000breathing}. Such variability is reflected in several components of the oscillatory signal. First, the breathing period varies from cycle to cycle, which is influenced by autonomic balance, sleep stage transitions, and reflex inputs such as pulmonary stretch receptor feedback. This characteristic will be modeled as IRR below.
Second, the magnitude of each breathing cycle also changes from cycle to cycle, reflecting changes in tidal volume and respiratory effort, which arise from fluctuations in neural drive from the brainstem and metabolic demands mediated by central and peripheral chemoreceptors. This characteristic will be modeled as {\em amplitude modulation} (AM) below.
Third, the cycle morphology also varies from cycle to cycle, reflecting body position, respiratory effort with different respiratory muscle recruitment, airway resistance, neuromechanical coupling within the respiratory system \cite{west2020west}, etc. 
This characteristic represented in TAA will be modeled as a {\em multivariable wave-shape function} (mWSF) below.

\subsection{Multivariable wave-shape function}

To model the TAA signal, we start by recalling the following definition.
\begin{definition}
A smooth function $s:\mathbb{R}\to \mathbb{R}^p$, where $p\in \mathbb{N}$, is called $\tau$-periodic for $\tau>0$ if $\tau=\min\{t>0;\,s(t+x)=s(x),\,\forall x\in \mathbb{R}\}$. 
\end{definition}
It is easy to check that if $e_j^\top s$ is $\tau_j$-periodic, where $\tau_j\in \mathbb{N}$ and $e_j\in \mathbb{R}^p$ is a unit vector with the $j$-th entry $1$, then $s(t)$ is $\tau$-periodic if the least common multiple of $\tau_1,\ldots,\tau_p$ is $\tau$.
However, when $s(t)$ is $1$-periodic, it does not imply that all $s_\theta(t):=\theta^\top s(t)$, where $\theta\in S^{p-1}$, are $1$-periodic. For example, if we write $e_i^\top s=\sum_{k=1}^\infty\alpha_{ik}\cos(2\pi kt+\beta_{ik})$, where $\alpha_{i1}>0$ and $\alpha_{ik}\geq 0$ for $k\geq 2$, and $\beta_{ik}\in [0,2\pi)$, when $\alpha_{11}\cos(2\pi t+\beta_{11}),\ldots,\alpha_{p1}\cos(2\pi t+\beta_{p1})$ are linearly dependent, we could find a $\theta$ so that the first Fourier coefficient of $s_\theta$ is $0$ and hence $s_\theta$ may not be $1$-periodic.

Consider the following generalization of the wave-shape function (WSF) discussed in \cite{Wu:2013}.
\begin{definition}
We call a 1-periodic smooth vector valued function $r:\mathbb{R}\to \mathbb{R}^p$ a {\em multivariable wave-shape function} (mWSF) if $e_l^\top r$ is 1-periodic for $l=1,\ldots,p$ and it satisfies
\[
\int_0^1r(t)dt=0\ \mbox{ and }\ \int_{0}^1\|r(t)\|_2dt=1\,,
\]
where $\|\cdot\|_2$ is the standard Euclidean norm.
\end{definition}

\subsection{What constitutes an oscillation?}\label{section: What constitutes an oscillation}

Empirically, respiration-induced oscillation is not consistently present in the TAA signal; it may appear intermittently and then disappear. To model respiration-induced oscillation in TAA signals appropriately, a critical question is {\em how many cycles must be observed before we can confidently assert the presence of an oscillation}? In general, we must address a fundamental question: 
\begin{quote}
{\em What constitutes an oscillation?}
\end{quote} 
Suppose, hypothetically, that an oracle reveals the exact number of cycles within a given segment. Should a single cycle suffice to classify the segment as oscillatory, or would the answer change if there were three cycles, or ten? We call this problem the {\em oscillation characterization problem}. To the best of our knowledge, this problem has received limited attention in the literature, and no universally accepted definition exists, even under the slowly varying AM and IRR assumptions we will discuss below in (S1) and (S2). This is likely because, traditionally, when oscillations are discussed, they are {\em implicitly} assumed to persist over time. The closest related works arise in the context of {\em oscillation change-point detection} \cite{colominas2023iterative,wu2024frequency}. In \cite{wu2024frequency}, the authors assume harmonic oscillations that persist for a sufficiently long duration, although no minimal length is specified; similarly, in \cite{colominas2023iterative}, the authors allow time-varying amplitude and frequency, but again, do not quantify the minimal duration required.

\begin{figure}[hbt!]
    \centering
\includegraphics[width=\textwidth]{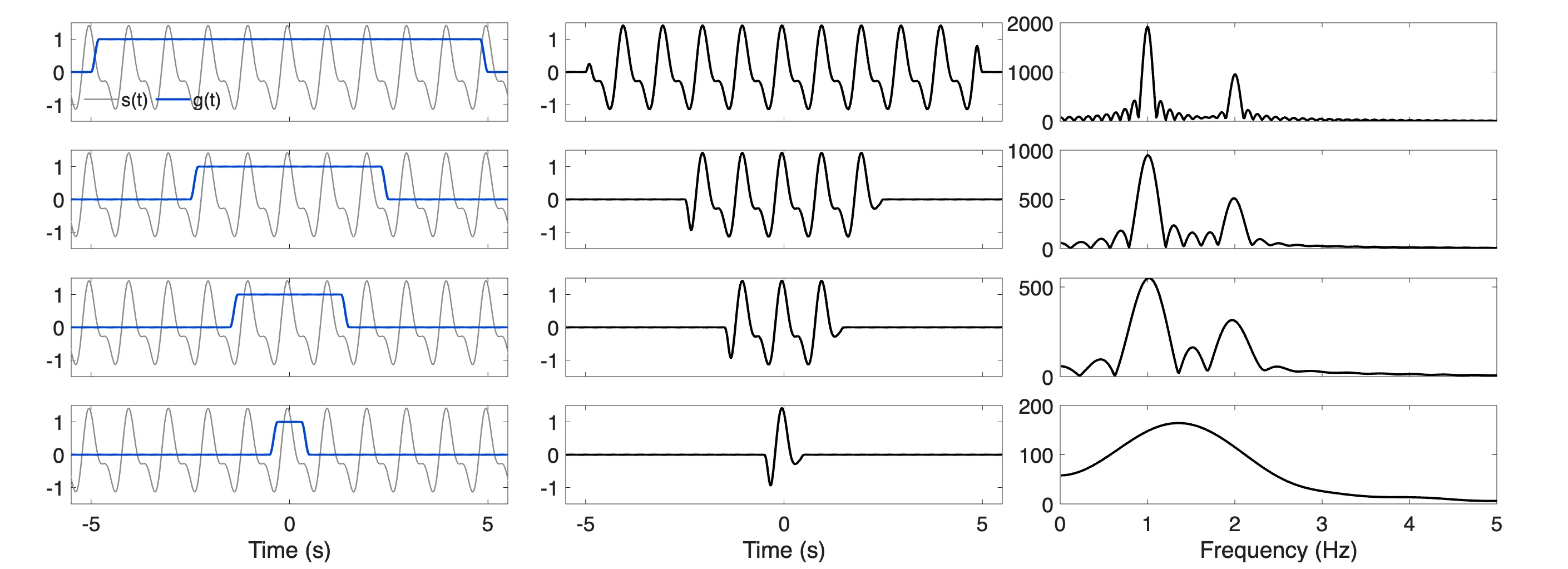}
    \caption{Illustration of how spectral spreading of the first harmonic is impacted by the observation window. The signal $s(t)=\cos(2\pi t)+0.5\cos(2\pi 2t+1)$ and $g(t)$ modeling the observation window are shown in the left column. The observed signal $h(t):=s(t)g(t)$ is shown in the middle column.} The Fourier magnitudes of $h(t)$ are shown in the right column. From top to bottom, $g$ is compactly supported on $[-L,L]$ with $L=5,2.5,1.5, 0.5$. In each case, $g(t)=1$ on $[-L+0.2, L-0.2]$ and tapers smoothly and monotonically to 0 near the boundaries. 
    \label{fig:HowManyCycles}
\end{figure}

We resolve this oscillation characterization problem by giving a mathematical criterion about what it means by {\em the signal is locally oscillatory over a finite period centered at $\upsilon$}. 
Consider a simple model where the signal oscillates with fixed frequency $1$ and amplitude $1$ but nonsinusoidally; that is, $s(t)$, where $s$ is a mean 0, unit $L^2$ norm, and 1-periodic function on $\mathbb{R}$. Suppose $s(t)$ is ``observed'' via a window function $g$, called an observation window, centered at $\upsilon\in \mathbb{R}$.  What is the behavior of $s(t)g(t-\upsilon)$, called an observed signal, when $g$ has varying widths? See Figure \ref{fig:HowManyCycles} for an illustration of how the window width impacts the signal in both time and frequency domains. As the window shortens, it becomes less reliable to infer the presence of an oscillatory structure from the time-domain signal, and more difficult to resolve the two spectral peaks at frequencies 1 and 2.

To quantify this, assume $g$ is smooth, symmetric at $0$, and compactly supported on $[-1,1]$. Clearly, its Fourier transform $\hat{g}$ decays fast. Denote $g_L(t):=g(t/L)$, where $L>0$ models the window width. Without loss of generality, assume $\upsilon=0$. Denote $h_L(t):=s(t)g_L(t)$.
Since $s(t)=\sum_{l=1}^\infty a_l\cos(2\pi lt+\beta_l)$, where $a_1>0$, $a_l\geq 0$ for $l>1$, and $\beta_l\in [0,2\pi)$, 
we have 
\[
\widehat{h_L}(\xi)=\frac{L}{2}\sum_{l=1}^\infty a_l \left(\hat{g}(L(\xi-l))e^{i\beta_l}+\hat{g}(L(\xi+l))e^{-i\beta_l}\right)\,.
\] 

Clearly, to capture local dynamics, $L$ cannot be too large, and to identify harmonics in the frequency domain, $L$ cannot be too small.  We say that the signal is oscillatory locally over $[-L,L]$ if {\em the spectral spreading of the first harmonic in $h_L$ is negligible on the other harmonics}. Since the signal is real, the first harmonic has conjugate components in the frequency domain centered at $\xi_0$ and $-\xi_0$. The definition requires $\hat{g}(L(\xi-1))$ and $\hat{g}(L(\xi+1))$ have negligible overlap at $0$.

\begin{definition}
Assume $g$ is real, smooth, symmetric at $0$, and compactly supported on $[-1,1]$, and for $L>0$, denote $g_L(t):=g(t/L)$. Suppose $s(t)$ is a smooth $1$-periodic real function.
For a small constant $\varepsilon>0$, the truncated signal $h_L(t):=s(t)g_L(t-\upsilon)$ is {\em oscillatory locally} over the period $[\upsilon-L,\upsilon+L]$ with precision $\varepsilon$, where $L>0$, if $g_{L}$ satisfies $\int_{|\xi|>1}|\widehat{g_L}(\xi)|d\xi \leq \varepsilon$.
\end{definition}

This definition requires that the spectral distribution associated with each harmonic is essentially concentrated around its frequency with minimal and controllable leakage. To derive a practical rule of thumb for $L$, or equivalently observing approximately $2L$ oscillatory cycles, consider the Gaussian window $g(t)=e^{-\pi^2t^2}$. Numerically, $g(t)$ behaves as if it were compactly supported on $[-1,1]$, and $\widehat{g_L}(\xi)=\frac{L}{\sqrt{\pi}}e^{-L^2\xi^2}$. Using properties of the complementary error function, we obtain $\int_{|\xi|>1}|\widehat{g_L}(\xi)|d\xi \leq \frac{2}{L\pi}e^{-L^2}$. Therefore, for this choice of window, achieving a tolerance of $\varepsilon\sim 1.8\times 10^{-8}$ requires observing approximately $8$ cycles, or $L=4$. Empirically, this rule of thumb extends well to other commonly used window functions. A similar argument applies when the signal satisfies the ANHM condition, since locally such signals can be well approximated by non-sinusoidal functions with fixed amplitude and frequency (see \cite{DaLuWu2011} for details).

\subsection{Phenomenological model for TAA signals}

With the above mathematical background, we now provide our phenomenological model for TAA signals during sleep. We model it as a $3$-dim random process
\begin{align}\label{ACCmodel ver1}
Y(t)=\begin{bmatrix}a_x(t)\\ a_y(t)\\ a_z(t)\end{bmatrix}=R(t)\left(\sum_{j=1}^J A_j(t)r_j(\phi(t))\chi_{I_j}+\begin{bmatrix}0\\ 0\\ G\end{bmatrix}\right)+\zeta(t)\in \mathbb{R}^3\,,
\end{align}
where $t\in \mathbb{R}$ denotes time in seconds, $I_1,\ldots,I_J\subset \mathbb{R}$, $I_i\cap I_j=\emptyset$ when $i\neq j$ and $|I_j|>0$, $A_j\in C^1(\mathbb{R})$ is a nonnegative function with strictly positive values over $I_j$, $R(t)\in C^2(\mathbb{R},SO(3))$,  $r_1,\ldots,r_J:\mathbb{R}\to \mathbb{R}^3$ are mWSFs, $\phi\in C^2(\mathbb{R})$ is a monotonically increasing function with $\phi'(t)>0$, $G>0$ is the gravitation as a trend, and $\zeta(t)$ is a vector random process with mean $0$ and finite covariance at all time. 
We further impose two conditions. There exists a small $\epsilon\geq 0$ so that the following {\em slowly varying} assumptions are satisfied:
\begin{enumerate}
\item [(S1)] $|A_j'(t)|\leq \epsilon A_j(t)$ over $I_j$, $j=1,\ldots,J$;
\item [(S2)] $|\phi''(t)|\leq \epsilon \phi'(t)$ over $I_j$, $j=1,\ldots,J$\,.
\end{enumerate}

In \eqref{ACCmodel ver1}, $I_1,\ldots,I_J$ denote the segments in which respiration-induced oscillation is detected in the fingertip-mounted TAA, which will be further elaborated below. 
The function $A_j(t)$, referred to as the amplitude modulation (AM), models the strength of the detected respiration-induced oscillation over $I_j$. Over the period $\mathbb{R}\backslash (\cup_{j=1}^J I_j)$, no respiration-induced oscillation is detectable by the fingertip-mounted TAA. 
The function $\phi(t)$ represents the phase of respiratory dynamics, with $\phi'(t)>0$ corresponding to the IRR. 
The function $R(t)$ describes the rotation of the fingertip-mounted TAA  around the fingertip pad at time $t$. Note that fingertip rotation in general may be different from the body torso rotation around the body's center of mass, which is used to determine the body position. During normal sleep, $R(t)$ may change occasionally but typically not frequently. The mWSF $r_j(t)$ characterizes the respiratory motion pattern recorded during $I_j$. 

The model \eqref{ACCmodel ver1} is too general. For practical purposes, we encode further observations. First, when body position changes, it is reasonable to expect weak or no respiration-induced oscillation \cite{bates2010respiratory}. We can therefore assume during $I_j$, $R(t)=R_j\in SO(3)$ is constant. Second, while the respiration-induced oscillation in the TAA signal is non-harmonic,  we observe that its oscillatory pattern is relatively simple in the sense that ``almost always'' only the top two Fourier modes of $\theta^\top R_j r_j$, for any $\theta\in S^2$, are nontrivial. This is probably because the mechanical oscillation associated with respiratory effort has been damped via the transmission to the peripherals. Therefore, we simplify \eqref{ACCmodel ver1} to
\begin{align}\label{ACCmodel ver2}
Y(t)=\begin{bmatrix}a_x(t)\\ a_y(t)\\ a_z(t)\end{bmatrix}=\sum_{j=1}^J A_j(t)\begin{bmatrix} s_{x,j}(\phi(t)) \\ s_{y,j}(\phi(t)) \\ s_{z,j}(\phi(t))\end{bmatrix}\chi_{I_j}+R(t)\begin{bmatrix}0\\ 0\\ G\end{bmatrix}+\zeta(t)\in \mathbb{R}^3\,,
\end{align}
where 
\[
s_{x,j}(t):=e_1^\top R_jr_j(t)=\sum_{k=1}^2 \alpha_{x,j,k}\cos(2\pi kt+\beta_{x,j,k})\,, 
\]
$\alpha_{x,j,1}\geq \alpha_{x,j,2} \geq0$, $\sum_{k=1}^2 \alpha_{x,j,k}^2=2$, $\beta_{x,j,k}\in [0,2\pi)$, and
$s_{y,j}:=e_2^\top R_jr_j$, and $s_{z,j}:=e_3^\top R_jr_j$ have the same expansion. $s_{x,j}$, $s_{y,j}$, and $s_{z,j}$ are the WSFs associated with the x-axis, y-axis, and z-axis of the TAA signal.

To complete our phenomenological model, we need to determine the minimal length of $I_j$. Based on the discussion of the oscillation characterization problem in Section \ref{section: What constitutes an oscillation}, the segment should be long enough to include at least eight cycles, corresponding to approximately 40 s. However, given the nature of {\em recycling} information from the TAA signal, a more stringent requirement by considering longer segments is necessary. Moreover, to simplify subsequent analysis, particularly the labeling process, and to accommodate the wide range of breathing rates, a segment is considered to contain a {\em detectable respiration-induced oscillation} only if $I_j$ exceeds 60 s.  We therefore impose the following assumption.
\begin{enumerate}
\item [(S3)] $|I_j|\geq 60$ for $j=1,\ldots,J$.
\end{enumerate}

Assumptions (S1)--(S3) jointly say that while the respiratory dynamics may have time-varying AM and IRR, the AM and IRR do not change too fast, and if the sensor detects respiration-induced oscillation, it should last sufficiently long before we view it as the true detection. Since the oscillatory pattern might not be sinusoidal, we say the model \eqref{ACCmodel ver2} satisfies the {\em multivariable adaptive non-harmonic model (mANHM)}, generalizing the univariate version \cite{Wu:2013}. 

\begin{remark}
Our perspective of solving the oscillation characterization problem is consistent with our empirical rule of thumb for selecting the window length in time-frequency analysis. When approximate prior knowledge of the cycle length is available, relatively clean signals are well analyzed using windows spanning about $6\sim10$ cycles, whereas noisier signals benefit from longer windows, on the order of 15 cycles. In the absence of such prior information, one may instead employ R\'enyi entropy-based criteria for window length selection \cite{Sheu_Hsu_Chou_Wu:2017}. We also note that this observation aligns with physiologically motivated considerations reported in \cite{wu2023application} in the context of cadence analysis.
\end{remark}

\section{Methodology}\label{section algo}

We now introduce the proposed algorithm to obtain the TAA-resp signal and a respiratory motion index (RMI) to quantify the reliability of respiratory information encoded in TAA-resp. The RMI is derived from phase information associated with the putative respiration-induced oscillation \cite{alian2022reconsider}, estimated using the nonlinear-type time-frequency analysis method, SST \cite{DaLuWu2011}. For clarity, a brief review of SST is deferred to Section \ref{section app review sst} in the Appendix.

\subsection{Proposed TAA-resp signal}

Unlike prior work that extracts respiratory information from TAA sensors positioned on the torso \cite{jin2009performance,bates2010respiratory,bucklin2010inexpensive,liu2011estimation,mann2011simultaneous,dehkordi2011validation,fekr2014tidal,hafezi2020sleep}, respiration-induced oscillation is substantially weaker in TAA signals recorded at distal locations such as the fingertip. Consequently, existing approaches might be limited in this setting. Here, we exploit the unique characteristics of fingertip TAA signals and develop a tailored algorithm to transform the raw accelerometry data into a respiratory signal. The pseudocode for producing TAA-resp from TAA signals is provided in Algorithm \ref{alg:taa_resp}.

First, we consider the {\em velocity signal} obtained via evaluating the antiderivative of the TAA signal \cite{titterton2004strapdown}. See the Discussion section for a discussion of this antiderivative approach compared with existing literature. Take the x-axis in \eqref{ACCmodel ver2} as an example. The antiderivative of the clean part of $a_x(t)$, denoted as $v_x(t)$, over $I_j:=[b_j,e_j]$ is
\begin{align}
v_x(t)-v_x(b_j)&\,:=\sum_{k=1}^2 \alpha_{x,j,k}\int^t_{b_j} A_j(s)\cos(2\pi k\phi(s)+\beta_{x,j,k}) ds \nonumber\\
&\,\approx  \sum_{k=1}^2\frac{\alpha_{x,j,k}A_j(t)}{2\pi k\phi'(t)} \sin(2\pi k\phi(t)+\beta_{x,j,k}) \,,\label{eq: appendix: justification}
\end{align}
which enhances the lower frequency oscillation. 
We can see from \eqref{eq: appendix: justification} that the antiderivative behaves like a lowpass filter reweighting the spectrum by $1/\xi$. 
Then, the slowly varying trend, including the gravity component, can be removed by basic de-trend tools, and 
%
we can write down the pre-processed velocity signal as
\begin{align}\label{ACCmodel ver3}
Z(t)=\begin{bmatrix}v_x(t)\\ v_y(t)\\ v_z(t)\end{bmatrix}+\bar{\zeta}(t)\,,
\end{align}
where $\bar{\zeta}(t)$ is a mean 0 vector random process with finite covariance structure, 
\[
v_x(t)=\sum_{j=1}^J  \sum_{k=1}^2\frac{A_j(t)\alpha_{x,j,k}}{2\pi k\phi'(t)} \sin(2\pi k\phi(t)+\beta_{x,j,k})\chi_{I_j}, 
\]
with a slight abuse of notation, and $v_y,v_z$ are expressed similarly. See Section \ref{section math justification of TAA-resp.} for a mathematical justification of the above formula. 

Second, we enhance the putative respiration-induced oscillation by finding the optimal combination of $v_x$, $v_y$, and $v_z$ under physiological constraints. Divide the signal into segments of $60$ s long, where $60$ is chosen to coincide with (S3). Over each segment $I$, evaluate
\begin{align}\label{equation: determine the optimal direction}
\theta^*:=\arg\max_{\theta\in S^2} \frac{\sum_{j\in R} |\widehat{Z_\theta}(\xi_j)|}{\sum_{j} |\widehat{Z_\theta}(\xi_j)|}\,,
\end{align}
where $R$ is the spectral range from $0.1$ to $0.4$ Hz, and $Z_\theta(t):=\theta^\top Z(t)$. The spectral range from 0.1 to 0.4 Hz is determined based on normal respiratory physiology in adult \cite{coote1982respiratory}.  The existence of the maximizer comes from the compactness of $S^2$ and the continuity of the functional. Note that we intentionally select the projection that gives the combination with the highest possibility of encoding respiration-induced oscillations. 
The optimal axis derivation in \eqref{equation: determine the optimal direction} may remind readers of the widely used principal component analysis (PCA) approach for extracting respiratory information from TAA signals \cite{jin2009performance,liu2011estimation}. The relationship between the proposed method and PCA is discussed in detail in the Discussion section.

Set 
\begin{align}\label{ACCmodel ver4}
r_0(t):= Z_{\theta^*}(t)=\sum_{j=1}^J  \sum_{k=1}^2 A_{j,k}^*(t) \cos(2\pi k\phi(t)+\beta_{j,k})\chi_{I_j}+ \zeta^*(t)\,,
\end{align}
where $A_{j,k}^*(t):=\frac{A_j(t)\alpha_{j,k}}{2\pi k\phi'(t)}$ and $\zeta^*(t):={\theta^*}^\top\bar{\zeta}(t)$.
Finally, postprocess $r_0$ by 
\[
r(t) = \frac{\texttt{sign}(r_0(t))\times \min\{|r_0(t)|,\, 500\}}{Q_{99.9}(\{|r_0(s)|\chi_{|r_0|\leq 500};\,s\in I\})}\,,
\]
where $\texttt{sign}$ is the sign function, $Q_{99.9}$ is the $99.9\%$ percentile of $r_0(t)$ over $I$, and $\min\{|r_0(t)|,\, 500\}$ removes the unexpected shock-like artifacts. Here $500$ is determined by empirical observations. We call $r(t)$ the {\em TAA-resp signal}.

\algrenewcommand\algorithmicrequire{\textbf{Input:}}
\algrenewcommand\algorithmicensure{\textbf{Output:}}

\begin{algorithm}[htbp!]
\caption{Pseudocode for TAA-resp construction from 3-axis IMU data}
\label{alg:taa_resp}
\begin{algorithmic}[1]
\Require 3-axis IMU signal $X(t)=(x(t),y(t),z(t))^\top$ sampled at $f_s$ Hz
\Ensure TAA-resp sampled at $5$ Hz every 60 s
\For{each IMU channel $x,y,z$}
    \State Estimate and remove trend using 10s median filter
    \State \textit{(Antiderivative)} Integrate using cumulative summation \eqref{eq: appendix: justification}
    \State Again estimate and remove trend using 10s median filter
    \State Downsample from $f_s$ Hz to $5$ Hz
    \State Set values with greater than $500$ to $500$ and those less than $-500$ to $-500$ 
\EndFor

\State Generate random unit vectors $V:=\{v_i\}_{i=1}^{m}\subset S^2$, where we set $m=500$

\For{each $60$s segment $X_k=(x_k,y_k,z_k)^\top \in \mathbb{R}^{3\times 300}$}
    \State Remove mean and compute FFT for each channel. Get $\hat{X}_k\in \mathbb{R}^{3\times 300}$
    \State Evaluate
    \[
    v^*=\argmax_{v_i\in V}
    \max_{0.1<\xi_j<0.4}
    \left|v_i^\top \widehat{X}(\xi_j)\right|
    \]
    \State $\boldsymbol{r}_{k0}:={v^*}^\top X_k$
    \State Apply fourth-order Butterworth filter with passband $0.1$--$1$ Hz using zero-phase filtering to $\boldsymbol{r}_{k0}$ to get the TAA-resp $\boldsymbol{r}_{k}$
\EndFor
\end{algorithmic}
\end{algorithm}

\begin{remark}
While it is possible, we do not consider taking a second antiderivative to obtain {\em position} information. The primary reason is that repeated integration leads to significant error accumulation, particularly low-frequency bias, resulting in severe drift. Empirically, we observe that this low-frequency bias tends to contaminate the spectral band associated with respiration-induced oscillation, thereby negating the noise-reduction benefits gained from a single integration.
\end{remark}

\subsection{Proposed respiratory motion index (RMI)}

We now propose our index quantifying if there exists respiration-induced oscillation. First, we estimate the respiratory phase from one realization of $r(t)$ sampled at $f_s>0$ Hz over a finite interval $I$, denoted as $\boldsymbol{r}\in \mathbb{R}^N$, where $\boldsymbol{r}(i)=r(i/f_s)$. Second, we unwrap the signal using the estimated respiratory phase and evaluate the quality of the respiration-induced oscillation. We refer readers with interest to Section \ref{section:math interpretation of RMI} for mathematical interpretation of RMI. The pseudocode for RMI calculation for TAA-resp is provided in Algorithm \ref{alg:taa_resp}.

\subsubsection{Step 1: obtain respiratory phase}\label{section REI step 1}

Step 1 depends on the sharpened TFR determined by SST and a ridge detection algorithm to estimate the IRR and the phase function, denoted as $\tilde{\phi}(t)$, so that $\tilde{\phi}(t)\approx \phi(t)$ over $I_j$.
See Section \ref{section app review sst} for details. 

\subsubsection{Step 2: obtain respiratory motion index as detection statistics}\label{section REI step 2}

We apply a simple yet effective unwrapping approach to quantify the strength of respiration-induced oscillation. 
The input signal $\boldsymbol{r}$ is truncated into nonoverlapping 60-second segments, denoted as $\boldsymbol{r}_1,\ldots,\boldsymbol{r}_K$, where $K\geq J$. Each segment is then unwrapped using the associated estimated phase $\tilde{\phi}$ from Step 1, yielding signals $\boldsymbol{z}_1,\ldots,\boldsymbol{z}_K$, with Fourier transforms $\hat{\boldsymbol{z}}_1,\ldots,\hat{\boldsymbol{z}}_K$. For each segment, $k=1,\ldots,K$, the RMI is defined as 
\begin{equation}\label{definition REI}
     \texttt{RMI}_k = \frac{\sum_{i\in I_R} |\hat{\boldsymbol{z}}_k(i)|^2 }{\sum_{l\in I_A} |\hat{\boldsymbol{z}}_k(l)|^2}\,,
\end{equation}
where $I_R$ denotes the spectral bands $0.9-1.1$ Hz, and $I_A$ denotes the range $0.1-3$ Hz.
Figure~\ref{fig:IMUresp visualization unwrapping} illustrates TAA-resp and its spectrum before and after the unwrapping process.

\begin{algorithm}[htbp!]
\caption{Pseudocode for RMI computation from TAA-resp}
\label{alg:rmi}
\begin{algorithmic}[1]
\Require Preprocessed TAA-resp signal $\boldsymbol{r}$ sampled at $5$ Hz
\Ensure RMI value for each $60$s window
\For{each $60$s TAA-resp segment, $\boldsymbol{r}_k$,}
    \State Identify dominant frequency
    
        \State Estimate respiratory phase $\phi(t)$ using SST and ridge detection
        \If{Number of cycles $\phi(\mathrm{end}) < 4$}
            \State Set $\texttt{RMI}_k=0$
        \Else
            \State Unwrap $\boldsymbol{r}_k$ using $\phi(t)$
        \State Compute the FFT of the unwrapped signal, denoted as $\boldsymbol{z}_k$
        \State Set
        \[
        \texttt{RMI}_k:=
        \frac{
        \sum_{0.9<\xi<1.1}
        |\widehat{\boldsymbol{z}}_k(\xi)|^2
        }{
        \sum_{0.1<\xi<3}
        |\widehat{\boldsymbol{z}}_k(\xi)|^2
        }
        \]
        \EndIf
\EndFor
\end{algorithmic}
\end{algorithm}

\section{Material and statistics}\label{section material}

\subsection{Material}

We retrospectively analyzed data from two prospective observational studies conducted at Taipei Veterans General Hospital (VGHTPE), Taiwan, between June and December 2023, and Duke University Hospital (DUKE), NC, USA, between February 2024 and August 2024. Both are tertiary medical centers. The VGHTPE study was approved by the Medical Ethics Committee (IRB No. 2023-04-003A) and the DUKE study was approved by the Duke University Health System (DUHS) Institutional Review Board (Approval number Pro00113559). Both studies adhered to the ethical principles of the 1975 Declaration of Helsinki.

The following protocol was carried out in both studies. Subjects aged $\geq$20 years, eligible for overnight in-laboratory PSG study or referred directly from outpatient clinics, were recruited. Written informed consent was obtained before enrollment. During the study period, the subject can withdraw from the study at any time. Exclusion criteria included heart transplantation, New York Heart Association (NYHA) class III-IV heart failure, chronic obstructive lung disease with Global Initiative for Chronic Obstructive Lung Disease (GOLD) categorization 3 or 4, chronic opioid use, severe stroke (modified Rankin Scale $\geq 4$), tracheostomy, inability to provide informed consent or complete questionnaires, or inability to comply with instructions for use of the TipTraQ device (e.g., advanced Alzheimer's disease, unconsciousness, or severely impaired cognitive function). Participants could withdraw at any time.
The TipTraQ system is a home sleep apnea testing platform consisting of a fingertip-worn miniaturized device equipped with a TAA sensor that communicates via Bluetooth with a smartphone, along with a cloud-based artificial intelligence system for automated scoring of sleep apnea events and sleep stages \cite{chen2025validation}.

Physiological signals from the TipTraQ device and PSG were collected simultaneously throughout the overnight sleep study, typically lasting over 6 hours. TipTraQ TAA signals were sampled at 50 Hz. Synchronization between TipTraQ and PSG recordings was performed using instantaneous heart rates derived from PSG ECG and TipTraQ PPG signals. For quality control, recordings were excluded if any PSG channel was deemed uninterpretable by a sleep expert, if ECG waveforms or annotations were missing, or if the TipTraQ device detached or malfunctioned during sleep. 
PSG annotations, including sleep stages, arousals, and sleep-related respiratory events, were visually scored according to the  AASM criteria \cite{berry2017aasm} by the sleep technicians, verified by a sleep specialist with disputes or uncertainties resolved through discussion in meetings to reach a consensus.

In this work, we analyzed 39 whole-night recordings from the VGHTPE database and 10 whole-night recordings from the DUKE database. All exceeded 4 hours in duration. In the VGHTPE database, all subjects are Asian, the mean age and BMI are $55.1\pm 14.2$ year-old and $27.7\pm 4.1$ kg/m${}^2$, 10 subjects are female, the recording time is $7.15\pm 0.47$ hours, and apnea-hypopnea index (AHI) is $19.82\pm 17.02$ events/hour. In the DUKE database, 7 subjects are white and 3 are African Americans, the mean age and BMI are $54\pm 18.4$ year-old and $32.1\pm 5.8$ kg/m${}^2$, 6 subjects are female, the recording time is $7.37\pm 0.87$ hours, and AHI is $6.99\pm 7.70$ events/hour.

\subsection{Label procedure}\label{sec:label procedure}

The fingertip is not a usual location to record TAA signals and extract respiratory information. We therefore need a criterion to decide if there is a detected respiration-induced oscillation so that experts can follow and label additional segments. The rule is listed in Figure \ref{figure: label rule and criteria}. In binary analysis, we call moderate and good as {\em high-quality}, and poor as {\em low-quality}.

\begin{figure}[hbt!]
\centering
\begin{tikzpicture}[
    >=stealth,
    box/.style={rectangle, draw, rounded corners, align=center,
                font=\small, minimum width=2.4cm, minimum height=0.8cm},
    poor/.style={box, fill=red!12,    draw=red!55!black},
    mod/.style={box,  fill=orange!16, draw=orange!70!black},
    good/.style={box, fill=green!14,  draw=green!55!black},
]
\node[box] (a) at (-0.75, 5) {1 min TAA-resp};

\node[box] (b1) at (-4.5, 3) {Not $99$th percentile\\magnitude $> 10$};
\node[box] (b2) at ( 2.0, 3) {$99$th percentile\\magnitude $> 10$};

\node[mod,dashed]  (f1) at (-6.3, 1.6) {No artifacts,\\clear $8$–$20$ cycles,\\stable amplitude};
\node[poor,dotted] (f2) at (-3.2, 1.6) {Otherwise};

\node[box]  (c2) at (1.0,  1.6) {Artifact $\leq 6$\,s};
\node[poor,dotted] (c1) at (4.2,  1.6) {Artifact $> 6$\,s};

\node[box]  (d2) at (1.0, 0.0) {\#visible cycles\\$\geq 8$ and $\leq 20$};
\node[poor, dotted] (d1) at (4.2, 0.0) {\#visible cycles\\$< 8$ or $> 20$};

\node[poor, dotted] (e1) at (-2.0, -2) {Irregular fast\\amplitude/WSF\\changes};
\node[mod, dashed]  (e2) at (1.0, -2) {otherwise};
\node[good] (e3) at (4.2, -2) {No artifacts,\\regular amplitude\\and WSF changes};

\draw[->] (a)  -- (b1);
\draw[->] (a)  -- (b2);
\draw[->] (b1) -- (f1);
\draw[->] (b1) -- (f2);
\draw[->] (b2) -- (c1);
\draw[->] (b2) -- (c2);
\draw[->] (c2) -- (d1);
\draw[->] (c2) -- (d2);
\draw[->] (d2) -- (e1);
\draw[->] (d2) -- (e2);
\draw[->] (d2) -- (e3);

\node[font=\small\bfseries, anchor=west] at (-7.1,-0.1) {Expert label:};
\node[poor, dotted, minimum width=0.5cm, minimum height=0.32cm] (sw1) at (-6.8,-0.8) {};
\node[anchor=west] at (sw1.east) {\,Poor};
\node[mod, dashed,  minimum width=0.5cm, minimum height=0.32cm] (sw2) at (-6.8,-1.4) {};
\node[anchor=west] at (sw2.east) {\,Moderate};
\node[good, minimum width=0.5cm, minimum height=0.32cm] (sw3) at (-6.8,-2.0) {};
\node[anchor=west] at (sw3.east) {\,Good};
\end{tikzpicture}
\caption{The expert's labeling criteria. Terminal nodes are colored by the resulting label (see legend).  \label{figure: label rule and criteria}}
\end{figure}

We emphasize the typically low SNR associated with respiration-induced oscillation in the TAA signal. This motivates the use of relatively stringent thresholds, including a magnitude threshold of 10 and an artifact duration threshold of 6 s. The term ``visible cycle'' refers to a qualitative assessment of airflow-like cyclic patterns as determined by expert reviewers. ``Artifact'' refers to shock-like or noise-like patterns.
``Irregular amplitude changes from cycle to cycle'' denotes abrupt, non-predictable variations in amplitude between successive cycles, with differences exceeding 30\% in peak amplitudes of adjacent cycles. In contrast, ``irregular amplitude changes slowly'' refers to amplitude variations that may also exceed 30\% but evolve gradually and in a more structured manner across cycles, following an identifiable trend (e.g., progressively increasing or decreasing).
``Irregular WSF changes from cycle to cycle'' denotes rapid, non-patterned fluctuations between successive cycles. In contrast, ``WSF changes slowly'' refers to gradual cycle-to-cycle variation that follows a consistent progression, allowing an underlying trend to be identified. Whether fluctuations are non-patterned or trend-like is determined by expert assessment.
We further note that oscillatory signals characterized as having ``visible oscillatory cycles with irregular WSF change'' but slow amplitude variation can often be approximated by a fixed-WSF oscillation with higher-order harmonics contaminated by noise. 
To facilitate a consistent and practical labeling process, we adopt these qualitative criteria and rely on expert judgment for their interpretation, rather than enforcing rigid quantitative definitions in all cases. 

The labeling procedure in this study was performed by the authors H.-T. Wu, who has over 10 years of experience in respiratory signal analysis, and C.-W. Chen, who has over 5 years of industrial experience in sleep medicine. During labeling, the expert evaluated only the constructed TAA-derived respiratory signal within 1-minute windows. Both annotators were blinded to all other signals, metadata, and outcomes, and based the assessment solely on the TAA-resp signal. Labels were generated on a subject-by-subject basis and subsequently stored for further analysis.  
All annotations were reviewed for completeness and consistency prior to analysis. The label agreement between two annotators was analyzed using Cohen's kappa. All disagreed labels were discussed by the two annotators to determine the final label. The MATLAB-based interface, allowing visualization and manual annotation of each segment, is available at \url{https://github.com/hautiengwu2/TAA-resp.git}.

\subsection{Statistics}

To quantify whether TAA-resp better captures respiratory effort or airflow dynamics, we evaluate the ``synchronized correlation'' between TAA-resp and THO, ABD, and airflow. The synchronized correlation of two functions $f$ and $g$ over $[a,b]$ with window size $w>0$ is defined as 
\[
\texttt{corr}_s(f,g):=\max_{h\in[-w,w]} \frac{\int_{I(h)} f(x)g(x+h)dx}{\|f\|_{L^2(I(h))}\|g\|_{L^2(I(h))}}\,,
\]
where $I(h)=[\max(a,a+h), \min(b,b+h)]$. In this work, we set $b-a=60$ s and $w=5$ s, reflecting the fact that one breathing cycle is roughly $5$-seconds long. The synchronization step is necessary since the phases among TAA-resp and other channels are usually not consistent. 

To visualize and compare the statistical distributions across different groups, we generated violin plots using a kernel density estimator with a Gaussian kernel and a bandwidth of $h=0.02$.

For the application of RMI in a binary classification task, we applied receiver operating characteristic (ROC) analysis to learn the optimal decision threshold. Model performance was validated using a leave-one-subject-out cross-validation (LOSOCV) scheme. To address class imbalance, when the ratio between the two groups exceeded 3:1, the majority class was downsampled to achieve a 3:1 ratio. Performance was primarily evaluated in terms of sensitivity and specificity, with overall accuracy and F1 score also reported.
When any row or column of the confusion matrix contains only zeros, we assign the corresponding metric a value of 0 and report the number of such occurrences and results excluding these cases.

Reported values are means $\pm$ standard deviation (SD), unless otherwise stated. In all instances of hypothesis testing, $p < 0.05$ is considered statistically significant.

\section{Results}\label{section result}

For parameters used in calculating the TAA-resp  algorithm, the window length for the median filter is 10 s, which is chosen to remove trend oscillating slowly, and the bandpass filter is the 4th order Butterworth filter with spectral band 0.1--1 Hz. When we calculate RMI, the window length for SST  is 30 s, which is chosen to be sufficiently long to cover at least 6 breathing cycles in adults, the ridge extraction smoothness penalty $\lambda$ is set 0.05 based on previous experience, and the bandwidth for SST reconstruction is 0.05 Hz. 
While to our knowledge there does not exist any respiration extraction algorithm specifically designed for TAA installed on the fingertip, for the sake of completeness, we compare TAA-resp with various existing algorithms designed for extracting respiratory signals from a TAA sensor installed on the torso or wrist and report the results in Section \ref{section:method comparisons}.

\subsection{Visualization}

\begin{figure}[hbt!]
    \centering
\includegraphics[width=0.325\textwidth]{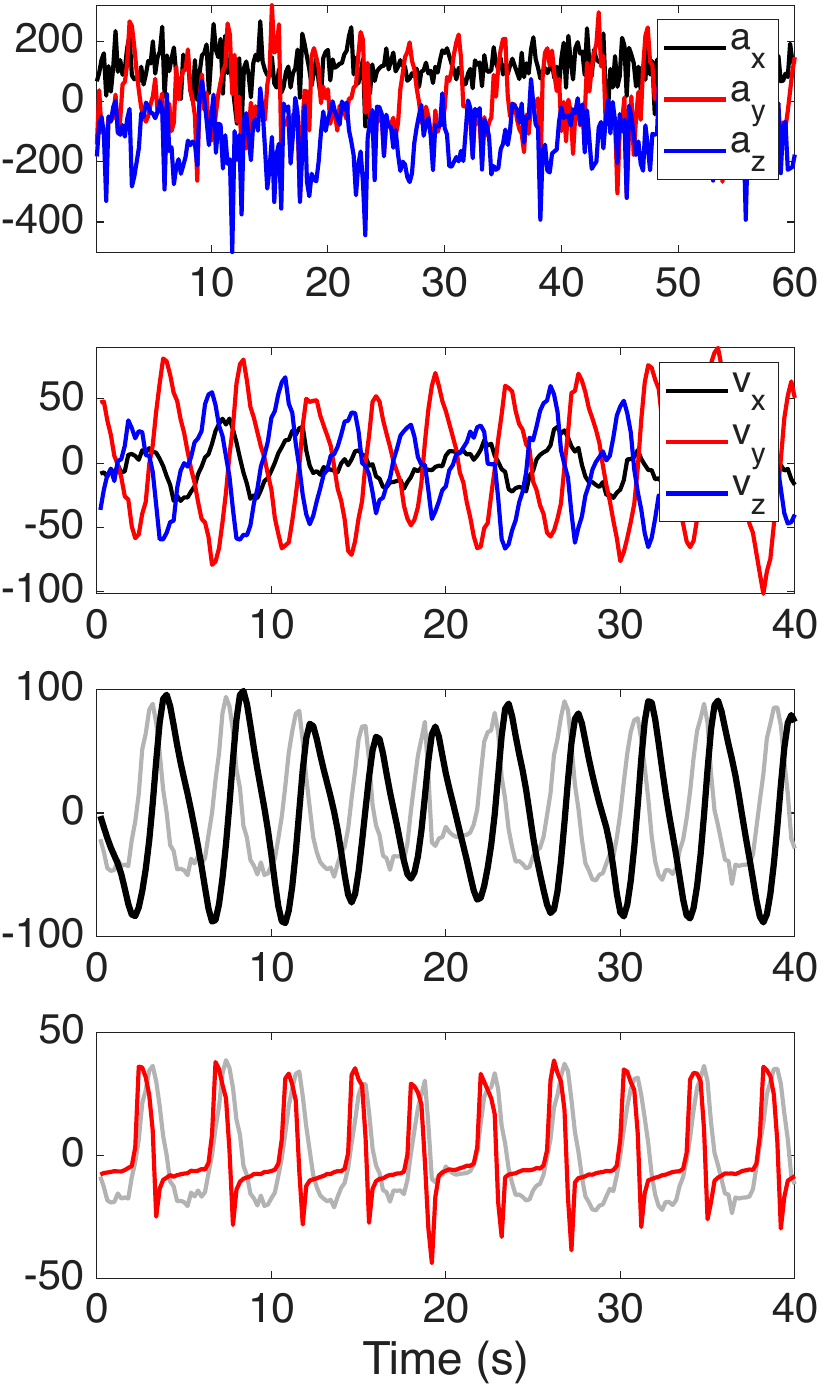}
\includegraphics[width=0.325\textwidth]{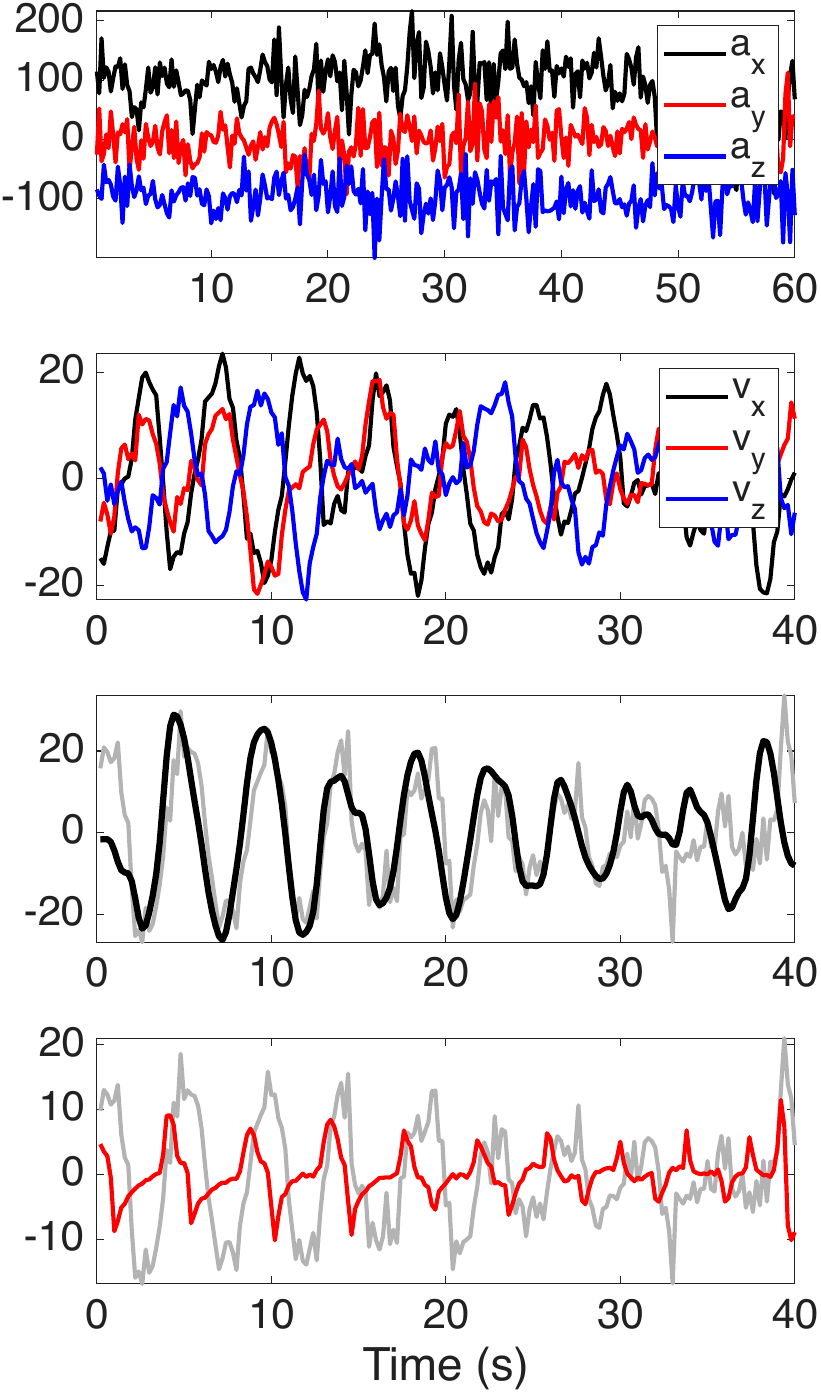}
\includegraphics[width=0.325\textwidth]{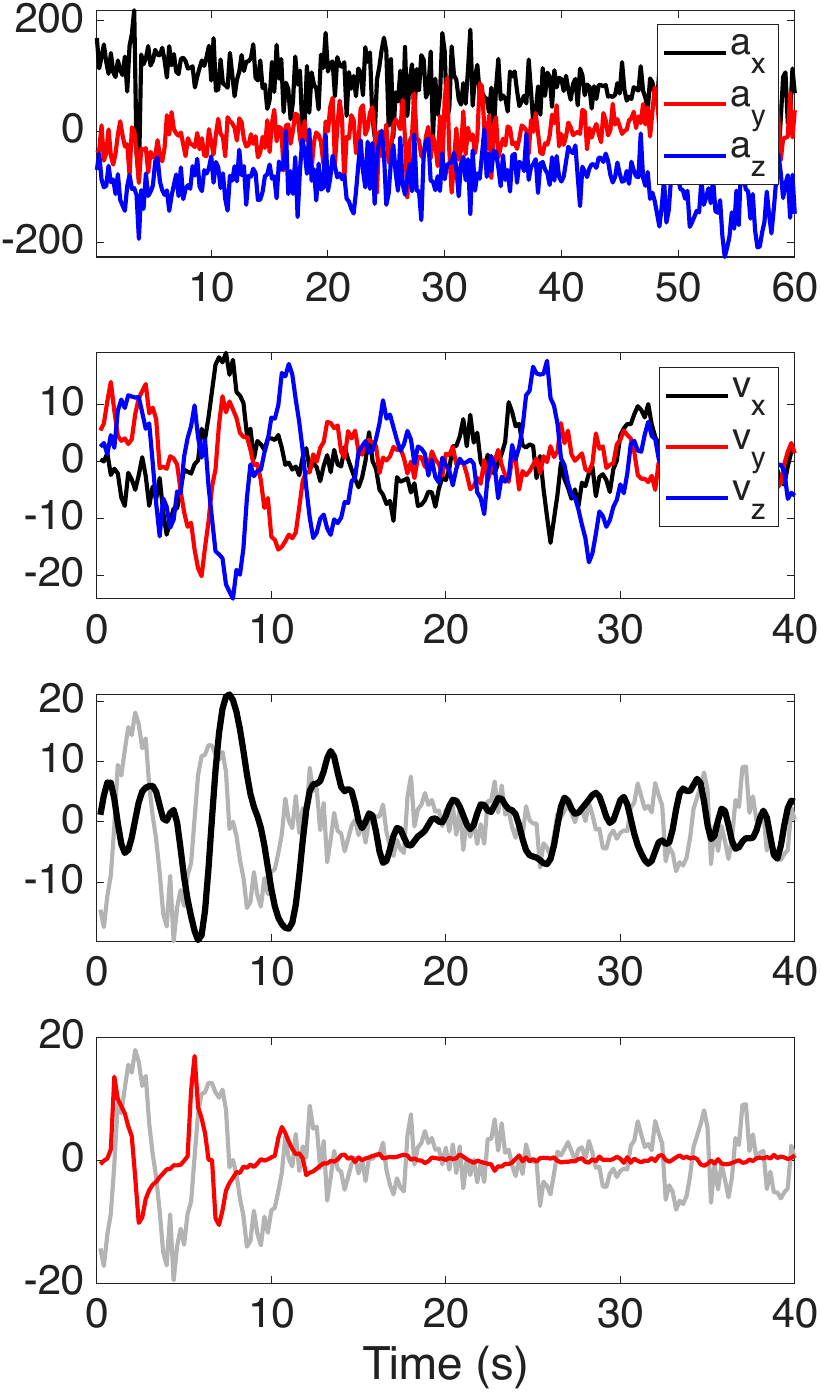}
    \caption{From left to right column: illustration of typical TAA-resp signals labeled as good (RMI=0.88), moderate (RMI=0.68), and poor (RMI=0.31). The top row shows the raw acceleration signals after removing the mean (unit: $m/s^2$), where $a_x$ is shifted above by $2\times \texttt{SD}(a_x)$ and $a_z$ is shifted down by $2\times \texttt{SD}(a_z)$ to enhance the visualization, the second row shows the integrated acceleration signals after removing the mean (unit: $m/s$). The third row shows the proposed TAA-resp  (black curve, arbitrary unit) superimposed with THO (gray curve, arbitrary unit), where THO is scaled to match TAA-resp for visualization purposes, and the bottom row shows the airflow signal (red curve, unit: $L/m$) superimposed with THO, where THO is scaled to match airflow for visualization purposes.
    \label{fig:IMUresp visualization}}
\end{figure}

\begin{figure}[hbt!]
    \centering
\includegraphics[width=0.8\textwidth]{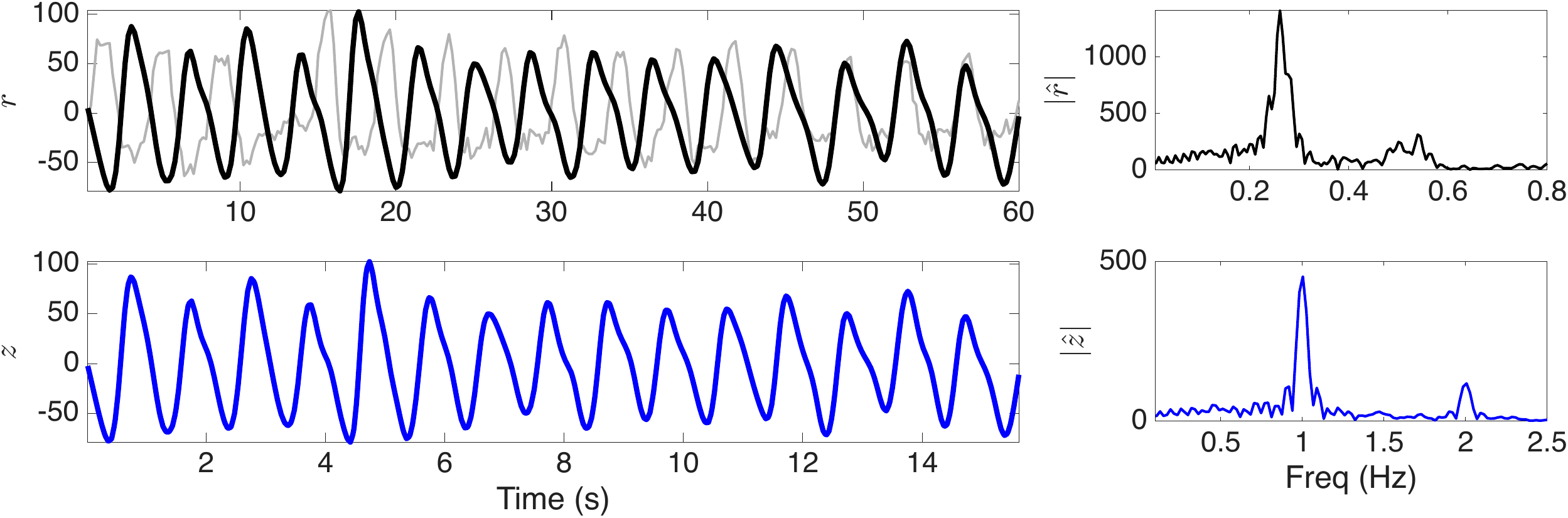}
\includegraphics[width=0.8\textwidth]{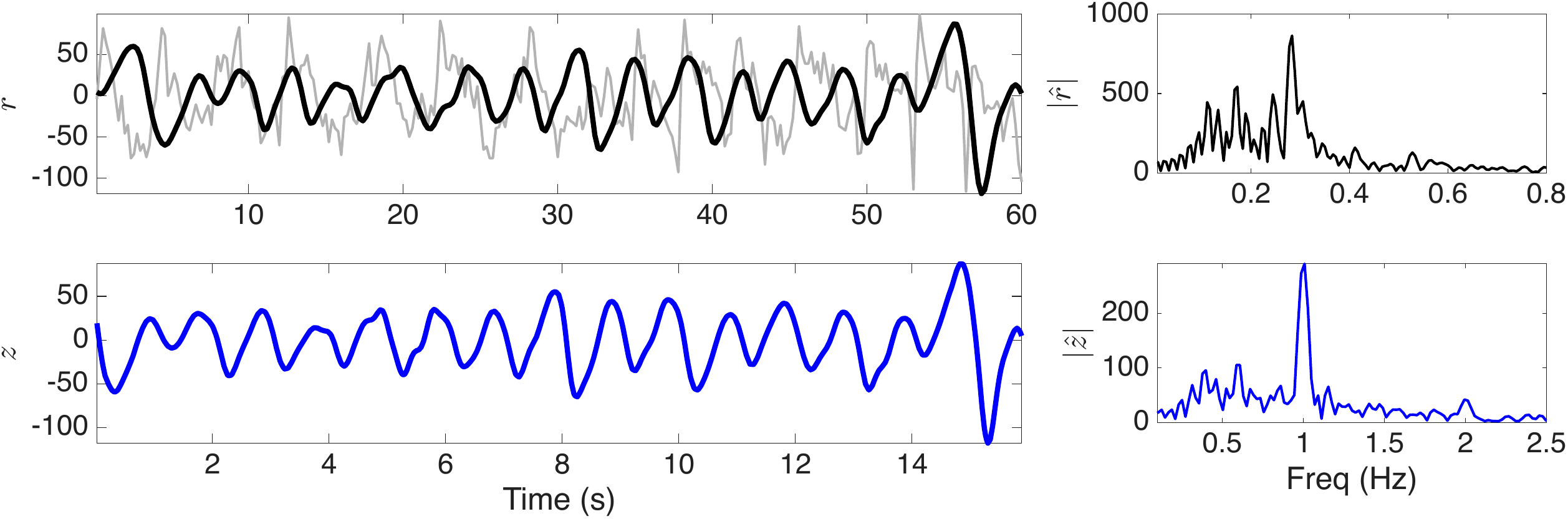}
\includegraphics[width=0.8\textwidth]{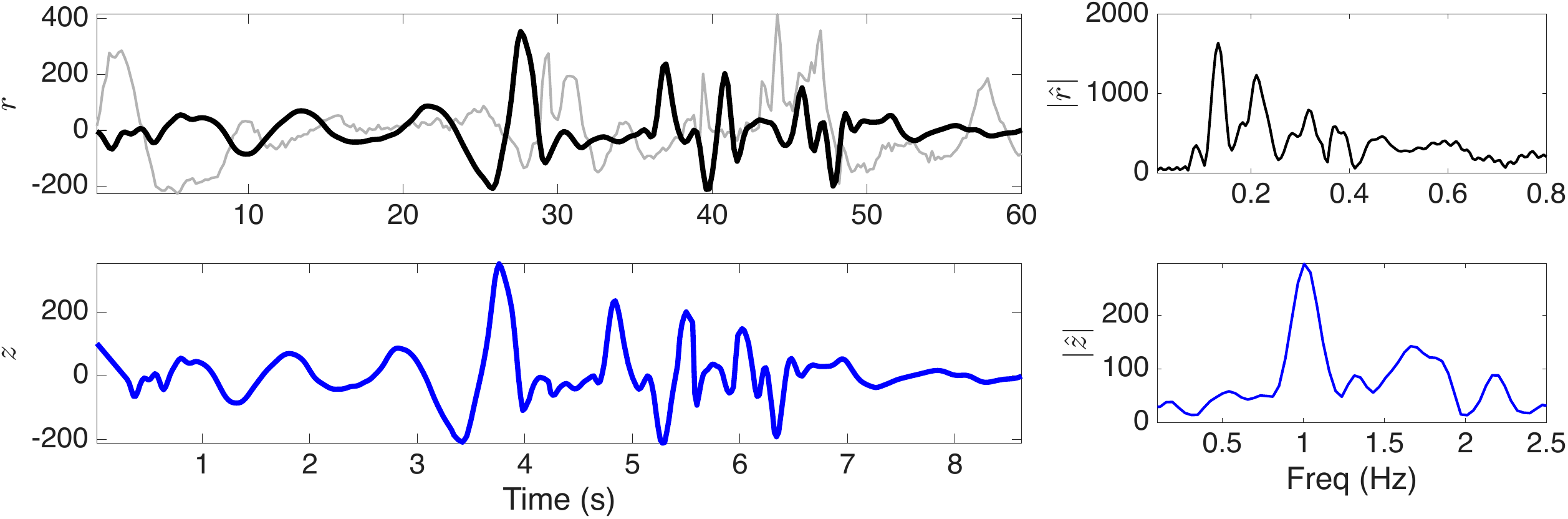}   
 \caption{Illustration of the effect of unwrapping and the corresponding spectrum. The top panel presents a TAA-resp segment labeled as high quality (RMI = 0.83), the middle panel presents a segment labeled as moderate quality (RMI = 0.64), whereas the bottom panel presents a segment labeled as poor quality (RMI = 0.44). The simultaneously recorded THOs are superimposed as gray curves.
    \label{fig:IMUresp visualization unwrapping}}
\end{figure}

We start with a visualization of TAA-resp under different labeled conditions. See Figure \ref{fig:IMUresp visualization}, where we show the first 40 seconds of a one-minute segment to enhance the visualization. It is clear that TAA-resp captures respiratory cycles in THO.  
Although oscillatory behavior can be observed in the raw TAA signal, it is substantially obscured by noise. After integration, the velocity signals exhibit clearer and more regular oscillations. For example, in the left panel, $a_x$ shows minimal oscillatory structure and $a_z$ displays noisy oscillations, whereas $v_x$ and $v_z$ exhibit pronounced periodic patterns. This contrast becomes even more evident in the middle panel. In the right panel, corresponding to an obstructive sleep apnea event, no clear regular oscillation is observed in the TAA signal, its integration, or TAA-resp. 
Note that the WSF of TAA-resp is not fixed, and THO does not necessarily lead TAA-resp in phase; this relationship depends on the subject and torso position. Figure \ref{fig:IMUresp visualization unwrapping} further illustrates TAA-resp before and after the unwrapping process. It is clear that before the unwrapping, the spectrum is spreading due to IRR, and the spectral energy becomes more concentrated after unwrapping.

\subsection{Experts' label}

Among the 39 subjects, 21 had their entire overnight recordings annotated by the first annotator (H.-T. Wu), while for the remaining subjects, one of every three consecutive segments was annotated. The second expert (C.-W. Chen) independently annotated one of every three consecutive segments from all subjects.

Across the overlapping $5,593$ segments in the VGHTPE database , the two annotators assigned different labels to 575 segments, including 22 segments with a discrepancy between moderate and good and 553 segments with a discrepancy between bad and moderate.
The two annotators demonstrated 94.8\% linear-weighted observed agreement across the three ordinal categories (good, moderate, and bad). Accounting for the ordinal structure of the ratings and agreement expected by chance, linear-weighted Cohen's $\kappa$ \cite{cohen1968weighted} was 0.539 (95\% CI, 0.505-0.572), which reaches moderate agreement according to \cite{landis1977measurement}.

Across the overlapping $1,524$ segments in the DUKE database, the two annotators assigned different labels to 125 segments, including 20 segments with a discrepancy between moderate and good, 104 segments with a discrepancy between bad and moderate, and 1 segment with a discrepancy between bad and good.
The two annotators demonstrated 95.87\% linear-weighted observed agreement across the three ordinal categories (good, moderate, and bad). Accounting for the ordinal structure of the ratings and agreement expected by chance, linear-weighted Cohen's $\kappa$ was 0.619 (95\% CI, 0.559-0.675), which reaches moderate agreement according to \cite{landis1977measurement}. 

See Section \ref{section disagreement segments} in the appendix for examples of disagreement. 
The disagreed labels were discussed and decided by both annotators.
Finally, among 5,863 one-minute segments in the VGHTPE database, $4,683$ were labeled as poor, 1,094 as moderate, and 86 as good. 
Among 1,512 one-minute segments in the DUKE database, 1,194 were labeled as poor, 295 as moderate, and 23 as good. 
All analyses below were carried out with these final labels.

\subsection{Distribution of quality labels}

\begin{figure}[hbt!]
    \centering
\includegraphics[width=0.6\textwidth]{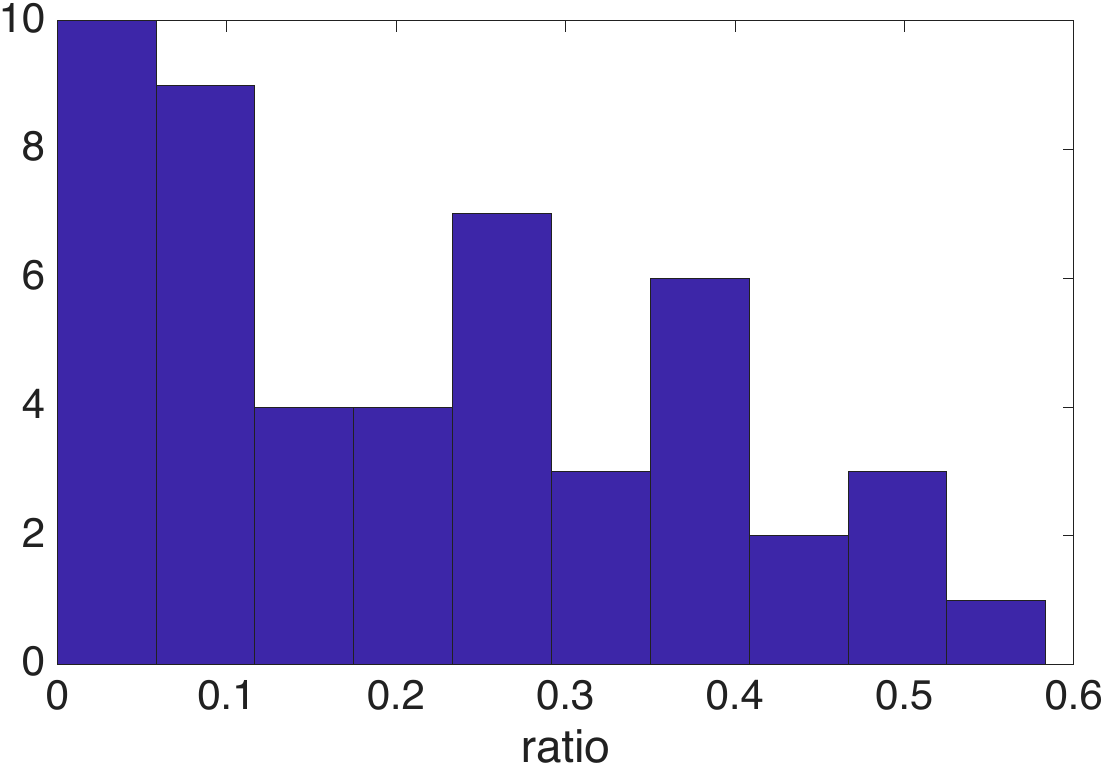}
    \caption{Ratio of TAA-resp segments labeled as high-quality (moderate or good) over 39 subjects.}
    \label{fig:GoodRatiohistogram}
\end{figure}

Figure \ref{fig:GoodRatiohistogram} shows the distribution, across subjects, of the proportion of segments labeled as high-quality. The mean$\pm$SD and median$\pm$MAD of this proportion are $21.25\pm 16.11$\% and $18.99\pm 13.69$\%, with a maximum of 58.33\% and a minimum of 0\%. Notably, for some subjects, at least one-third of the night exhibits high-quality TAA-resp. 

\subsection{Relationship between TAA-resp and airflow, THO, and ABD}

\begin{figure}[hbt!]
    \centering
\includegraphics[width=0.6\textwidth]{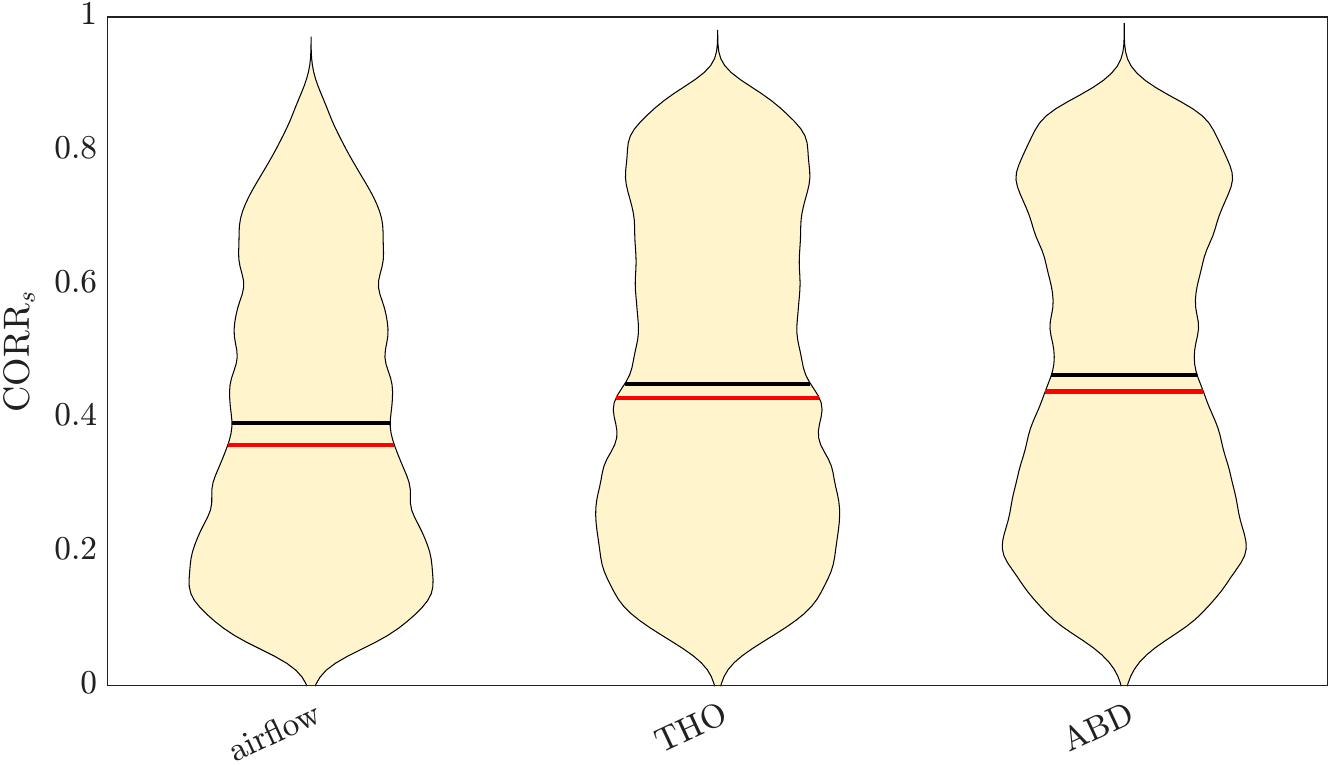}
\includegraphics[width=0.6\textwidth]{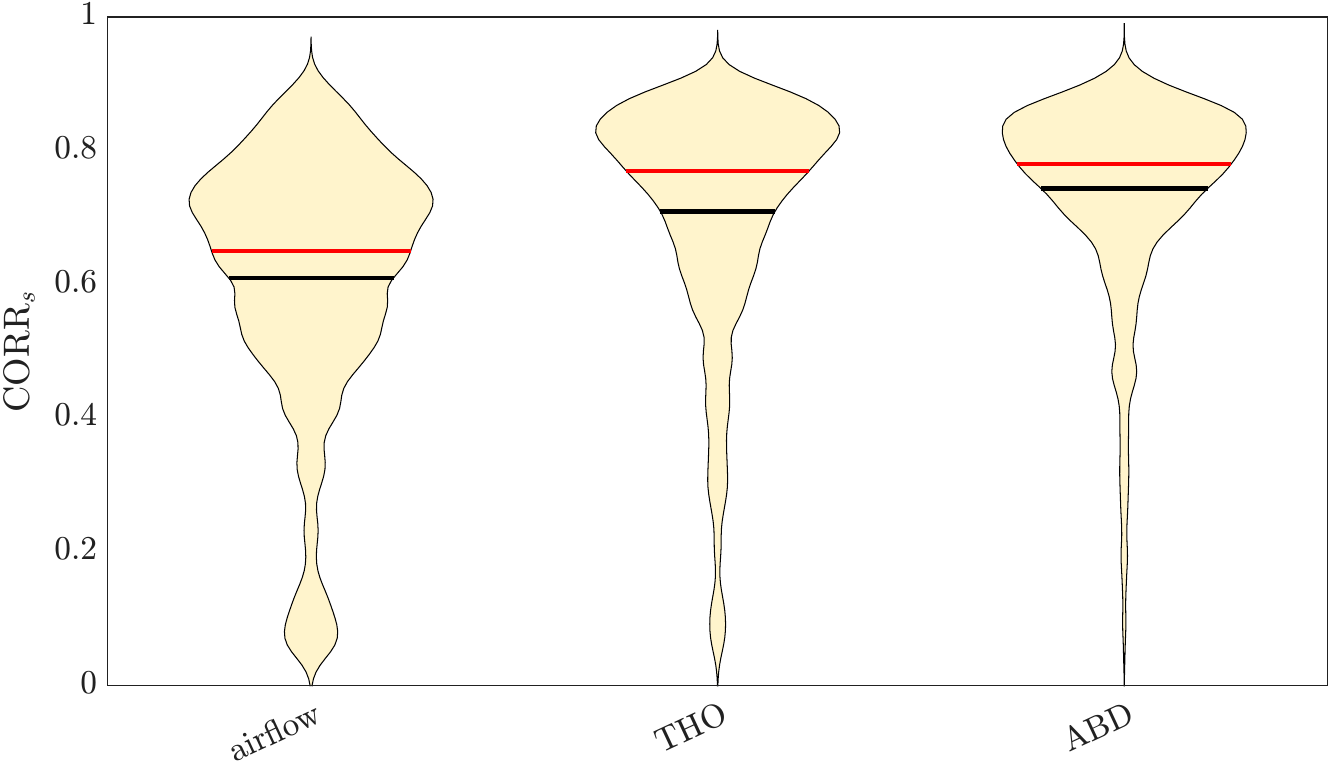}
    \caption{Distributions of correlations between TAA-resp and THO, ABD, and airflow over all segments (top) and those labeled as high-quality (moderate or good). The mean and median of each group are marked as red and black vertical lines.}
    \label{fig:REIvsCORR}
\end{figure}

Distributions of synchronized correlations between TAA-resp and THO, ABD, and airflow over all segments and segments labeled as high-quality are shown in Fig. \ref{fig:REIvsCORR}. 
Across all segments, the synchronized correlation (mean $\pm$ SD) with THO, ABD, and airflow was $0.46 \pm 0.24$, $0.45 \pm 0.24$, and $0.39 \pm 0.21$, respectively; corresponding median $\pm$ MAD values were $0.43 \pm 0.21$, $0.42 \pm 0.20$, and $0.36 \pm 0.18$. 
Over high-quality segments, the synchronized correlations are $0.72 \pm 0.14$, $0.76 \pm 0.16$, and $0.64 \pm 0.17$ (mean $\pm$ SD), with median $\pm$ MAD of $0.79 \pm 0.10$, $0.76 \pm 0.14$, and $0.65 \pm 0.12$ for THO, ABD, and airflow, respectively.
In both cases, paired one-sided Wilcoxon signed-rank tests with Bonferroni correction demonstrate that, within these segments, TAA-resp exhibits significantly higher synchronized correlation with ABD and THO compared with airflow (adjusted $p < 10^{-10}$). These findings suggest that TAA-resp predominantly captures respiratory effort, consistent with established physiological understanding.

\begin{figure}[hbt!]
    \centering
\includegraphics[width=\textwidth]{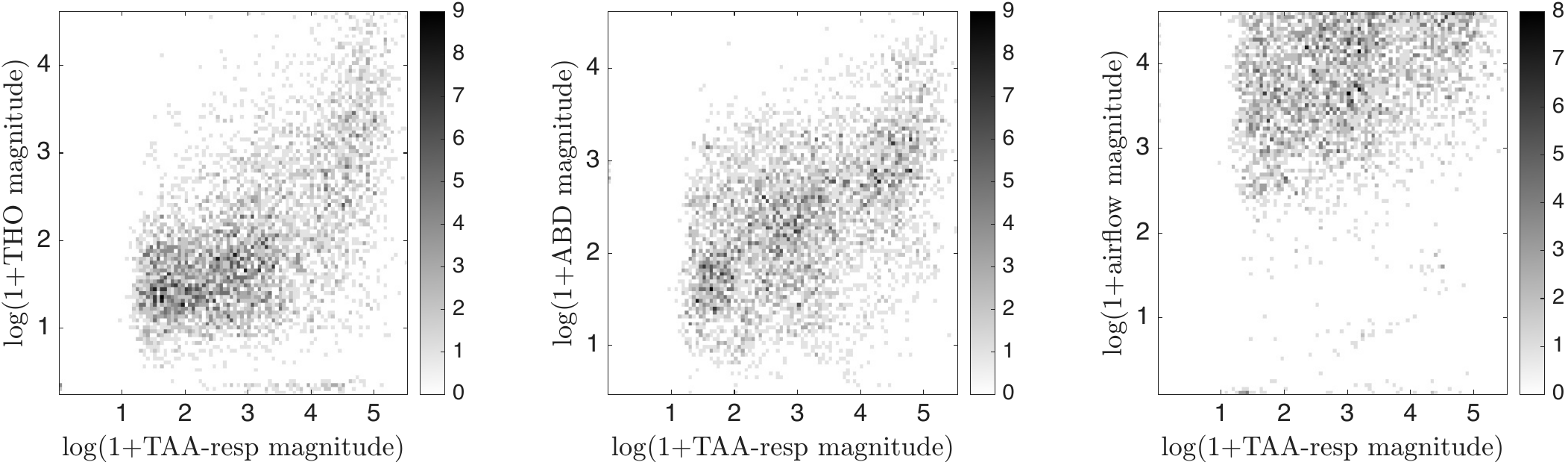}
\includegraphics[width=\textwidth]{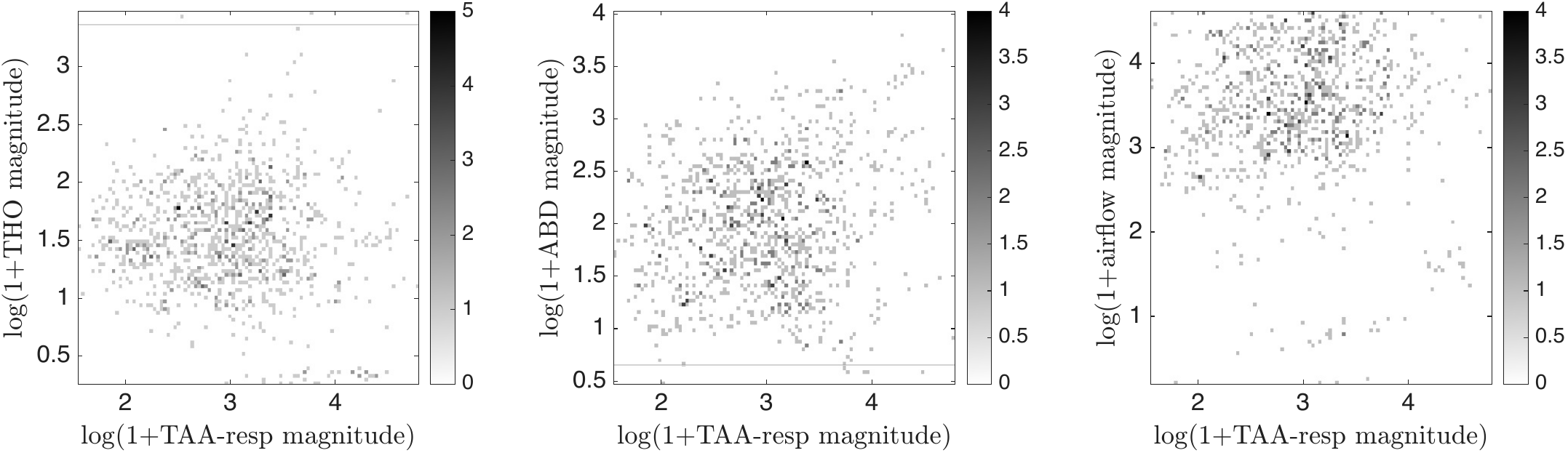}
    \caption{Two-dimensional histograms of $\log(1+$magnitude of TAA-resp$)$ versus $\log(1+$magnitude of THO$)$ (left), $\log(1+$magnitude of ABD$)$ (middle), and $\log(1+$magnitude of airflow$)$ (right), shown for all segments (top) and high-quality segments (bottom).}
    \label{fig:amplitudeRelationship}
\end{figure}

Note that the unit of TAA-resp is arbitrary. We define the {\em magnitude} of a signal over a 1-minute segment as the 99\% quantile of its absolute value. The relationships between TAA-resp magnitudes and the corresponding magnitudes of THO, ABD, and airflow over all segments and those labeled as high-quality are illustrated as scatter plots in Fig. \ref{fig:amplitudeRelationship}, all in log scale. Since VGHTPE and DUKE databases are collected from different PSGs and their airflow, THO, and ABD are not comparable, here we only report results on the VGHTPE database. When all segments are considered, a clear nonlinear relationship is observed between TAA-resp and THO magnitudes; this relationship is less pronounced for ABD and becomes diffuse for airflow. To quantify this nonlinear relationship, we apply the Akaike information criterion (AIC) to select the optimal polynomial order when fitting $\log(1+$magnitude of TAA-resp$)$ against $\log(1+$magnitude of THO$)$, $\log(1+$magnitude of ABD$)$, and $\log(1+$magnitude of airflow$)$ over all segments. The maximal order under consideration is 3. 
The optimal polynomial order against THO (ABD and airflow resp.) is 3 (3 and 2 resp.), with adjusted $R^2=0.38$ (0.25 and 0.10 resp.) and root mean squared error (RMSE) 0.72 (0.66 and 0.73 resp.). This result provides quantitative evidence for the above finding that TAA-resp predominantly captures respiratory effort. However, such a relationship does not exist when we only focus on the high-quality segments.

\subsection{How accurately can we estimate IRR from high-quality TAA-resp?}

\begin{figure}[hbt!]
    \centering
\includegraphics[width=\textwidth]{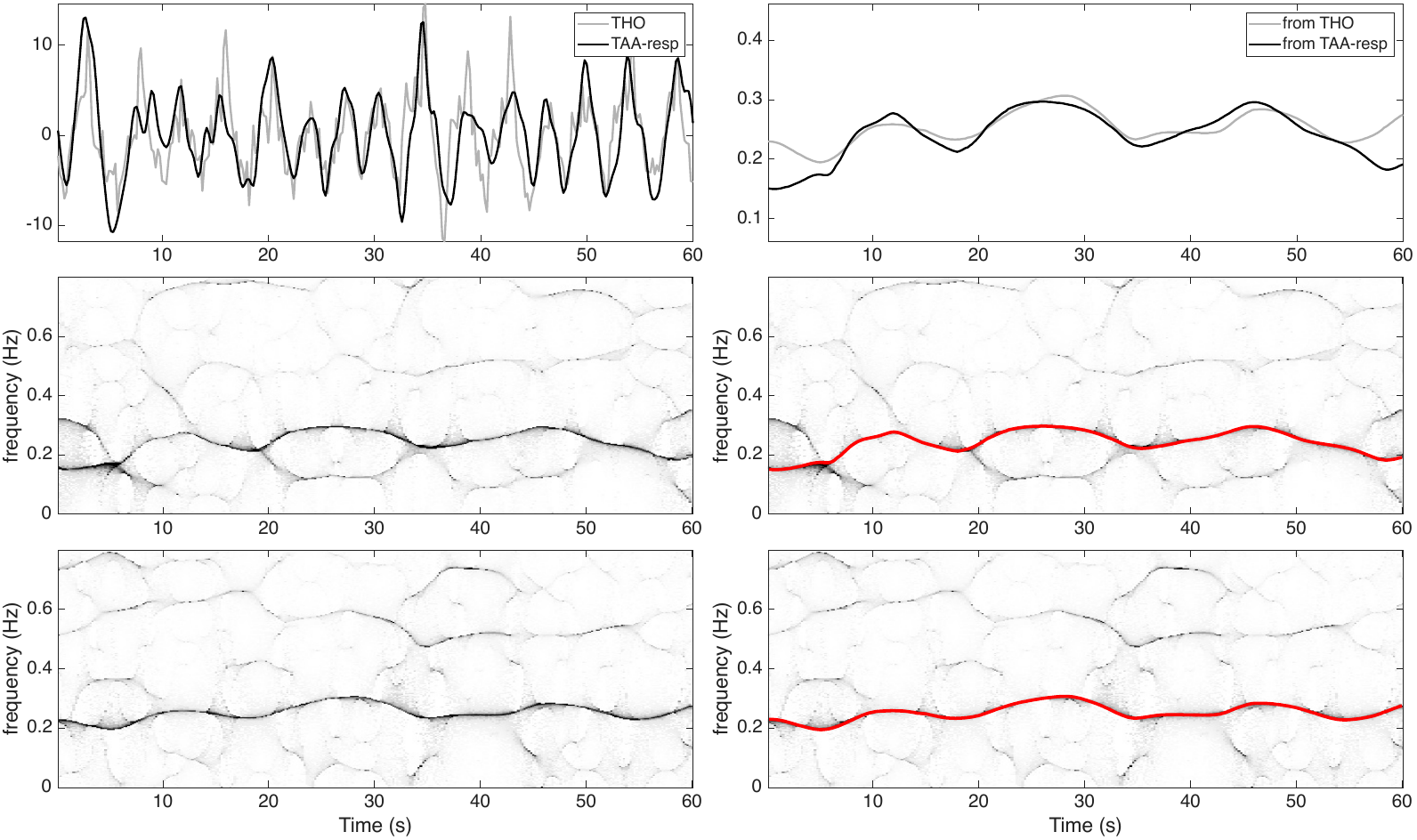}
    \caption{TAA-resp and THO are shown in the top left panel, and the associated TFRs are shown in the middle and bottom rows. The THO is scaled to have the same magnitude as TAA-resp to enhance the visualization. The estimated IRRs are shown in the top right panel and superimposed in each TFR as red curves. The RMSE is 0.02.}
    \label{fig:TFRexampleModerate}
\end{figure}

\begin{figure}[hbt!]
    \centering
\includegraphics[width=\textwidth]{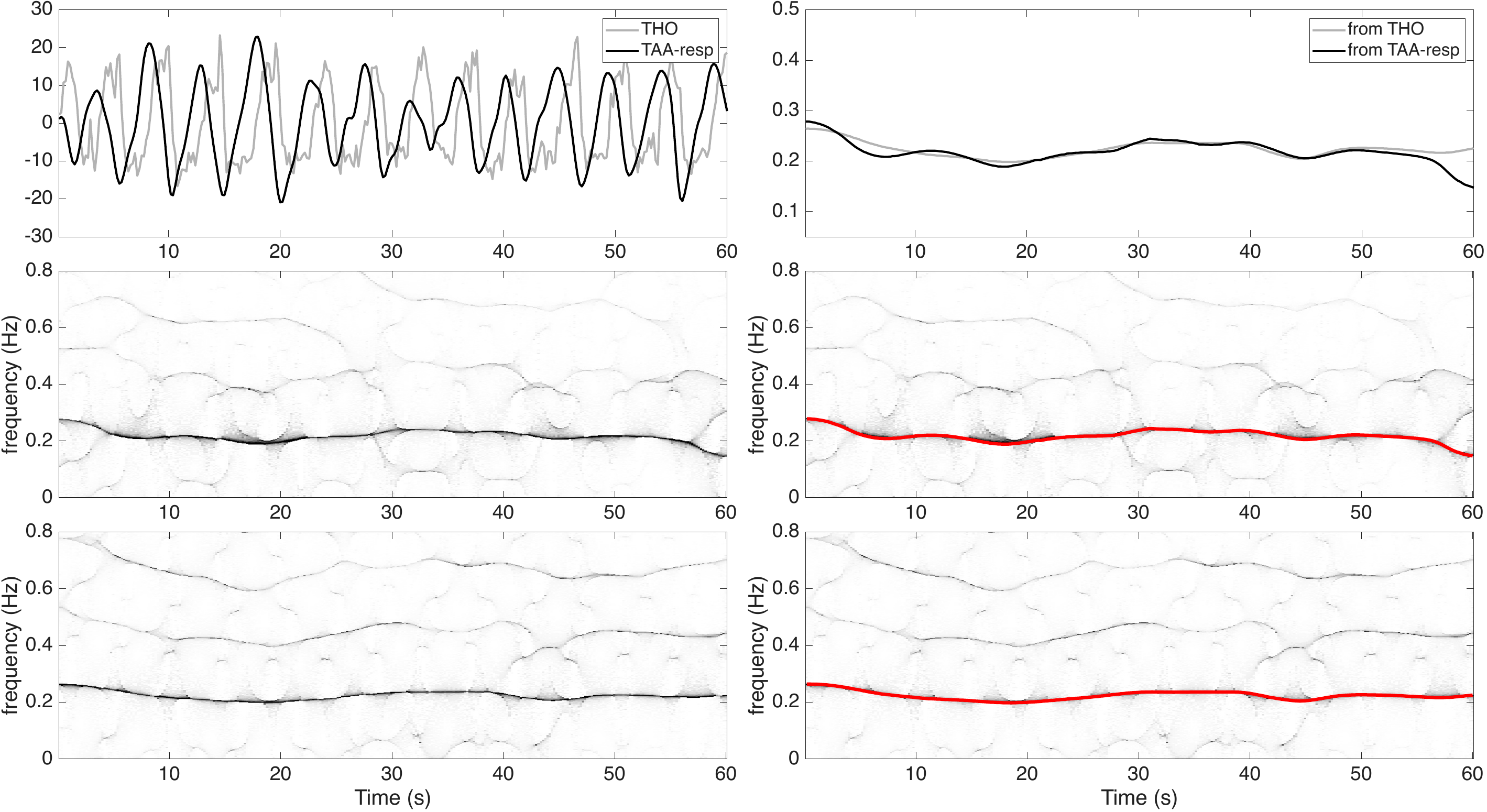}
    \caption{TAA-resp and THO are shown in the top left panel, and the associated TFRs are shown in the middle and bottom rows. The THO is scaled to have the same magnitude as TAA-resp to enhance the visualization. The estimated IRRs are shown in the top right panel and superimposed in each TFR. The RMSE is 0.015.}
    \label{fig:TFRexampleGood}
\end{figure}

We evaluate the accuracy of IRR estimation from high-quality TAA-resp signals, using the IRR derived from THO signals via SST as the ground truth. See Figure \ref{fig:TFRexampleModerate} for an illustration when the TAA-resp is labeled as moderate. It is clear that the respiratory rate changes from time to time, and the IRR estimated from TAA-resp can accurately approximate that of THO, with the root mean square error (RMSE) of 0.02 Hz. See Figure \ref{fig:TFRexampleGood} for an illustration when the TAA-resp is labeled as good, where the RMSE is 0.015 Hz.

The root mean square error (RMSE) of IRR over 1-minute segments is $0.027 \pm  0.022$ Hz across all high-quality segments, $0.026  \pm  0.019$ Hz for high-quality segments labeled REM, N2, or N3, $0.029  \pm  0.026$ Hz for high-quality segments labeled wake or N1, $0.027 \pm 0.022$ Hz for high-quality event-free segments, and $0.026 \pm 0.023$Hz  for high-quality apnea or hypopnea segments. The RMSE for high-quality segments labeled REM, N2, or N3 was significantly lower than that for segments labeled wake or N1 ($p = 0.02$, Wilcoxon rank-sum test with Bonferroni correction). No significant difference was observed between event-free and apnea/hypopnea segments ($p = 0.21$).

The average respiratory rate (RR) error over each 1-minute segment is $0.007 \pm   0.019$ Hz across all high-quality segments, $0.006 \pm   0.015$ Hz for high-quality segments labeled REM, N2, or N3, $0.008 \pm   0.023$ Hz for high-quality segments labeled wake or N1, and $ 0.007 \pm 0.018$ Hz for high-quality event-free segments. No significant difference was observed between segments labeled REM, N2, or N3 and those labeled wake or N1 ($p=0.19$) and between event-free and apnea/hypopnea segments ($p = 0.75$).

\subsection{High-quality TAA-resp versus apnea events}
Among 1,498 1-min segments with high-quality labels, 67 and 24 segments were labeled as hypopnea and apnea, respectively, corresponding to 16.2\% and 2.7\% of all hypopnea- and apnea-labeled segments. The Chi-square test shows that the proportions of segments are significantly different with $p < 10^{-10}$.
The Fisher’s exact test shows that the proportion of segments with hypopnea is significantly higher than that of apnea with the odds ratio 6.85 (95\% CI=[4.23, 11.11]) and $p <10^{-5}$. This result suggests that high-quality TAA-resp more often appears during event-free segments.

\subsection{High-quality TAA-resp versus sleep stage}
Among 1,498 1-min segments with high-quality labels, 116, 189, 59, 1,042, and 92 segments were labeled as wake, REM, N1, N2, and N3, respectively, corresponding to 6.8\%, 18.4\%, 6.3\%, 29.9\%, and 43.2\% of all wake-, REM-, N1-, N2-, and N3 labeled segments. 
The Chi-square test shows that the proportions of segments are significantly different with $p<10^{-10}$. 
The post-hoc comparisons with Bonferroni correction using two-proportion tests show that the proportion of segments with N3 is not significantly higher than that of N2 with $p=0.06$ and significantly higher than that of wake, REM, and N1 labels with $p<10^{-10}$, $2.3\times 10^{-8}$, and $p<10^{-10}$.

\subsection{Summary statistics of RMI}
\begin{figure}[hbt!]
    \centering
\includegraphics[width=0.6\textwidth]{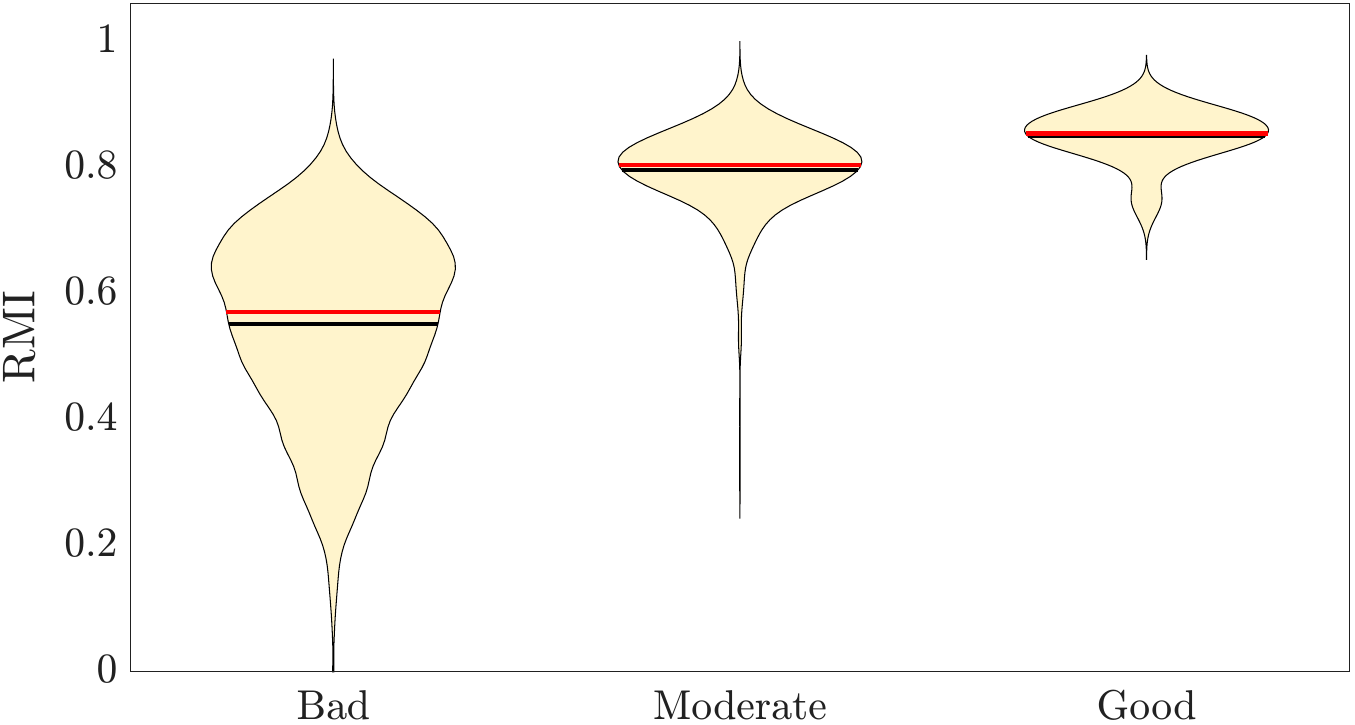}
    \caption{Distribution of RMI over three groups. The mean and median of each group are marked as red and black vertical lines.}
    \label{fig:REIdistribution}
\end{figure}

Figure \ref{fig:REIdistribution} shows violin plots of the RMI distributions for these groups. The mean $\pm$ standard deviation of the good, moderate, and poor groups are $0.85\pm 0.04$, $0.80\pm 0.06$, and $0.55\pm 0.15$. 
The median $\pm$ median absolute deviation of the good, moderate, and poor groups are $0.86\pm 0.03$, $0.81\pm0.05$, and $0.57\pm0.12$. 
Paired one-sided Wilcoxon signed-rank tests with Bonferroni correction indicate that RMI is significantly higher in the good group than in the moderate group, and higher in the moderate group than in the poor group (adjusted $p < 10^{-5}$).

\subsection{RMI distribution over different sleep stages}

\begin{figure}[hbt!]
    \centering
\includegraphics[width=\textwidth]{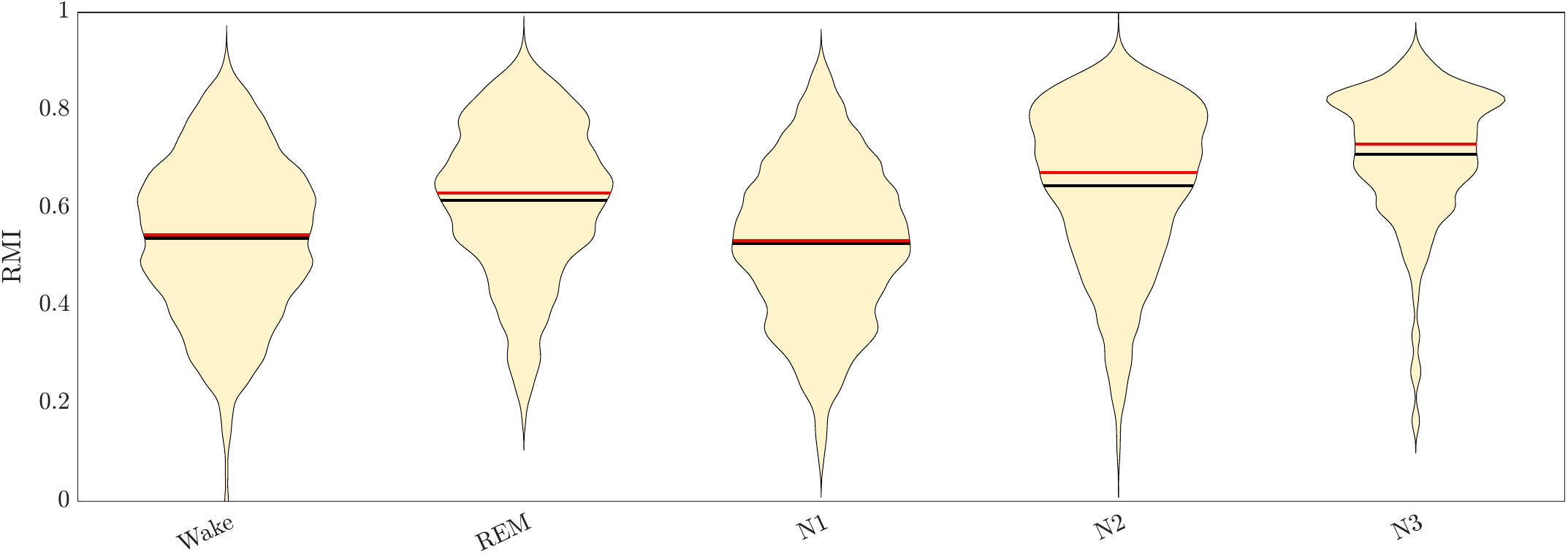}
    \caption{Violin plots of RMI over different sleep stages. The mean and median of each group are marked as red and black vertical lines.}
    \label{fig:REIvsSleepStage}
\end{figure}

Figure \ref{fig:REIvsSleepStage} shows violin plots of RMI distributions across sleep stages. We see that the median RMI decreases from N3, N2, REM, N1, to wake. 
The mean$\pm$ standard deviation (median$\pm$median absolute deviation) of RMI over wake, REM, N1, N2, and N3 are $0.54\pm 0.16$ (0.55$\pm$0.13), $0.62\pm 0.15$ ($0.63\pm 0.12$), $0.53\pm 0.16$ ($0.53\pm 0.13$), $0.65\pm 0.16$ ($0.67\pm 0.13$), and $0.71\pm 0.13$ (0.73$\pm$0.10).

Among 1,716 (1,025, 940, 3,480, and 213 respectively) segments labeled as wake (REM, N1, N2, and N3 respectively), there are 116 (189, 59, 1,042, and 92 respectively) segments labeled high-quality, or equivalently 6.76\% (18.44\%, 6.28\%, 29.94\%, and 43.19\% respectively). 
We run the Kruskal-Wallis test and show that RMIs are significantly different across sleep stages ($p<10^{-10}$). To identify which stage differed, we performed post-hoc pairwise comparisons of mean ranks using the Dunn test with Bonferroni correction for multiple comparisons. The results showed that the median RMI during N3 was significantly higher than others; during N2, the median RMI was significantly higher than wake, REM, and N1 and lower than N3; 
during N1, the median RMI was significantly lower than REM, N2, and N3, and higher than wake; 
during REM, the median RMI was significantly lower than in N2 and N3 and higher than in wake and N1; 
and during wake, the median RMI was significantly lower than others except N1. This result overall coincides with our physiological knowledge. More details can be found in the Discussion section.

\subsection{RMI distribution over different apnea events}

Next, we study the relationship between labels, RMI, and apnea events. Among the 7,375 1-minute segments with respiratory event labeling, 6,085 had neither hypopnea nor apnea, 414 had hypopnea, and 876 had apnea. In segments without hypopnea or apnea, 4,678, 1,298, and 109 were labeled as poor, moderate, and good, respectively (23.12\% high-quality). In segments with hypopnea, 347, 67, and 0 were labeled as poor, moderate, and good (16.18\% high-quality). In segments with apnea, 852, 24, and 0 were labeled as poor, moderate, and good (2.74\% high-quality). These statistics suggest that a larger ratio of segments are labeled as high-quality when there is no hypopnea or apnea.

\begin{figure}[hbt!]
    \centering
\includegraphics[width=0.6\textwidth]{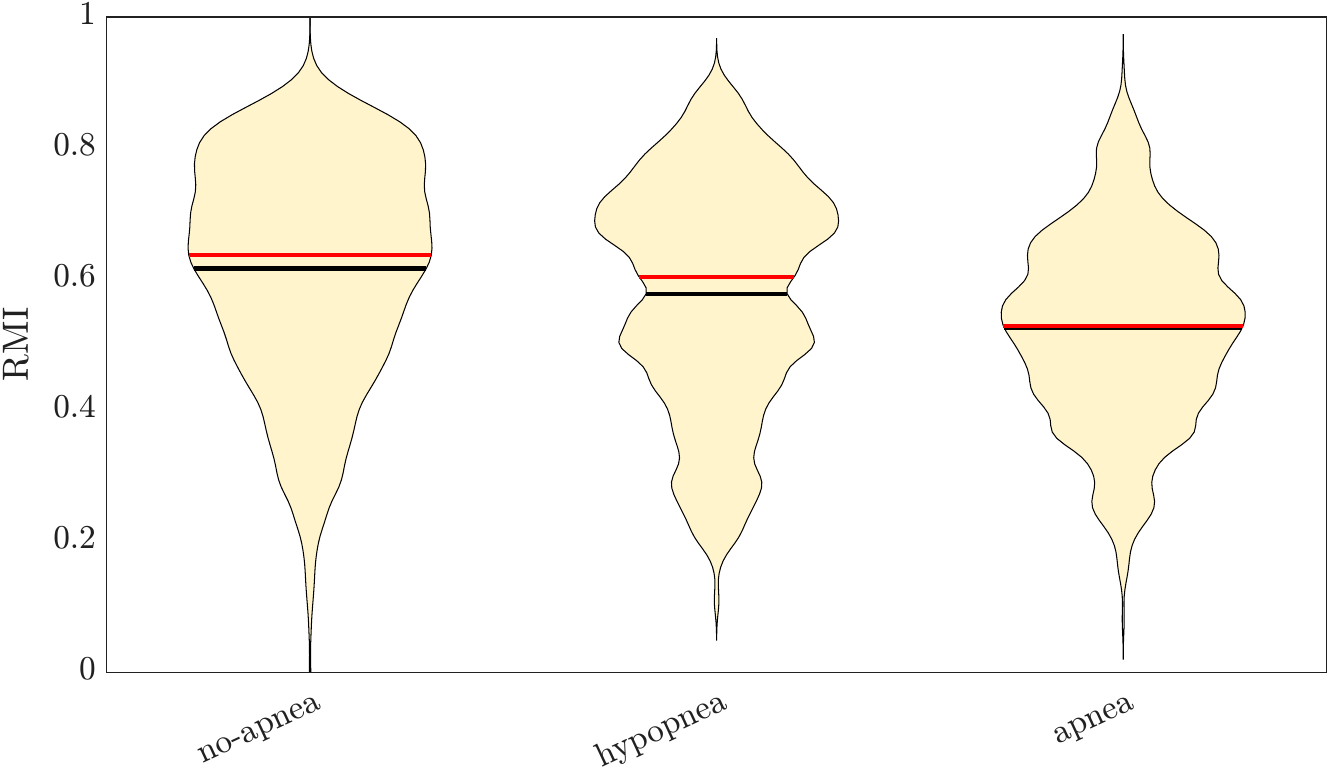}
    \caption{RMI distributions for segments with and without sleep apnea. The mean and median of each group are marked as red and black vertical lines.}
    \label{fig:REIvsApnea}
\end{figure}

Figure \ref{fig:REIvsApnea} shows RMI distributions for event-free segments. Visually, the median RMI was slightly higher in event-free segments than in those with hypopnea or apnea. 
The mean$\pm$standard deviation (median$\pm$median absolute deviation) of RMI over event-free segments, segments with hypopnea, and segments with apnea are $0.62\pm 0.17$ (0.64$\pm$0.14), $0.58\pm 0.17$ ($0.60\pm 0.15$), and $0.53\pm 0.14$ (0.53$\pm$0.12).

A Kruskal-Wallis test confirmed significant differences across these conditions ($p< 10^{-10}$). Post-hoc pairwise comparisons of mean ranks using the Dunn test with Bonferroni correction revealed that event-free segments had a significantly higher median RMI than those with hypopnea; segments with hypopnea also had a significantly higher median RMI than those with apnea. This result suggests the potential of using RMI to distinguish event-free segments from others when respiration-induced oscillation is detected. 

\subsection{RMI as a predictor}
For practical application, we evaluated how well RMI aligns with expert labels. LOSOCV with ROC analysis gives sensitivity and specificity of 0.84$\pm$0.21 and 0.88$\pm$0.08, with overall accuracy 0.89$\pm$0.05 and F1 0.67$\pm$0.24 if we replace NaN quantities associated with either row or column of the confusion matrix containing only zeros by 0. If we remove 2 cases with any row or column of the confusion matrix containing only zeros, sensitivity and specificity are 0.88$\pm$0.12 and 0.88$\pm$0.08, with overall accuracy 0.89$\pm$0.05 and F1 0.70$\pm$0.20. These results suggest that RMI, despite being a scalar metric, captures meaningful information on the quality of TAA-resp and might be a useful preliminary quality indicator. 

We further consider the potential of predicting event-free segments using LOSOCV. To balance the group size, we reduce the event-free segments by 3 in the training dataset to match the size of segments with apnea or hypopnea.  
LOSOCV with ROC analysis gives sensitivity and specificity of 0.49$\pm$0.19 and 0.54$\pm$0.32, with overall accuracy 0.53$\pm$0.16 and F1 0.61$\pm$0.17 if we replace NaN quantities associated with either row or column of the confusion matrix containing only zeros by 0. If we remove 5 cases with any row or column of the confusion matrix containing only zeros, sensitivity and specificity are 0.49$\pm$0.19 and 0.60$\pm$0.27, with overall accuracy 0.52$\pm$0.16 and F1 0.60$\pm$0.18.
These results are considered negative from the perspective of prediction. See below for more discussion.

\section{Discussion and conclusion} \label{section discussion}

We demonstrate that body motion captured by a TAA placed on the fingertip, a location not previously studied, encodes respiratory information intermittently.
We propose an algorithm involving nonlinear transformation and optimal projection to derive TAA-resp from the TAA signal. Second, we introduce a quantitative index, RMI, to assess the quality, or strength, of respiration-induced oscillation. 
Using a dataset comprising 39 full-night recordings, we characterize several key properties of TAA-resp. 
We found that over all segments, it has a higher synchronized correlation with ABD or THO than with airflow (Figure \ref{fig:REIvsCORR}), and its amplitude is more strongly and consistently correlated with THO or ABD motion than with airflow (Figure \ref{fig:amplitudeRelationship}). However, although the synchronized correlation of TAA-resp signal quality with THO and ABD signals is significantly stronger than with airflow in high-quality segments, this relationship is not preserved when signal amplitudes are compared. This observation supports our claim that TAA-resp predominantly reflects respiratory effort rather than airflow, but the magnitude of high-quality TAA-resp cannot be reliably used to infer respiratory effort.
Respiration-induced oscillation is detectable for 21.2\% of total sleep time on average, reaching up to 58\% in some cases. High-quality TAA-resp can accurately estimate IRR with RMSE around 0.027 Hz. RMI is higher for high-quality segments and lower for poor-quality segments, and its distribution aligns with physiology, with higher values during REM, N2, and N3 sleep and in the absence of apnea or hypopnea events. RMI can also be applied to predict expert quality labels with 84\% sensitivity and 88\% specificity, which is based on a LOSOCV. 

We shall first discuss the proposed labeling criteria in Section \ref{sec:label procedure}. Due to the recycling nature of this work, we do not claim optimality of either the labeling procedure or the criteria employed. It was designed purposely to be stringent so that we can confirm the existence of recyclable information.
While we do not pursue such standardization here, the development of a more reliable and widely accepted labeling framework would likely require either large-scale studies involving multiple expert annotators or, more in line with established clinical practice, a process combining expert consensus, systematic literature review, and iterative revision, as exemplified by the established apnea event scoring guidelines of the American Academy of Sleep Medicine (AASM). Moreover, from those segments shown in Figures \ref{fig:disagreement 0 and 1} and \ref{fig:disagreement 2 and 1} in the Appendix, we see that the label discrepancy between the two annotators comes from the subjective justification of ``irregularity'', ``change fast'', or other qualitative descriptions, and these discrepancies usually occur when the magnitude is less than 10. Due to the nature of weak respiratory information in the fingertip, we could expect a more unstable signal, which lowers the agreement rate. 
 
TAA-resp is not intended to replace direct measurements of respiratory dynamics or those derived from TAA recordings acquired on the torso \cite{jin2009performance,bates2010respiratory,bucklin2010inexpensive,liu2011estimation,mann2011simultaneous,dehkordi2011validation,fekr2014tidal,hafezi2020sleep}. Rather, it provides a means of {\em recycling} otherwise underutilized information contained in signals already acquired by commonly used home-care devices. 
The relatively low proportion of high-quality segments reflects an inherent physical limitation of this opportunistically repurposed sensing modality. Although this limitation may constrain its standalone utility, the approach may nevertheless be valuable in sensor-limited settings, such as level 4 home sleep apnea testing \cite{collop2007clinical}, where the number of available channels is minimal and extracting complementary information from each signal is essential. In this setting, every bit counts.
When respiration-induced oscillation is reliably captured, TAA-resp may serve as a complementary feature for downstream analysis. 
For instance, in patients undergoing continuous positive airway pressure (CPAP) therapy, airflow signals acquired during device use enable estimation of the residual apnea-hypopnea index (AHI) for clinical assessment. However, during periods when CPAP is detached, respiratory information recovered from TAA-resp, when of sufficient quality, may act as a surrogate measure and thereby support clinical decision-making.
Moreover, it can be integrated with other physiological signals or derived features to improve the performance of sleep apnea detection or prediction models. Such integration may enhance robustness, especially in environments where traditional respiratory measurements are unavailable or of low quality. 
We leave a systematic exploration of these applications to future work.

When the respiration-induced oscillation is detectable by the fingertip-mounted TAA, we shall discuss the signal properties from the physiological perspective.
Respiratory dynamics vary across sleep stages \cite{coote1982respiratory,rostig2005nonrandom}. Compared with stable breathing during non-rapid eye movement (NREM) sleep, rapid eye movement (REM) sleep shows greater variability in both IRR and AM in the airflow signal. This is due to collectively reduced respiratory motor neuron activity associated with REM-related muscle atonia, irregular brainstem respiratory output, and transient bursts of autonomic and phasic motor activity. 
Intuitively, this seems to conflict with our observations of RMI; that is, RMI during REM sleep is significantly higher than during wakefulness and N1, and is even comparable to that observed during N2 or N3. This elevated RMI in REM can be attributed to muscle atonia, a hallmark of REM sleep, which suppresses most skeletal muscle activity and reduces gross body movements \cite{schenck1993rem,ayappa2003upper}. As a result, the torso and limb become relatively stable, 
which allows the TAA to capture cleaner, more coherent respiration-induced oscillations transmitted from the torso to the fingertip. Similarly, there are more high-quality labeled segments, and RMI is overall higher during deep sleep (N2 or N3) due to gross body movement. 

Regarding respiratory effort, it is known that it increases during sleep apnea, particularly in OSA, where paradoxical thoracic and abdominal movements are often observed due to upper airway obstruction. One might therefore expect larger torso excursions during OSA events compared to event-free periods, and consequently a stronger respiratory signature in TAA-resp. However, our results do not support this intuition. Specifically, during OSA events, a smaller proportion of 1-minute segments contain detectable respiratory oscillations compared to hypopnea, and an even smaller proportion compared to event-free segments.
Although we do not have direct evidence, we hypothesize that this discrepancy arises from physiological instability during apnea events. Increased sympathetic activation \cite{roche1999screening}, recurrent arousals \cite{kim2019effect}, and compensatory body movements \cite{amatoury2018new} during apnea events may destabilize the torso, introducing motion artifacts that degrade the coherence of vibration transmission to the fingertip. As a result, the respiratory-related signal captured by the TAA becomes less regular and less detectable, despite the increased underlying respiratory effort. More study is needed to confirm this hypothesis, which will be our future research direction.

The morphology of TAA-resp deserves discussion. Although thoracic motion is the primary driver of respiration, and might therefore be expected to impose a consistent phase relationship between TAA-resp and THO, our results do not support this intuition. This is not surprising, as the transmission of respiratory motion to the fingertip, which is a distal and mechanically indirect measurement site, is mediated by complex biomechanical pathways, including limb articulation, soft tissue coupling, and orientation mismatch between the fingertip and torso. These effects are further influenced by posture, collectively obscuring any global phase relationship. This contrasts with proximal sensor placements. For instance, abdominal TAAs have been shown to produce respiratory signals that lead airflow in phase \cite{gollee2007real}, reflecting closer coupling to respiratory mechanics.  
Overall, while high-quality TAA-resp reliably encodes IRR and its amplitude correlates well with that of THO, their morphology provides limited information about the temporal structure of the respiratory cycle. In particular, they are less suitable for identifying the precise timing of inspiration and expiration or for capturing clinically relevant waveform features such as flow limitation.

In this work, we focus on a single TAA sensor placed at a distal body location. In contrast, prior studies have explored multi-sensor configurations using multiple TAAs or inertial measurement units (IMUs)\footnote{An IMU comprises a TAA and a gyroscope, and may also include a magnetometer.} to better characterize respiratory dynamics. For example, dual-TAA sensor setups have been deployed on the chest and upper abdomen \cite{doheny2020estimation}, or on the chest and back \cite{de2022differential}. Dual-IMU sensor setups have been considered on the chest and back \cite{beck2020measurement}, at the level of the 10th rib along the mid-axillary line \cite{lapi2014respiratory}, or on the chest and abdomen \cite{elfaramawy2017wireless}. Multi-sensor arrangements, such as three IMUs positioned on the abdomen, chest wall, and coccyx \cite{cesareo2018assessment}, or on the abdomen, chest wall, and lower back \cite{angelucci2023imu} have also been considered. This list is not exhaustive; the cited works provide broader coverage of existing approaches. Although not directly evaluated here, the proposed algorithms are likely extensible to such multi-sensor configurations and may further enhance respiratory characterization. We will explore this possibility in our future work.

To establish our mANHM, we first address a fundamental question: {\em What constitutes an oscillation?} This question provides the conceptual foundation for introducing the notion of a high-quality respiratory signal, which in turn informs the development of the expert labeling criteria and the definition of RMI. Importantly, these subsequent constructs should not be interpreted as deductive consequences of the foundational question; rather, they are methodological constructs developed within the conceptual framework it establishes. Making these relationships explicit necessarily entails a trade-off between generality and operational specificity. Imagine, for example, we have an airflow signal recorded during sleep-speech. In this case, the signal may be dominated by airflow fluctuations associated with speech, which ``oscillates'' in a manner fundamentally different from normal breathing. Should such a signal be considered high-quality? Under our criteria, where regular oscillation is the key component, this airflow signal would be classified as poor quality, even though it faithfully encodes physiological information and is considered good from a sensing perspective. Similarly, airflow signals recorded during apnea events would also be deemed poor quality under our definition, clearly contradicting practical needs. 
In this sense, our signal-quality criterion is deliberately {\em stringent}. It is likely specific to our setup and may therefore require a different formulation when the objective shifts to detecting respiratory disturbances. However, this stringency is necessary in this exploratory context, where the goal is to recycle information from TAA signals recorded at the fingertip, which are particularly susceptible to voluntary movements, irrelevant motions, and other artifacts. We leave the exploration of respiratory disturbance from fingertip TAA to our future work.

Another technical aspect that warrants discussion is the construction of the TAA-resp. To enhance the weak respiratory component captured at a distal sensor location, we employ an antiderivative transformation, leveraging its inherent averaging property to improve signal quality. To our knowledge, while velocity has been widely considered in other applications \cite{foxlin2005pedestrian}, it has not been previously explored in the context of extracting respiratory information from TAA signals.
In contrast, prior work has considered the use of derivatives of TAA signals \cite{liu2011estimation,mann2011simultaneous}, primarily to amplify higher-frequency components and suppress low-frequency drift when sensors are placed proximally. Regarding the construction of the optimal axis in \eqref{equation: determine the optimal direction}, the proposed formulation is specifically designed to enhance respiratory oscillations within a targeted spectral range. In the absence of such spectral constraints, the approach is closely related to singular value decomposition (SVD), a widely used dimensionality reduction technique. While often associated, SVD and PCA \cite{jin2009performance,liu2011estimation} are not identical: SVD extracts dominant patterns in the data, whereas PCA identifies directions of maximal variance after mean removal.
In the present setting, both approaches may be influenced by undesired artifacts, which can dominate the signal given the weak micro-vibrations measurable at the fingertip. By construction, the optimal axis emphasizes respiratory components and may lead to undesired results in some special cases. For instance, during central sleep apnea (CSA), when respiratory effort is absent, the resulting TAA-resp signal may erroneously exhibit enhanced oscillations arising from incidental components within the selected spectral band, rather than remaining silent. A simple mitigation is to incorporate {\em harmonic consistency checking} by jointly tracking the fundamental frequency and its second harmonic. This reduces the likelihood of incidental components being misidentified as respiratory oscillations, as stochastic artifacts are unlikely to exhibit consistent energy at both the fundamental frequency and its harmonics. A systematic exploration of TAA-resp during CSA is an interesting future direction.

We now discuss RMI from a technical perspective. Its design is motivated by two key considerations: one rooted in classical periodicity detection, dating back to Fisher \cite{fisher1929}, and the other arising from the nonlinear relationship between IRR and the Fourier transform.
Fisher originally proposed testing for the presence of a sinusoidal component in i.i.d. Gaussian noise by examining the maximum of the periodogram over canonical frequencies. The underlying idea is that if a sufficiently strong oscillatory component at frequency $\xi_0$ is present, and the sampling rate is sufficiently higher than $2\xi_0$, then the Fourier-based power spectrum will exhibit a peak at $\xi_0$. This framework has since been extended to accommodate nonstationary and dependent noise, as well as multiple oscillatory components for the change-point detection mission \cite{wu2024frequency}  (see \cite{wu2024frequency} for more relevant work). However, real-world biomedical signals rarely exhibit purely sinusoidal behavior with fixed frequency and amplitude. Time-varying amplitude, frequency, and WSF are common, leading to spectral spreading of oscillatory components and, consequently, reduced power for Fisher-type tests and their generalizations.  In the design of RMI, TAA-resp is unwrapped so that its frequency becomes fixed, thereby reducing the spectral spreading caused by time-varying frequency. 

From those segments with bad labels but high RMI, as shown in Figures \ref{fig:LQ but high RMI-1} and \ref{fig:LQ but high RMI-2}, we observe not only the impact of stringent signal quality criteria but also the challenges associated with RMI, even when accounting for time-varying frequency. Developing a more robust RMI measure that better captures respiratory information is closely intertwined with defining appropriate signal quality criteria. These two aspects should therefore be developed jointly, which we will investigate in our future work.

To further illustrate this point, suppose we omit the unwrapping step described in Section \ref{section REI step 1}, and instead define an alternative index, denoted RMI0, directly from the non-unwrapped signal following the procedure in Section \ref{section REI step 2}; that is, 
$\texttt{RMI0}(j):= \frac{\sum_{\xi_0-2}^{\xi_0+2} |\hat{\boldsymbol{y}}_j(i)|^2 }{\sum_{k\in J_A} |\hat{\boldsymbol{y}}_j(k)|^2}$, where $\xi_0$ is the peak frequency in the range $[0.1,0.75]$, and $J_A$ is the spectral range $0.05-1$ Hz. Empirically, RMI0 exhibits inferior performance compared to RMI. However, RMI0 is computationally efficient and thus well-suited for low-power settings, particularly in home care environments. Notably, indices of this type have been previously proposed as respiratory quality metrics \cite{birrenkott2015respiratory} and applied to other physiological signals \cite{su2024model,chiu2024signal}.

This study has several limitations. As a preliminary exploration of a relatively underexplored problem, the proposed signal quality labeling criteria are not yet optimal and warrant further refinement. Due to labor constraints, only 49 full-night recordings from two hospitals were annotated. Although this dataset is sufficient to support the main findings, larger-scale studies will be necessary to establish broader consensus and improve generalizability. 
Motivated by the signal-recycling nature of this work, we developed a mathematical framework to formalize the stringent definitions of high- and low-quality segments and to evaluate RMI as a measure of respiration-induced oscillation quality over 60-s segments. This stringent framework explains the relatively low F1 when RMI is used to classify high-quality segments. Accordingly, RMI may serve as a useful preliminary quality indicator, but a fundamentally different approach is likely needed to achieve robust, fully automated quality assessment.
An important open question in this direction, both theoretically and practically, is whether oscillatory intervals can be directly delineated without relying on fixed-length segmentation. This problem is closely related to change-point detection; however, detecting change points in oscillatory components remains largely underexplored. Existing approaches typically assume fixed frequency and amplitude \cite{wu2024frequency}, whereas the present setting requires methods that can accommodate time-varying amplitude and frequency. Developing such methods would enable direct identification of oscillatory intervals and, in turn, provide a principled basis for optimizing indices such as RMI according to the target application and operating environment. This direction may ultimately lead to a more practical and robust framework for automated assessment of respiration-induced oscillations.
We leave these important questions for future investigation.

\bibliographystyle{elsarticle-num} 
\bibliography{referenceQ}

\appendix
\section{A quick review of synchrosqueezing transform} \label{section app review sst}

We begin with the short-time Fourier transform (STFT), which forms the foundation of the synchrosqueezing transform (SST) \cite{DaLuWu2011}. The STFT is a widely used tool for analyzing nonstationary time series. Its underlying principle follows a divide-and-conquer strategy: the signal is partitioned into short segments, within which it is assumed to be approximately stationary, allowing the Fourier transform to extract meaningful frequency information. By assembling the results across all segments, one obtains a time-frequency representation (TFR), i.e., a function defined on the time-frequency plane that captures the signal’s temporal dynamics. 

Mathematically, let $f$ be a proper signal, which in general may be a tempered distribution. Let $h$ be a smooth window function that decays sufficiently fast. In practice, a Gaussian window centered at the origin is commonly used. The STFT of $f$ with respect to $h$ is defined as 
\begin{equation}\label{eq:stft}
    V_f^{(h)}(t,\xi) = \int_{-\infty}^\infty f(\tau)h(\tau - t)e^{-i2\pi\xi(\tau-t)} \,d\tau
\end{equation}
where $t\in \mathbb{R}$ denotes time and $\xi\in\mathbb{R}$ denotes frequency. We call $\mathbb{R}^2$, in which the pair $(t,\xi)$ lies, the time-frequency domain. The product $f(\tau)h(\tau - t)$ corresponds to the truncation (or localization) step in the divide-and-conquer framework, while $V_f^{(h)}$ is the resulting TFR obtained by aggregating information across all time points $t$.  $|V_f^{(h)}|^2$ is known as the spectrogram of the signal $f$. In the statistical literature, $|V_f^{(h)} (t,\cdot)|^2$ is referred to as the tapered periodogram of the truncated signal centered at $t$.

The TFR determined via the STFT, or spectrogram, has been widely applied in scientific research. However, it exhibits several well-known limitations. In particular, its performance is sensitive to the choice of window function and its bandwidth. Owing to the uncertainty principle, different window widths induce trade-offs between time and frequency resolution, which may lead to undesirable effects such as spectral smearing and interference. The SST is a nonlinear time-frequency analysis technique designed to mitigate these limitations. At a high level, SST leverages the phase information of the STFT to reallocate (or ``sharpen'') the TFR. The resulting representation typically exhibits improved concentration and reduced dependence on the window choice. This enhanced sharpness facilitates more accurate ridge extraction for instantaneous frequency estimation, as well as improved reconstruction, denoising, and detection of oscillatory components, among others. 

Mathematically, SST augments the STFT with two additional steps. First, we define the {\em reassignment rule}, also known as an instantaneous frequency estimator: 
\[
\Omega_f^{(h,\nu)}(t,\xi) := 
\begin{cases}
    \frac{-i\partial_tV_f^{(h)}(t,\xi)}{2\pi V_f^{(h)}(t,\xi)}, & \quad |V_f^{(h)}(t,\xi)| > \nu\\
    -\infty, & \quad   |V_f^{(h)}(t,\xi)| \leq \nu\,,
\end{cases}
\]
where $t,\xi\in \mathbb{R}$ and $\nu>0$ is a threshold introduced to ensure numerical stability. This quantity estimates the instantaneous frequency associated with each STFT coefficient.
Second, we sharpen the STFT-based TFR via a reassignment (or synchrosqueezing) operation:
\[
S^{(h,\nu,\alpha)}_f(t,\eta)=\int_0^\infty V_f^{(h)}(t,\xi) g_\alpha\left(|\Omega_f^{(h,\nu)}(t,\xi)-\eta|\right)d\xi\,,
\]
where $\alpha>0$ is a small constant, $g_\alpha$ is a smoothing kernel satisfying $g_\alpha\to \delta_0$ weakly as $\alpha\to 0$, and $\eta\in \mathbb{R}$ denotes frequency. We call $\alpha$ the resolution of SST. Intuitively, SST redistributes STFT coefficients by mapping each at $\xi$ to its estimated instantaneous frequency $\Omega_f^{(h,\nu)}$. For a given $\eta$, the kernel $g_\alpha$ determines how strongly contributions from the associated reassigned frequencies are aggregated, thereby concentrating the energy of the TFR around true oscillatory components. In other words, we decide how much we trust the STFT coefficient using $g_\alpha$ and move all such STFT coefficients to the $\eta$ location. 

For a signal satisfying ANHM with control parameter $\epsilon>0$, reconstruction can be achieved directly from $S^{(h,\nu,\alpha)}_f(t,\eta)$. For simplicity, consider a single-component signal $f(t)=A(t)\cos(2\pi\phi(t))$ with sinusoidal oscillation. The reconstruction formula is given by
\[
\tilde{f}^{\mathbb{C}}(t):=\frac{1}{h(0)}\int_{I_t}S^{(h,\nu,\alpha)}_f(t,\eta)d\eta \,,
\]
where $I_t:=[\phi'(t)-\epsilon^{1/3},\phi'(t)+\epsilon^{1/3}]$. We have
\[
\tilde{f}^{\mathbb{C}}(t)=A(t)e^{i2\pi\phi(t)}+O(\epsilon) \,.
\]
Clearly, $\tilde{f}^{\mathbb{C}}(t)$ is the complex form of $f$, up to an error controlled by $\epsilon$.
With $\tilde{f}^{\mathbb{C}}(t)$, the amplitude modulation $A(t)$ can be recovered via its modulus, while the phase function $\phi(t)$ can be obtained through phase unwrapping.

From a computational perspective, SST is straightforward to implement. The variables $t$ and $\xi$ are discretized, and the STFT and the reassignment rule are efficiently computed using the Fast Fourier Transform. The frequency variable $\eta$ is discretized similarly to construct the synchrosqueezed representation. In practice, the reconstruction depends on the instantaneous frequency $\phi'(t)$, which is discretized similarly to construct the synchrosqueezed representation. In practice, the reconstruction depends on the instantaneous frequency estimated by ridge extraction algorithms \cite{su2024ridge}.

For more technical details, we refer readers to \cite{DaLuWu2011,alian2022reconsider,wu2025uncertainty}. Here, we provide a calculation that shows how SST performs. Take a function $f(t)=Ae^{i2\pi \xi_0 t}$, where $A,\xi_0>0$, which is a special ANHM with fixed amplitude and frequency with the cosine function as the wave-shape function (WSF). Suppose the window function $h$ is Gaussian, centered at 0. The STFT becomes
\[
V_f^{(h)}(t,\xi)=\int Ae^{i2\pi \xi_0 \tau}h(\tau-t)e^{-2\pi i \xi (\tau-t)}d\tau=Ae^{i2\pi \xi_0 t}\hat{h}(\xi-\xi_0)\,,
\]
and the reassignment rule with $\nu=0$ becomes
\[
\Omega_f^{(h,0)}(t,\xi) =  \xi_0\,.
\]
Here, we set $\nu=0$ to simplify the calculation. Therefore, if we normalize the window $h$ so that $\int_0^\infty\hat{h}(\xi-\xi_0)d\xi=1$, SST of $f$ becomes
\[
S^{(h,\nu,\alpha)}_f(t,\eta)=\int_0^\infty Ae^{i2\pi \xi_0 t}\hat{h}(\xi-\xi_0) g_\alpha(|\xi_0-\eta|)d\xi=Ae^{i2\pi \xi_0 t}g_\alpha(|\xi_0-\eta|)\,,
\]
which is centered at $\eta$ with small support depending on $\alpha$. In practice, $\alpha$ is much smaller than the width of $h$, which leads to the desired increased TFR contrast. The argument for functions satisfying ANHM is similar, since locally such a function can be well approximated by one with fixed amplitude and frequency. See \cite{DaLuWu2011} for details.

\subsection{Numerical implementation}

Numerically, assume the TFR of the input signal determined by SST is denoted as $\mathbf{R}=\mathbb{C}^{N\times M}$, where $M\in \mathbb{N}$ is the number of frequency bins. 
We apply the path-optimization algorithm \cite{su2024ridge} to extract ``ridge'' that represents IRR. Denote $[M]=\{1,2,\ldots,M\}$. The algorithm is:
\begin{equation}
c^*=\arg\max_{c:[N]\rightarrow[M]}
\sum_{\ell=1}^N
\big|\widetilde{\mathbf{R}}(\ell,c(\ell))
\big|
-
\lambda\sum_{\ell=1}^{N-1}\left|\Delta c(\ell)\right|^2\in [M]^N\,,
    \label{singleCurveExt:theory}
\end{equation}
where $\Delta c\in [M]^{N-1}:=\{(m_1,\cdots,m_{N-1}):m_k\in[M],\,1\leq k\leq N-1\}$ so that $\Delta c(\ell):=c(\ell+1)-c(\ell)$ for $\ell=1,\ldots,N-1$, which is the numerical differentiation of the curve $c$, $\lambda>0$ is the penalty term constraining the regularity of the fit curve $c^*$, and $\big|\widetilde{\mathbf{R}}(\ell,q)\big|=\log\frac{|\mathbf{R}(\ell,q)|}{\sum_{i=1}^N\sum_{j=1}^M|\mathbf{R}(i,j)|}$ is a normalization of the matrix $\mathbf{R}$. While there are other ridge extraction algorithms, we focus on this one due to its numerical efficiency. We refer readers to \cite{su2024ridge} for other available ridge extraction algorithms and \cite{liu2026probabilistic} for existing theoretical analysis of ridges. Denote the estimated IRR as $\tilde{\phi}'(t)$ by
\[
\tilde{\phi}'(i/f_s)=\delta_fc^*(i)
\]
for $i=1,\ldots,N$, where $\delta_f>0$ is the frequency bin size. To obtain the respiratory phase, a na\"ive approach is applying the cumulative sum to estimate $\phi'(t)$'s antiderivative. However, we lose the initial phase, and the cumulative sum accumulates numerical errors. We thus consider a different approach via reconstruction. Denote $\tilde{h}_1\in \mathbb{C}^N$, with
\[
\tilde{h}_1(i)=\sum_{k=c^*(i)-L}^{c^*(i)+L}\mathbf{R}(i,k)\in \mathbb{C}\,,
\]
where $L\delta_f=0.1$. It is well known that this reconstruction formula is an accurate estimate of the first harmonic of the putative respiration-induced oscillation in the complex form up to a universal constant, even when the noise is nonstationary \cite{wu2025uncertainty}. In other words, over $I_j$, we have
\[
\tilde{h}_1(i)\approx CA_j(i/f_s)\alpha_{j,1}e^{i (2\pi \phi(i/f_s)+\beta_{j,1})}
\]
for a universal constant $C>0$.
As a result, we obtain the respiratory phase estimate $\tilde{\phi}(t)$ by unwrapping 
\[
\left\{\left(i/f_s, \frac{\tilde{h}_1(i/f_s)}{|\tilde{h}_1(i/f_s)|}\right)\right\}_{i=1}^N\,,
\] 
and hence $\tilde{\phi}(t)\approx \phi(t)$ over $I_j$.

\section{Mathematical justification of TAA-resp signal construction}\label{section math justification of TAA-resp.}

To simplify the justification of \eqref{eq: appendix: justification}, we focus on one harmonic and consider $a(t)\cos(\phi(t))$ over $[0,T]$ that satisfies the ANHM with precision $\epsilon\geq 0$ and $\phi(0)=\pi/2$. We aim to find a pair of functions $b,\psi$ so that $b(t)\cos(\psi(t))$ satisfies the ANHM, $b\approx a$, $\phi\approx \psi$, and
\begin{equation}\label{eq: desired relationship}
\int_0^ta(s)\cos(\phi(s))ds=\frac{b(t)}{\psi'(t)}\sin(\psi(t))
\end{equation}
for $t\in [0,T]$.
Differentiate \eqref{eq: desired relationship} and get
\[
a(t)\cos(\phi(t))=\left(\frac{b}{\psi'}\right)'(t)\sin(\psi(t))+b(t)\cos(\psi(t))\,.
\]
By a direct trigonometric identity, we have
\[
\left(\frac{b}{\psi'}\right)'(t)\sin(\psi(t))+b(t)\cos(\psi(t))=\sqrt{b(t)^2+c(t)^2}\cos(\psi(t)+\theta(t))\,,
\]
where $c(t)=(\frac{b}{\psi'})'(t)$ and $\cos(\theta(t))=b(t)/\sqrt{b(t)^2+c(t)^2}$.  By a direct calculation, we know $(\frac{b}{\psi'})'(t)\leq 2\epsilon (\frac{b}{\psi'})(t)$, which means when $\epsilon$ is sufficiently small, $\theta(t)=O(\epsilon)$ and $a(t)-b(t)=O(\epsilon^2)$. This justifies \eqref{eq: appendix: justification}.

On the other hand, the antiderivative of the noise part of $a_x(t)$  can be formally written as
\[
\Phi(t)=\int_0^t \zeta(s)ds\,.
\]
Suppose $\zeta(t)$ is Gaussian white noise. It is well known that $\Phi(t)$ is Brownian motion, whose power spectral density scales as $\xi^{-2}$ when $\xi\to 0$. Therefore, any realization of $\Phi(t)$ exhibits a strong low-frequency component resembling a trend, while the high-frequency components are attenuated.
Overall, integration improves the signal-to-noise ratio (SNR) within the spectral band associated with respiration-induced oscillation, as the contribution of high-frequency noise is suppressed; this benefit, however, comes at the cost of amplifying low-frequency noise and bias, which appear as a slowly varying trend, and sacrificing the high harmonic information.

\section{Mathematical interpretation of RMI}\label{section:math interpretation of RMI}
The RMI  has a clear mathematical foundation. With the estimated phase $\tilde{\phi}(t)$ from \eqref{ACCmodel ver4}, the unwrap gives us
\begin{align}
\tilde{r}(t):=\,&r(\tilde{\phi}^{-1}(t))\nonumber\\
=\,&\sum_{j=1}^J \sum_{k=1}^2 A^*_{j,k}(\tilde{\phi}^{-1}(t))\cos(2\pi k \phi(\tilde{\phi}^{-1}(t))+\beta_{j,k}) \chi_{\tilde{I}_j}+\xi^*(\tilde{\phi}^{-1}(t))\\
\approx\,&\sum_{j=1}^J \sum_{k=1}^{2} \tilde{A}_{j,k}(t)\cos(2\pi k t+\beta_{j,k}) \chi_{\tilde{I}_j}+\tilde{\xi}(t)\,,\nonumber
\end{align}
where $\tilde{A}_{j,k}(t):=A^*_{j,k}(\phi^{-1}(t))$, and the last approximation holds since $\tilde{\phi}$ is an accurate approximation of $\phi$, and hence $\phi(\tilde{\phi}^{-1}(t))\approx t$, $\tilde{I}_j=[\tilde{\phi}^{-1}(b_j),\tilde{\phi}^{-1}(e_j)]$, and $I_j=[b_j,e_j]$. Since $\phi'(t)$ changes slowly by assumption, $\tilde{A}_j(t)$ changes slowly, and $\tilde{\xi}(t)$ is still a mean 0 random process with finite covariance structure.
The main benefit we obtain after unwrapping is that $\tilde{r}(t)$ oscillates at a fixed frequency $1$, while the AM is not constant. The Fourier transform of the oscillatory component over $\tilde{I}_j$ hence has a simplified form:
\begin{align}
&\mathcal{F}\left(\sum_{k=1}^2 \tilde{A}_{j,k}(t)\cos(2\pi k t+\beta_{j,k})\right)(\xi)\nonumber\\
=&\,\left(\frac{1}{2}\sum_{k=1}^{2} \left(e^{i\beta_{j,k}}\delta_{\cdot-k}+e^{-i\beta_{j,k}}\delta_{\cdot+k} \right)\star\mathcal{F}(\tilde{A}_{j,k})\right)(\xi)\\
=&\,\frac{1}{2}\sum_{k=1}^{2} \left(e^{i\beta_{j,k}}\mathcal{F}(\tilde{A}_{j,k})(\xi-k)+e^{-i\beta_{j,k}}\mathcal{F}(\tilde{A}_{j,k})(\xi+k) \right)\nonumber\,.
\end{align}
Due to the smoothness and slowly varying assumption of $\tilde{A}_{j,k}$, $\mathcal{F}(\tilde{A}_{j,k})$ is supported near $0$ with the spreading depending on the variation of $\tilde{A}_{j,k}$. Hence, $\mathcal{F}\left(\sum_{k=1}^{2} \tilde{A}_{j,k}(t)\cos(2\pi k t+\beta_{j,k})\right)(\xi)$ is supported near frequency $1$ and $2$ on the positive axis with the weight and spreading depending on $\mathcal{F}(\tilde{A}_{j,k})$.
This computation suggests that if respiration-induced oscillation exists in fingertip TAA over $I_j$, the associated RMI should be large. Otherwise, its contribution is mainly from the noise. This fact motivates us to consider RMI as the target statistic.


\begin{remark}
When additional respiratory effort measurements are available (e.g., from a chest or abdominal band), standard breathing-cycle detection algorithms can be used to identify expiration termination times $t_1<t_2<\ldots<t_m$. Due to breathing rate variability \cite{benchetrit2000breathing}, these timestamps are generally nonuniformly spaced.
In the ideal, noise-free case, $t_l=\phi^{-1}(l)$ for $l=1,\ldots,m$. The respiratory phase can then be estimated by interpolating the nonuniform samples $\left\{\left(t_l, \, l\right)\right\}_{l=1}^m$, typically using cubic splines,
to obtain $\check{\phi}(t)$. Since the initial phase is unknown, we have $\check{\phi}(t)\approx \phi(t)+c$, where $c\in [0,2\pi)$ is an unknown constant phase offset that does not affect the final results.
In practice, $\check{\phi}(t)$ is generally more accurate than $\tilde{\phi}(t)$, particularly in regions with weak respiration-induced oscillation, reflecting the benefit of using more reliable measurements.
\end{remark}

\section{Segments with disagreed labels} \label{section disagreement segments}

\begin{figure}[bht!]
    \centering
    \includegraphics[trim = 0 0 0 0, clip, width=\textwidth]{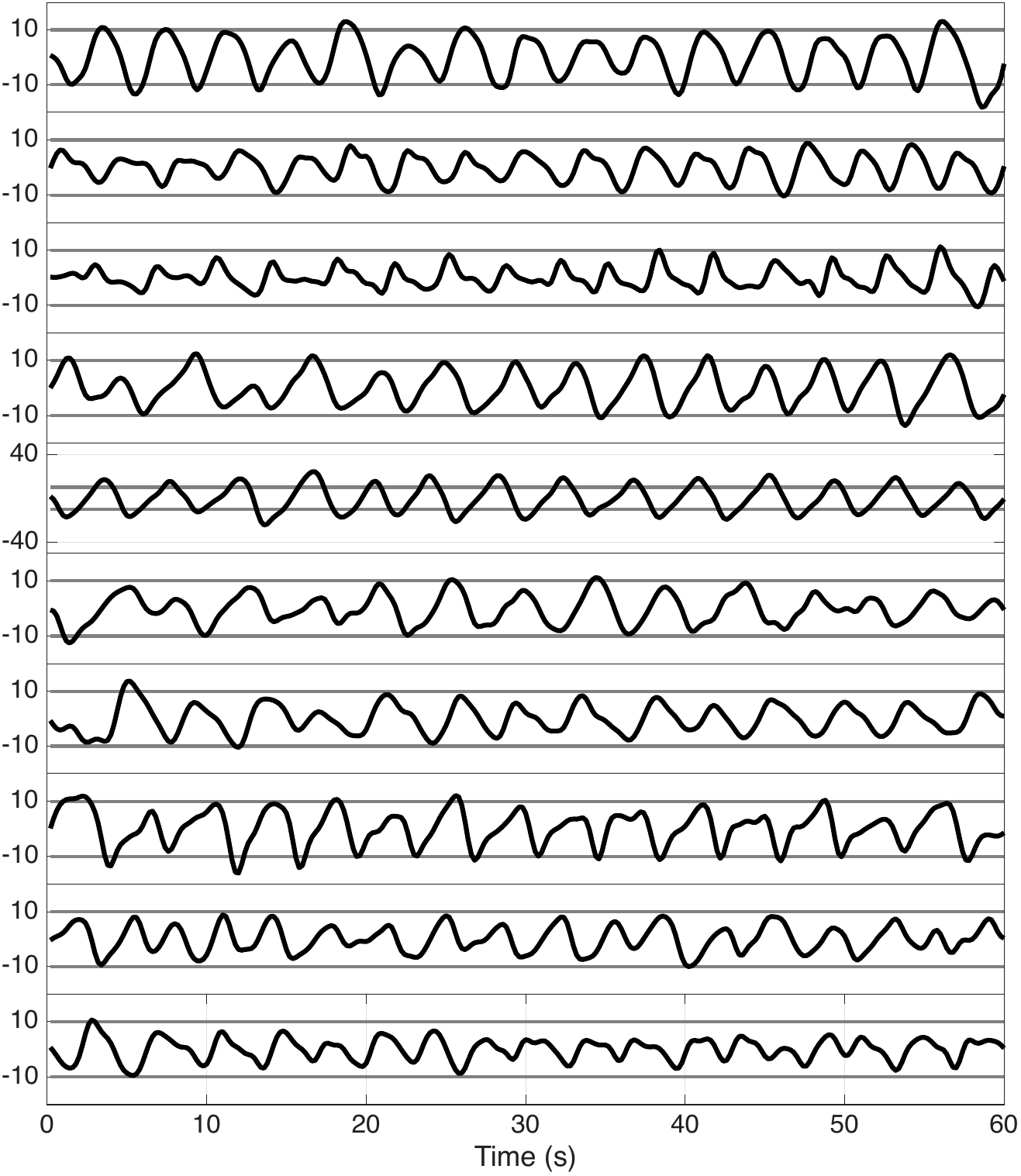}
    \caption{Ten randomly selected segments with different poor and moderate labels by two annotators. The horizontal gray lines are $\pm 10$.
     }
    \label{fig:disagreement 0 and 1}
\end{figure}

\begin{figure}[bht!]
    \centering
    \includegraphics[trim = 0 0 0 0, clip, width=\textwidth]{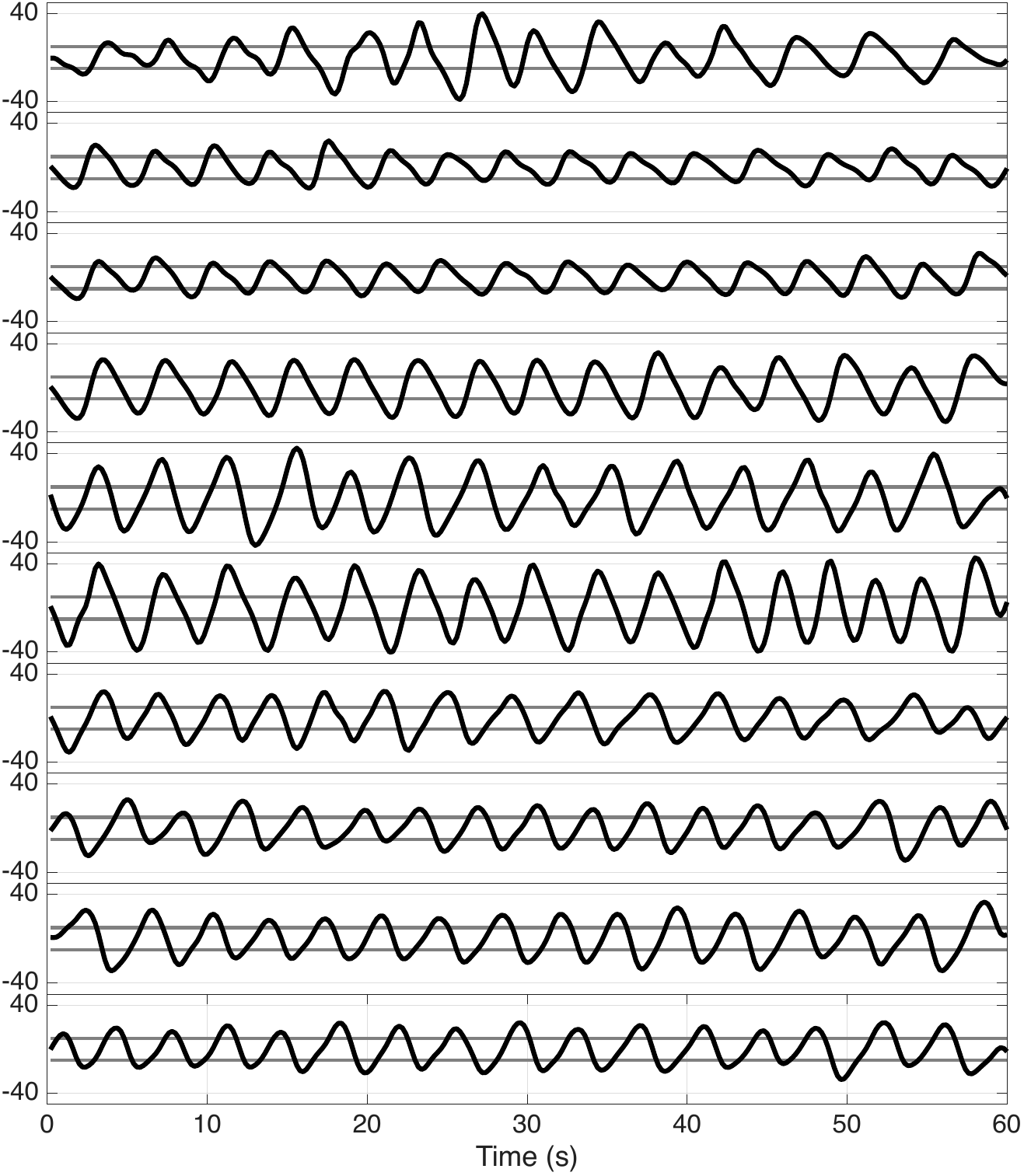}
    \caption{Ten randomly selected segments with different good and moderate labels by two annotators. The horizontal gray lines are $\pm 10$.
     }
    \label{fig:disagreement 2 and 1}
\end{figure}

\begin{figure}[bht!]
    \centering
    \includegraphics[trim = 0 0 0 0, clip, width=\textwidth]{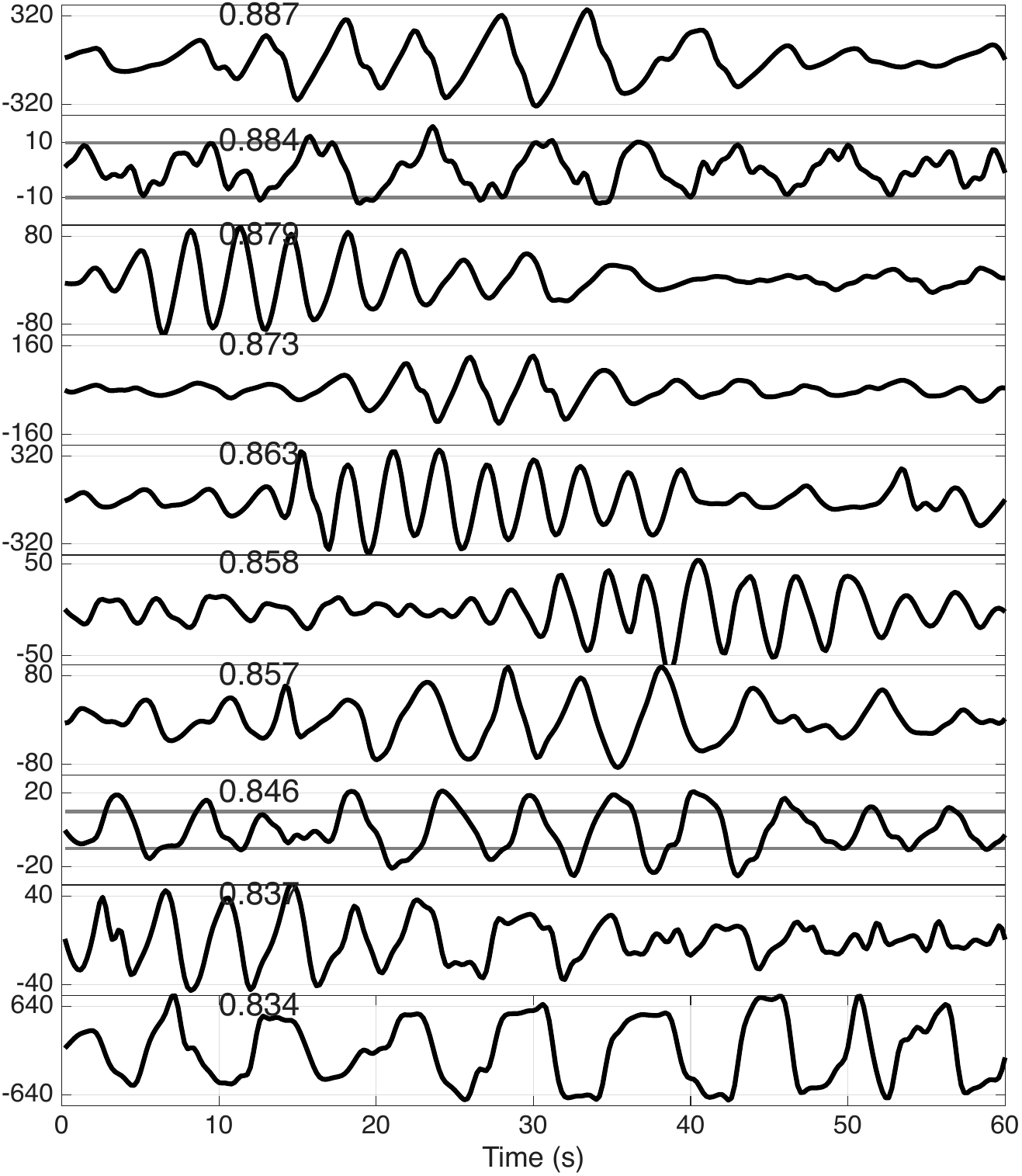}
    \caption{Ten segments with RMI $>0.7$ labeled poor by two annotators. The horizontal gray lines are $\pm 10$, which are shown when the $99$\% magnitude is less than 40.
     }
    \label{fig:LQ but high RMI-1}
\end{figure}

We demonstrate several segments where the two annotators gave different labels in Figures \ref{fig:disagreement 0 and 1} and \ref{fig:disagreement 2 and 1}. 
We also show 20 randomly selected segments labeled poor by both annotators but have RMIs higher than 0.7 in Figures \ref{fig:LQ but high RMI-1} and \ref{fig:LQ but high RMI-2}.

\begin{figure}[bht!]
    \centering
    \includegraphics[trim = 0 0 0 0, clip, width=\textwidth]{LQ_hRMI1.pdf}
    \caption{Ten segments with RMI $>0.7$ labeled poor by two annotators. The horizontal gray lines are $\pm 10$, which are shown when the $99$\% magnitude is less than 40.
     }
    \label{fig:LQ but high RMI-2}
\end{figure}

\clearpage
\section{Comparison with other methods}\label{section:method comparisons}

While no respiration extraction algorithms specifically designed for TAAs installed on the fingertip exist, for a comparison, we consider various existing algorithms for extracting respiratory signals from a TAA sensor installed on the torso, including the {\em fixed $z$-axis} (z-axis) algorithm \cite{fekr2014tidal}, {\em dominant-component} (Dom-com) algorithm \cite{bucklin2010inexpensive}, {\em vector-magnitude} (VM) algorithm \cite{dehkordi2011validation}, {\em principal component analysis (PCA) based axes-fusion} algorithm \cite{jin2009performance}, and {\em rotation-axis angular-rate} (Ro-axis) algorithm \cite{bates2010respiratory}. 

\begin{figure}[bht!]
    \centering
    \includegraphics[width=\textwidth]{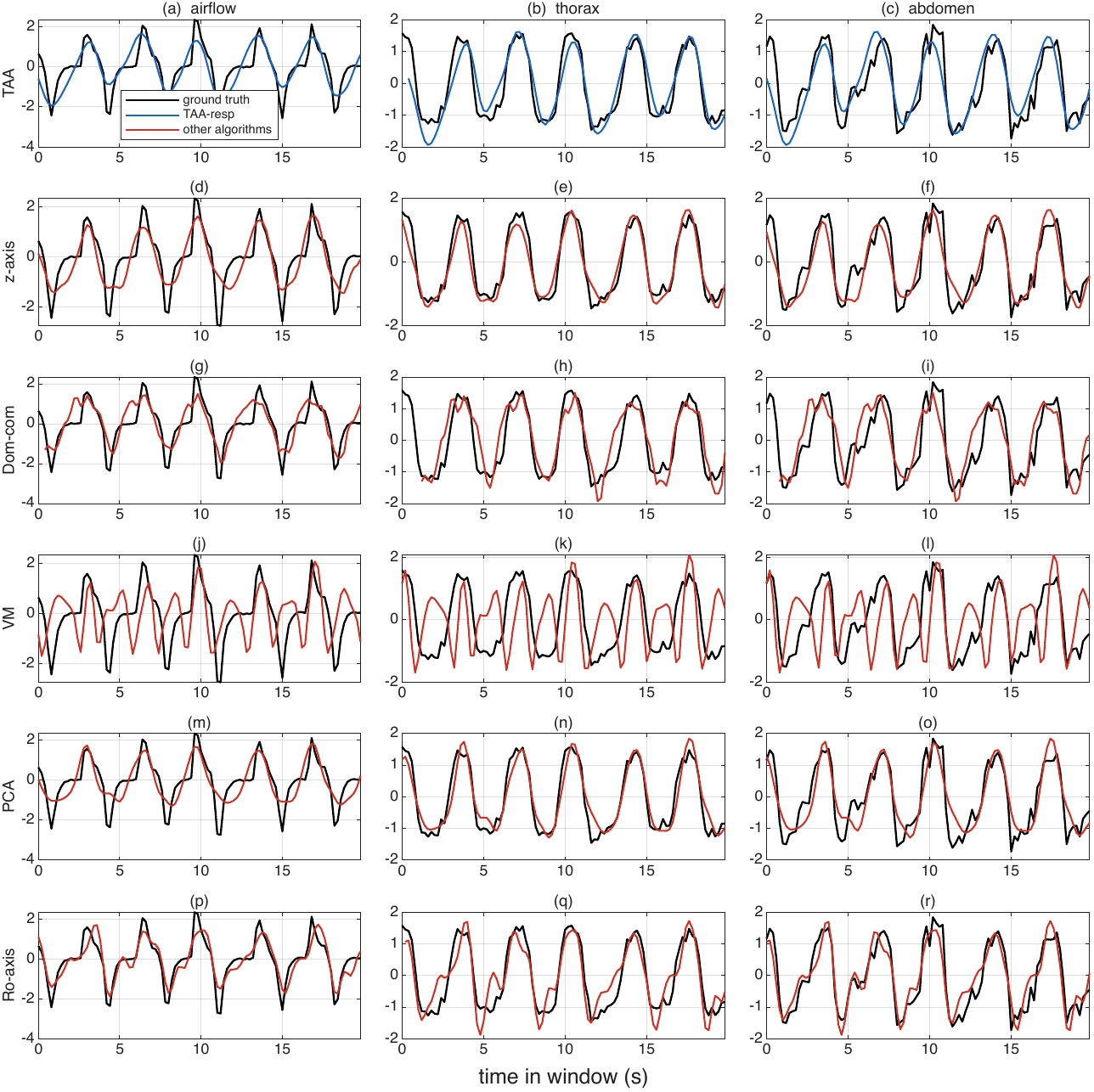}
    \caption{Comparison between TAA-resp (a)-(c), fixed z-axis (d)-(f), dominant-component (g)-(i), vector magnitude (j)-(l), PCA-based axes fusion (m)-(o), and rotational-axis angular rate (p)-(r) derived from one good fingertip TAA segment superimposed on simultaneously recorded ground truth over 20 seconds.}
    \label{fig:TAArespvsOthersGood}
\end{figure}

\begin{figure}[bht!]
    \centering
    \includegraphics[width=\textwidth]{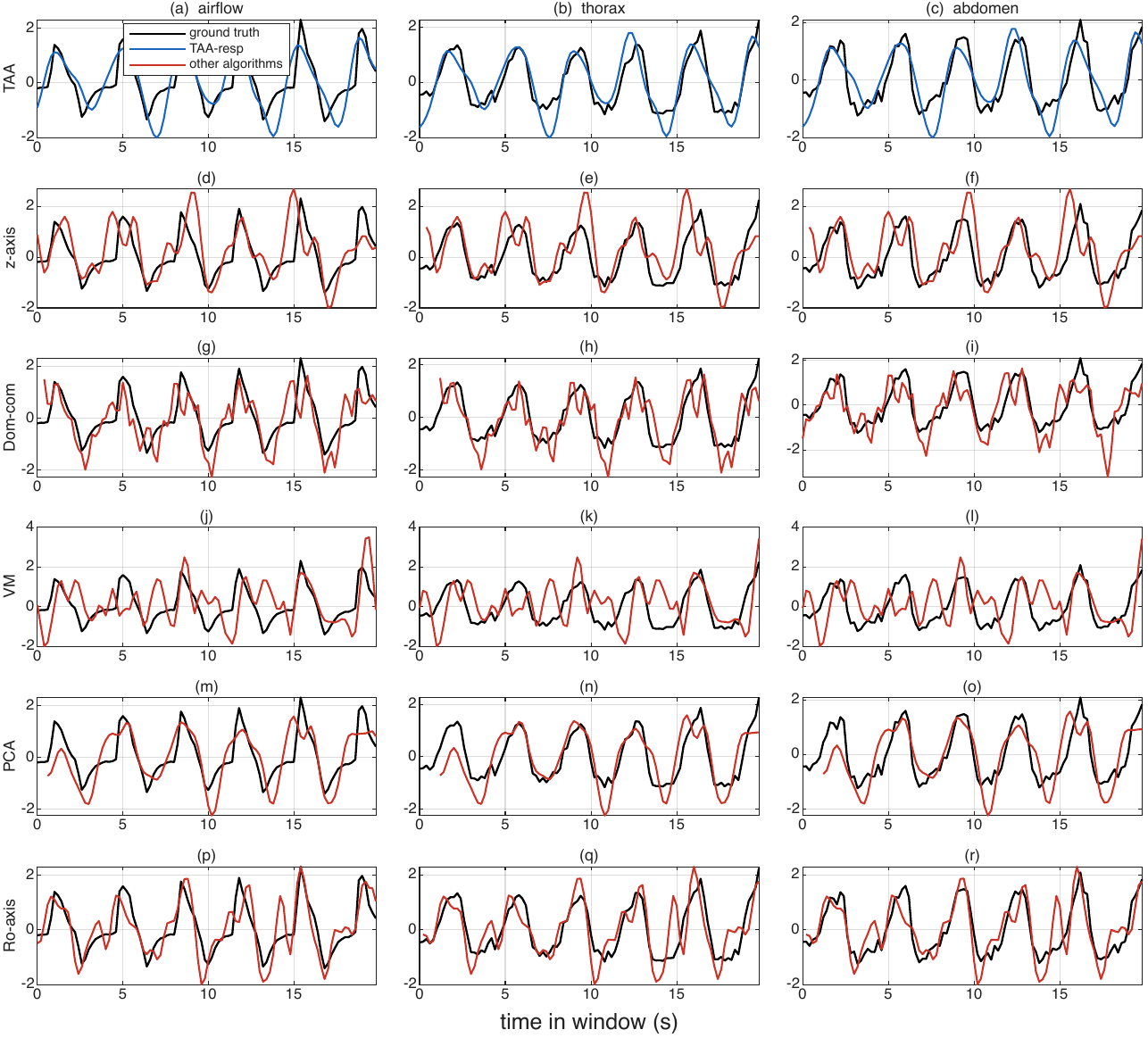}
    \caption{Comparison between TAA-resp (a)-(c), fixed z-axis (d)-(f), dominant-component (g)-(i), vector magnitude (j)-(l), PCA-based axes fusion (m)-(o), and rotational-axis angular rate (p)-(r) derived from one moderate fingertip TAA segment superimposed on simultaneously recorded ground truth over 20 seconds.}
    \label{fig:TAArespvsOthersMod}
\end{figure}

A visualization of these algorithms against the ground-truth over a good and moderate segment are shown in Figures \ref{fig:TAArespvsOthersGood} and \ref{fig:TAArespvsOthersMod}.

Next, we provide some quantification. For each algorithm, we evaluate the RMI of every annotated fingertip TAA recording in two databases for comparison. In addition, we also evaluate the synchronized correlation against the simultaneously recorded respiratory channels (airflow, thorax, abdomen). 

\begin{figure}[bht!]
    \centering
\includegraphics[width=.7\textwidth]{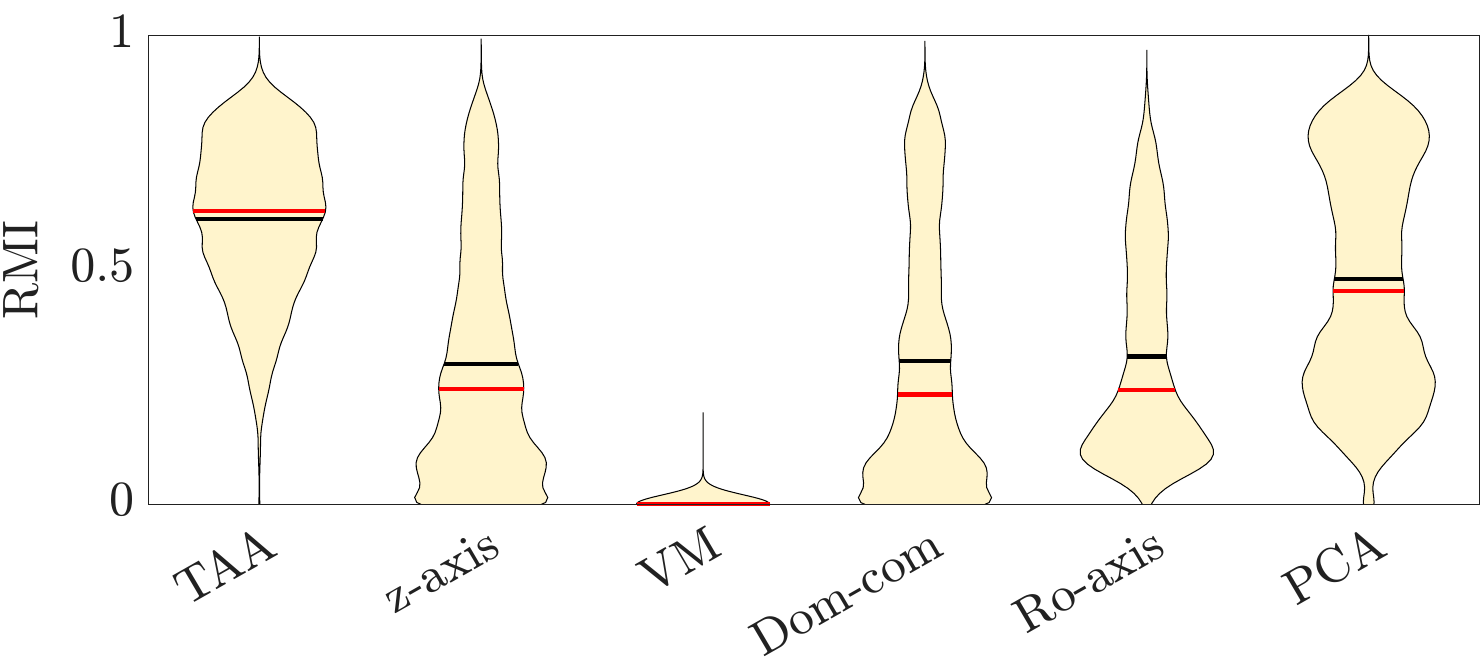}
    \includegraphics[trim=0 0 420 0, clip, width=\textwidth]{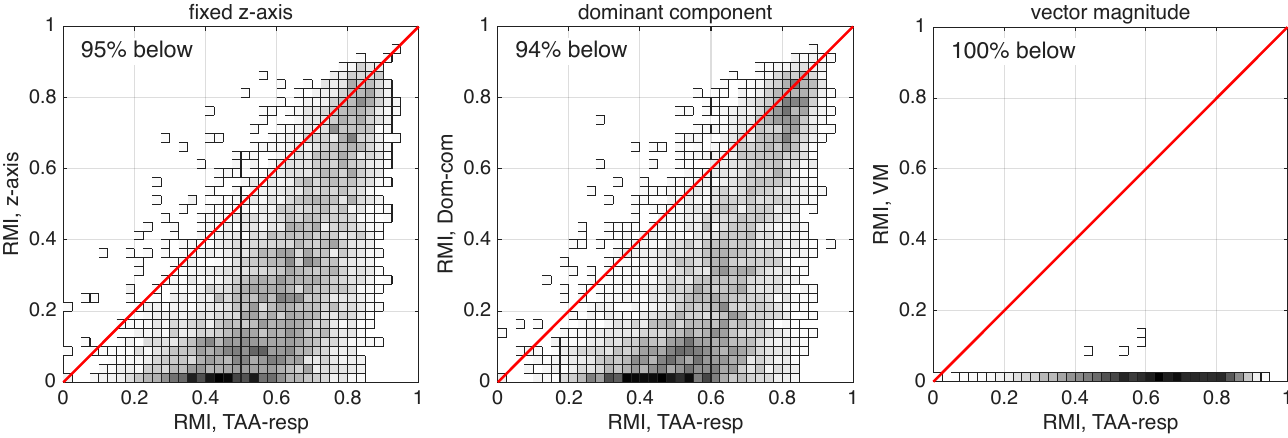}
    \includegraphics[trim=620 0 0 0, clip, width=0.66\textwidth]{RMI_vs_TAAresp_hist2_all_eq.pdf}
        \caption{Distribution of RMI over 7112 labeled segments of the 50 recordings, for TAA-resp and the five comparison algorithms. The mean and median of each distribution are marked as red and black respectively.
RMI of each algorithm against RMI of TAA-resp (one point per labeled segment) shown as a bivariate histogram (darker = more segments). The red line represents the identity.}
    \label{fig:RMI-violin}
\end{figure}

In the top of Figure \ref{fig:RMI-violin}, we show the distribution of RMI for each algorithm. We observe that TAA-resp is unimodal and concentrated, with mean $\pm$ standard deviation of $0.61\pm 0.16$ and median of $0.62$. The fixed $z$-axis, dominant component, and rotation-axis angular rate algorithms have mean $\pm$ SD of $0.30\pm 0.24$, $0.31\pm 0.26$, $0.32\pm 0.22$ and medians $0.24$, $0.23$, $0.24$ respectively. The RMI of the vector magnitude surrogate is essentially zero on every segment (mean and median $0$). This is expected as the fingertip accelerometer registers respiration as a weak oscillatory signal, and the Euclidean magnitude can introduce distortions in its temporal structure as discussed below. The PCA surrogate is the only one whose distribution overlaps with that of TAA-resp in the high-RMI range, and it is bimodal with one lobe in the high-RMI range and the other low RMI. The mean$\pm$ SD and median RMI of the PCA surrogate are $0.48\pm 0.24$ and $0.46$ respectively.
In the bottom of Figure \ref{fig:RMI-violin}, we show the 2-dim histogram of RMI of each algorithm against that of TAA-resp on the same segment, with the identity line marked. A segment below the line is one on which TAA-resp attained the higher RMI. That proportion is 95\% for the fixed z-axis, 94\% for the dominant component, 100\% for the vector magnitude, 74\% for the PCA-based axes fusion, and 95\% for the rotation-axis rate.

\begin{figure}
    \centering
     \includegraphics[width=.7\textwidth]{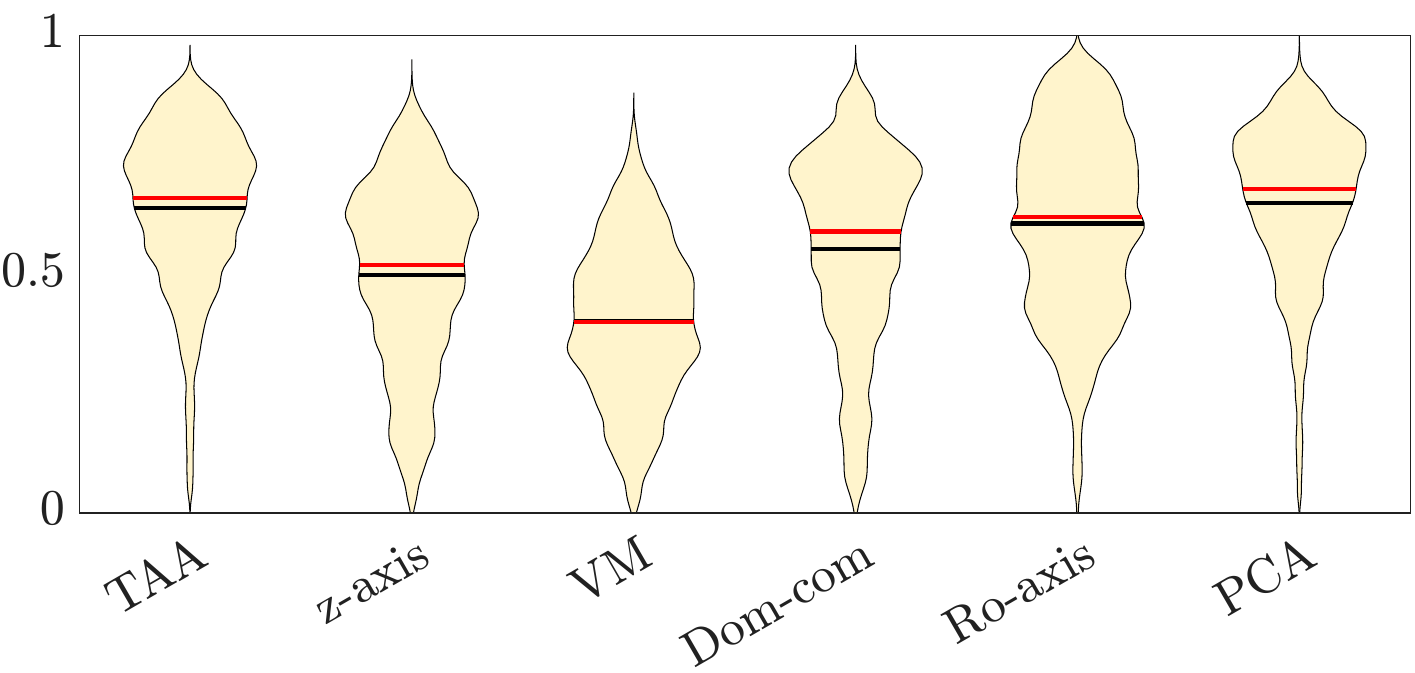}
     \includegraphics[width=.7\textwidth]{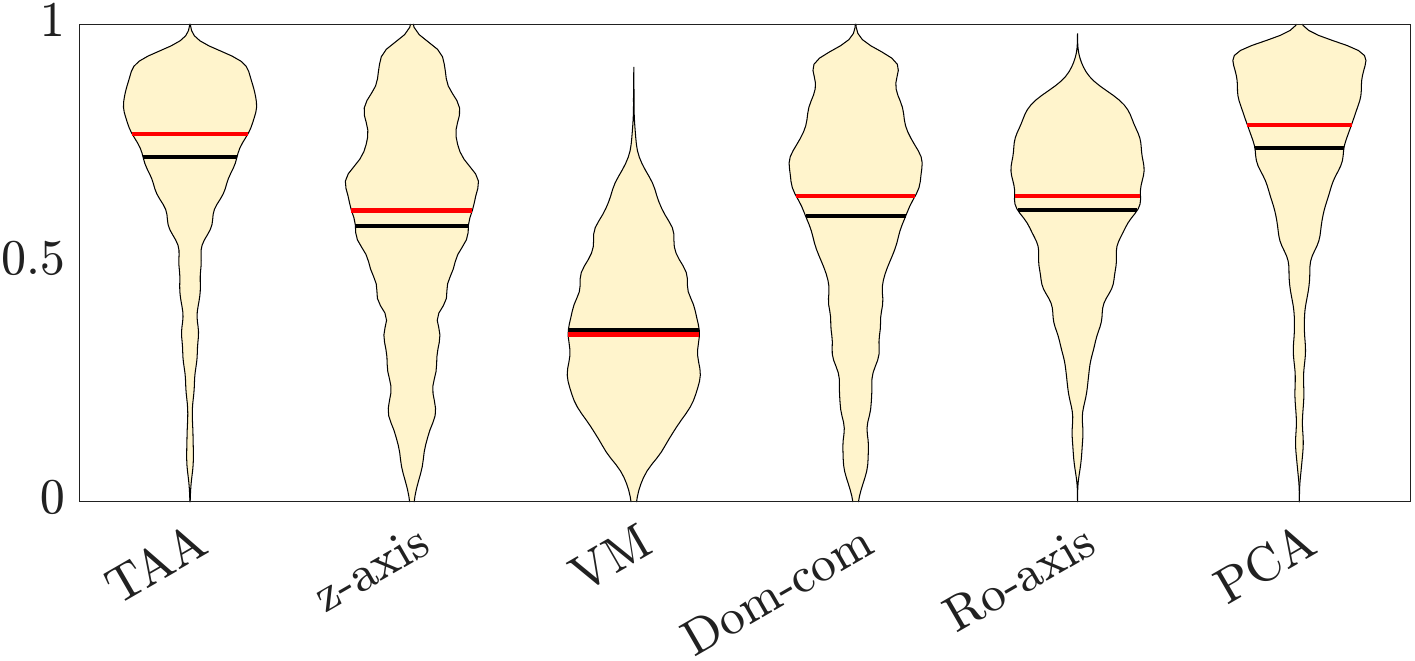}
     \includegraphics[width=.7\textwidth]{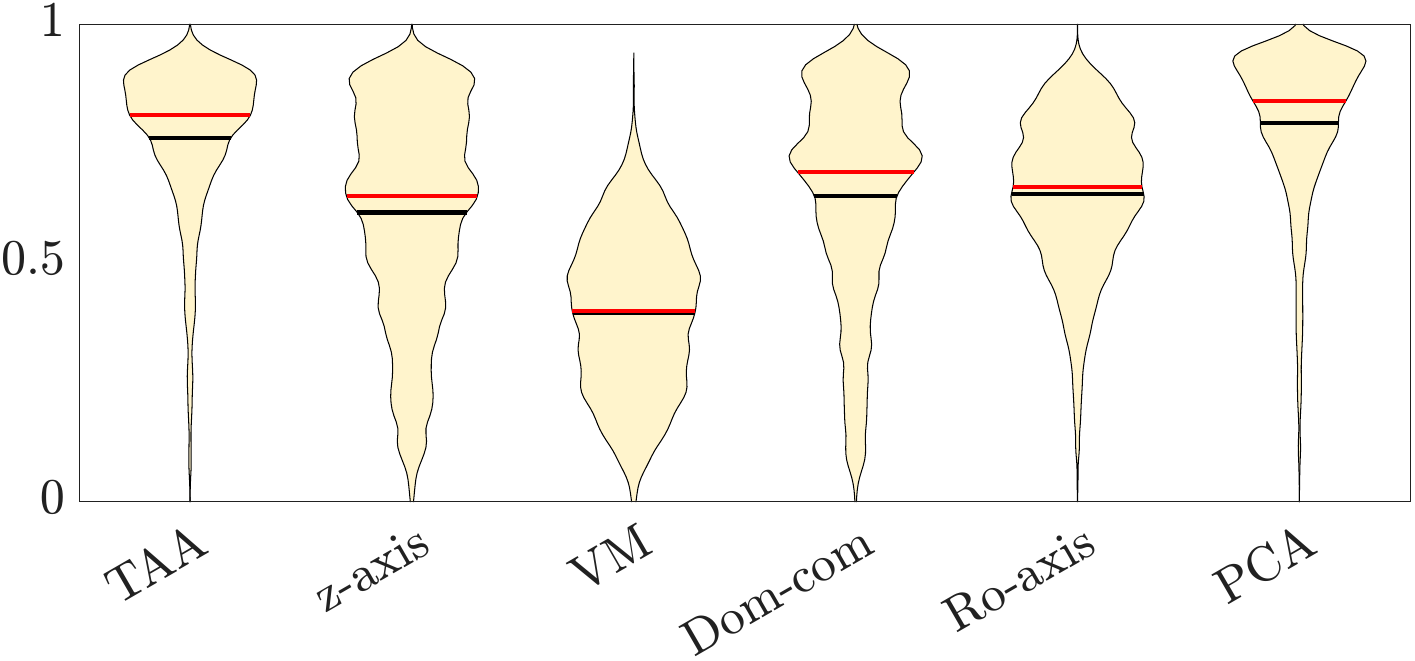}
   \caption{Distribution of synchronized correlation over each comparison algorithm with the airflow (top), THO (middle), and ABD (bottom) as ground truth. The mean and median of each algorithm are marked as red and black horizontal lines.
   }
    \label{fig:corr-all-violin}
\end{figure}

Violin plots of the synchronized correlation distributions of each algorithm with airflow, THO, and ABD are given in Figure \ref{fig:corr-all-violin}. We observe that the mean and median of the synchronized correlation of TAA-resp with the ground-truth are higher than those of other algorithms except the PCA-based axes-fusion, which is comparable.

Below, for the sake of self-containedness, we summarize these algorithms and report detailed summary statistics comparing them with TAA-resp, and provide interpretations. For each algorithm, except the steps precisely mentioned below, we follow the algorithm as faithfully as possible for the sake of fair comparison.

\subsection{Fixed z-axis algorithm}

The first algorithm is referred to as the \textit{fixed z-axis algorithm} (z-axis) \cite{fekr2014tidal}. In the setup of \cite{fekr2014tidal}, the TAA sensor is strapped to the sternum so that its $z$-axis is aligned with the anterior-posterior direction of the chest wall, which is the dominant direction of respiration-induced chest motion. As such, the $z$-axis signal is used as a representative of chest movement. In this algorithm, the $z$-axis signal is filtered through a 10th-order Butterworth low-pass filter with frequency cut off at $1$ Hz. The respiratory rate is subsequently estimated by counting the number of local maxima in the surrogate respiratory signal per minute. Note that the sensor orientation at the fingertip varies with hand posture, so the $z$-axis does not correspond to anterior-posterior chest motion as it does for a chest-mounted sensor. However, we follow the algorithm in \cite{fekr2014tidal} as faithfully as possible for the sake of fair comparison.
We note that the accelerometer in \cite{fekr2014tidal} was calibrated at 6 stationary positions, where a least-squares fit of a few seconds of raw data determined 12 calibration parameters of the accelerometer. Since our recordings were not acquired in this way, we cannot reproduce this step post hoc, and thus no calibration is applied.

The mean $\pm$ SD RMI of TAA-resp over 7112 segments and 50 recordings was $0.61\pm 0.16$ (median $0.62$) while for the $z$-axis, it was $0.30\pm 0.24$ (median $0.24)$. 74\% of the TAA-resp segments had an RMI greater than 0.5, while for the $z$-axis it was 23\%. The synchronized correlations (mean $\pm$ SD) of TAA-resp with airflow, thorax, and abdomen were $0.64\pm 0.17$, $0.72 \pm 0.18$, and $0.76\pm 0.16$ respectively, whereas for the $z$-axis, they were $0.50\pm 0.19$, $0.58 \pm 0.23$, $0.61 \pm 0.22$. TAA-resp achieved a higher synchronized correlation than the fixed $z$-axis algorithm in 83\% of airflow windows, 76\% of thorax windows, and 78\% of abdomen windows, and paired Wilcoxon signed rank tests rejected equality in all three cases ($p<10^{-223}$).

This algorithm is expected to work worse since fingertip position changes from time to time, even during sleep. The information carried in the z-axis is therefore uncertain.

\subsection{Vector-magnitude algorithm}
The VM algorithm was applied in \cite{dehkordi2011validation}, where the TAA sensor is mounted at the suprasternal notch, to sense airflow in the upper airway. In this algorithm, the raw TAA signal is detrended by subtracting the mean of each axis and then each axis is low-pass filtered with an $8$th-order Butterworth filter with frequency cut off at $1$ Hz. The VM algorithm then evaluates the standard Euclidean norm of the three filtered axes as the final respiratory surrogate. 
We apply a zero-phase $8$th-order Butterworth filter at the stated cut off in our setup.

The mean $\pm$ SD RMI of the VM algorithm was $0.00\pm 0.00$ (median $0.00$). The synchronized correlations (mean $\pm$ SD) with airflow, thorax, and abdomen were $0.40\pm 0.16$, $0.36\pm 0.16$, and $0.39\pm 0.16$ (TAA-resp values are the same as before). TAA-resp achieved a higher synchronized correlation than the VM algorithm in 87\% of airflow windows, 92\% of thorax windows, and 95\% of abdomen windows, and paired Wilcoxon signed rank tests rejected equality in all three cases ($p<10^{-309}$). This behaviour is expected from the constructed VM signal illustrated in Figure \ref{fig:TAArespvsOthersGood}.

Note that VM offers several advantages, including orientation invariance, reduced sensitivity to sensor placement, and a simple scalar representation of motion intensity across three axes. While TAA-resp can be interpreted as an alternative to VM, specifically designed to capture the weak oscillatory structure embedded in the TAA signal, they are fundamentally different.
Importantly, taking the magnitude of an oscillatory signal can introduce distortions in its temporal structure. For example, consider the signal $f(t)=\begin{bmatrix}\cos(2\pi t)\\ 0 \\ 0\end{bmatrix}$ with a sinusoidal oscillation at 1 Hz. Its magnitude is $\|f(t)\|_2 = |\cos(2\pi t)|$, which exhibits an oscillatory component at $2$ Hz and deviates from a purely sinusoidal waveform. Note that we can see this ``double pattern'' clearly in Figures \ref{fig:TAArespvsOthersGood} and \ref{fig:TAArespvsOthersMod}. This is reflected in the almost 0 RMI shown in Figure \ref{fig:RMI-violin}. We therefore also expect a lower performance of VM for the purpose of extracting respiratory information from the fingertip.

\subsection{Dominant-component algorithm}
The Dom-com algorithm in \cite{bucklin2010inexpensive}, where the TAA sensor was placed on the suprasternal notch, was developed to detect sleep-apnea. In this algorithm, the raw TAA signal is first detrended by subtracting the mean of each axis and then the dominant acceleration component is determined by performing PCA on the two components with the largest variances. We note that in \cite{bucklin2010inexpensive}, they also consider choosing the component with the largest variance as the dominant component but we follow the PCA method as it uses information from more than one axis and performed better overall in our test. The breathing signal is subsequently obtained by applying a zero-phase low-pass filter with frequency cut off at $5$ Hz in the frequency domain. The authors in \cite{bucklin2010inexpensive} note that the PCA computation for three axes would require a more powerful microprocessor for implementation, and thus they opt for one of the aforementioned simpler alternatives. 

We note that the $5$ Hz low-pass filter in \cite{bucklin2010inexpensive} is wider than the standard respiratory band (since the authors ultimately wanted to distinguish between breathing and non-breathing zones) and thus retains content above the Nyquist frequency of our $5$ Hz comparison clock. As such, we applied a $2$ Hz zero-phase low-pass filter instead. Another modification is that we apply the variance test within each 60-second window rather than once for the whole recording as fingertip orientation may change frequently over the course of the night.

The mean $\pm$ SD RMI of the dominant-component algorithm was $0.31\pm 0.26$ (median $0.23$) with 26\% of segments having an RMI over $0.5$. The synchronized correlations (mean $\pm$ SD) with airflow, thorax, and abdomen were $0.55\pm 0.20$, $0.60\pm 0.23$, and $0.64\pm0.22$ (TAA-resp values are the same as before). TAA-resp achieved a higher synchronized correlation than the dominant-component algorithm in 68\% of airflow windows, 77\% of thorax windows, and 74\% of abdomen windows, and paired Wilcoxon signed rank tests rejected equality in all three cases ($p<10^{-112}$).

Although this method can occasionally produce high-quality results similar to the z-axis algorithm shown in Figures \ref{fig:TAArespvsOthersGood} and \ref{fig:TAArespvsOthersMod}, the results are noisier, so we expect a lower performance in general. This fact is reflected in Figure \ref{fig:RMI-violin}.

\subsection{Rotation-axis angular-rate algorithm}
We refer to the algorithm developed in \cite{bates2010respiratory} as the \textit{rotation-axis angular-rate} algorithm. Unlike the preceding algorithms, \cite{bates2010respiratory} reconstructs the angular motion induced by breathing. At rest, the measured and normalized acceleration vector is similar to the gravity vector, and respiration rotates that vector within the coordinate frame of the device. The algorithm seeks to track the axis of this rotation over time. 

In this algorithm, the TAA signal is first low-pass filtered with frequency cut off at $1$ Hz and each sample is normalized to a unit direction vector. The angle and axis of the rotation between consecutive acceleration vectors are obtained from their dot and cross products respectively. Since the axis inverts whenever the direction of rotation reverses, the axes are reflected into a common hemisphere by comparison with a reference axis, taken to be the first principal component of the unnormalized rotation axes. The predominant axis of rotation is estimated as a sliding mean of these axes over a $30$-second window, in which each axis is weighted by its associated window change and a Hamming window. The mean direction of gravity is estimated by an unweighted sliding mean over the same window. The rotation angle about the predominant axis and the angular rate are subsequently computed, and the latter is validated against gyroscope measurements. The respiratory rate is obtained from the angular rate via a threshold-based breath-counting procedure. 

We note that \cite{bates2010respiratory} samples at $64$ Hz and does not specify which filter design they used. As such, we apply our $4$th-order zero-phase Butterworth filter at the same cut-off to our $50$ Hz recording. Moreover, \cite{bates2010respiratory} partitions each recording into static periods separated by movement, processes each period as a separate dataset with its own reference axis, and ensures that the sliding averaging windows never contain vectors recorded in different orientations. We treat each $60$-second window as such a period, computing the reference axis within the window and truncating the averaging at its boundaries. In addition, the movement detection step of \cite{bates2010respiratory} (which discards periods in which the rotation between consecutive samples exceeds a threshold) is not applied, as we only compare on high-quality segments. We note that \cite{bates2010respiratory} reports the respiratory waveform for 67\% of an overnight recording under this gating. 

The mean $\pm$ SD RMI of the rotation-axis angular rate algorithm was $0.32\pm 0.22$ (median $0.24$) with 24\% of segments having an RMI over $0.5$. The synchronized correlations (mean $\pm$ SD) with airflow, thorax, and abdomen were $0.61\pm 0.19$, $0.61\pm 0.17$, and $0.65\pm 0.16$ respectively (TAA-resp as before). TAA-resp achieved a higher synchronized correlation than the rotation-axis angular-rate algorithm in 57\% of airflow windows, 84\% of thorax windows, and 82\% of abdomen windows, and paired Wilcoxon signed rank tests rejected equality in all three cases ($p<10^{-22}$).

\subsection{PCA-based axes-fusion algorithm}
The PCA-based axes-fusion algorithm developed in \cite{jin2009performance} was applied to the setup where the TAA sensor was placed on the diaphragm. Axes-fusion refers to the reconstruction of a single respiratory signal from the three axes and in \cite{jin2009performance}, they evaluate three approaches: an analytical approach based on sensor orientation (Full Angle), PCA, and a hybrid of the two. We compare against the PCA approach since it performed the best in \cite{jin2009performance} (along with the hybrid approach) and it requires no knowledge of the sensor orientation, which is unavailable in our recordings.
In this algorithm, the raw TAA signal is first band-pass filtered and then the three filtered axes are fused by PCA. Specifically, the sensor coordinate system is rotated to the orientation in which one axis accounts for the largest variance in the data, and the signal along that axis is taken as the reconstructed respiration-induced movement. 

We note that the parameters of the band-pass filter are not specified in \cite{jin2009performance}. As such, we apply the same $0.1$-$1$ Hz zero-phase $4$th-order Butterworth filter used by our own method. Moreover, the sensor in \cite{jin2009performance} is mounted on the diaphragm, whereas the fingertip signals we utilize contain considerably weaker respiration-induced motion, so the direction of largest variance is not guaranteed to be the direction carrying respiratory motion. We also do not apply manual exclusion of artifacts (such as speech, eating, etc.) as done in \cite{jin2009performance} since we utilize sleep recordings. 

The mean $\pm$ SD RMI of the PCA-based axes-fusion algorithm was $0.48\pm 0.24$ (median $0.46$) with 46\% of segments having an RMI over $0.5$. The synchronized correlations (mean $\pm$ SD) of the PCA axes-fusion algorithm were $0.65 \pm 0.16$, $0.74 \pm 0.19$, and $0.79\pm 0.16$ (TAA-resp values are the same as before). TAA-resp achieved a higher synchronized correlation than the PCA axes-fusion algorithm in 49\% of airflow windows, 37\% of thorax windows, and 31\% of abdomen windows, and paired Wilcoxon signed rank tests rejected equality in all three cases ($p<10^{-3}$). 

Among all algorithms we compare, this PCA axes-fusion algorithm is the closest one to our proposed algorithm. The main differences are the antiderivative and spectrum-based projection direction determination. Visually in Figures \ref{fig:TAArespvsOthersGood} and \ref{fig:TAArespvsOthersMod}, while the PCA axes-fusion algorithm gives nice results, they tend to be ``noisier'' although a low-pass filter has been applied. Therefore, we expect the performance of this algorithm to be close to but lower than ours, which is evident from Figures \ref{fig:RMI-violin} and \ref{fig:corr-all-violin}, where the synchronized correlations are almost the same.

\begin{figure}[hbt!]
    \centering
    \includegraphics[width=\textwidth]{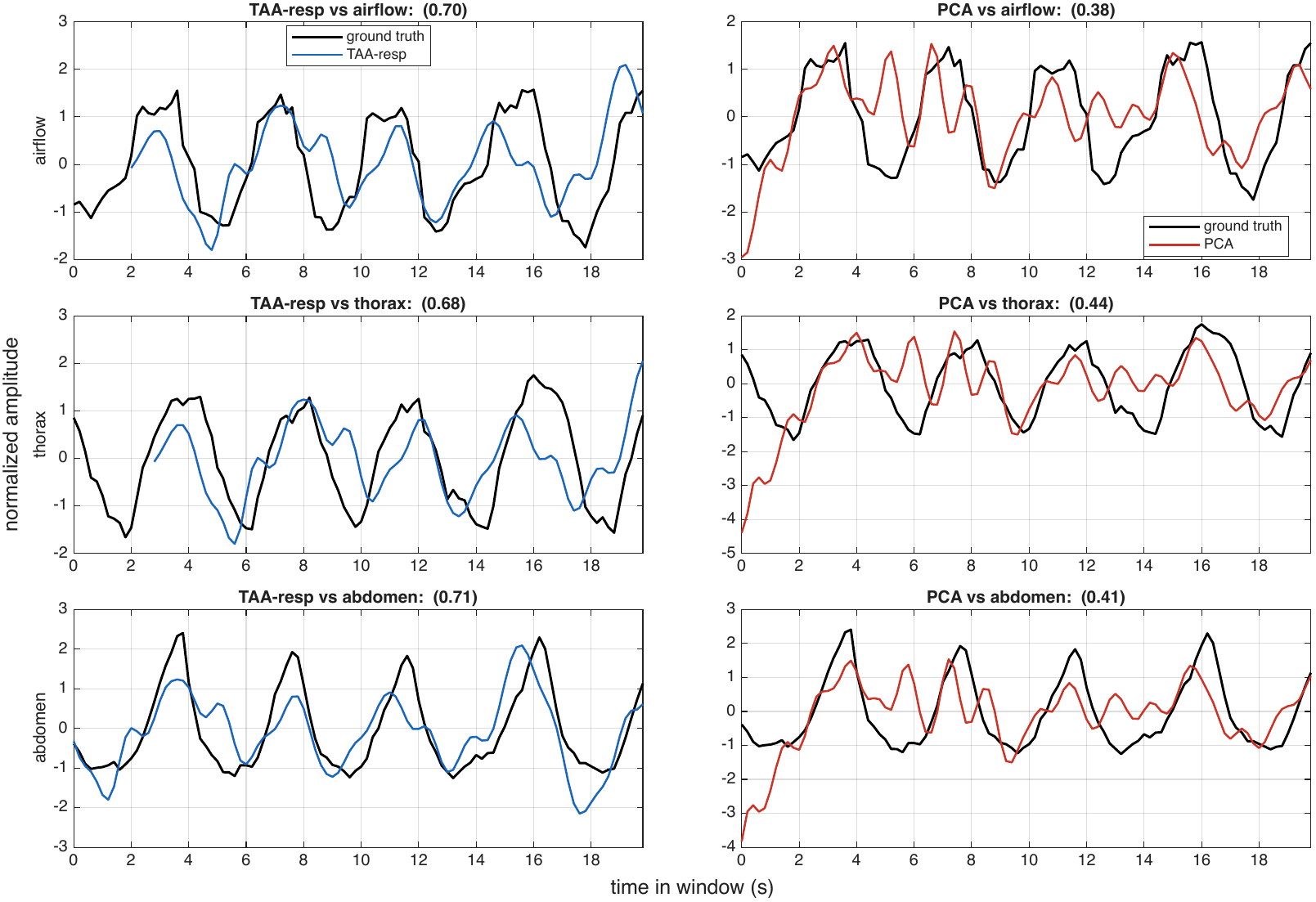}
    \includegraphics[width=\textwidth]{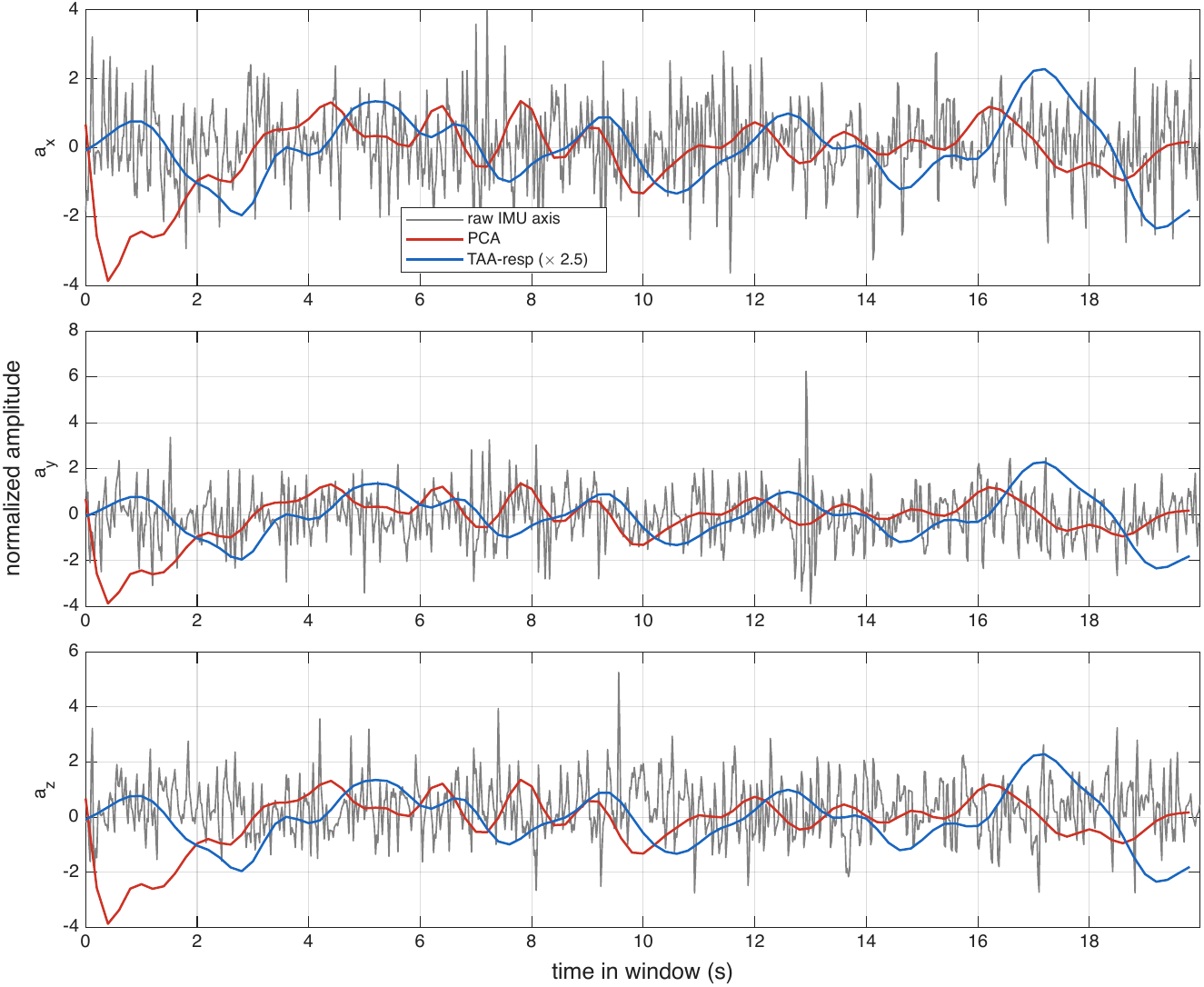}
    \caption{Comparison between respiratory surrogates derived via TAA-resp and PCA-based axes fusion over a 20 second moderate TAA segment superimposed with the ground truth (top) and both surrogates against the raw IMU data (bottom) where each surrogate is demeaned and scaled to unit standard deviation over the displayed interval.}
    \label{fig:PCAvsTAAresp}
\end{figure}

\begin{figure}[hbt!]
    \centering
    \includegraphics[width=\textwidth]{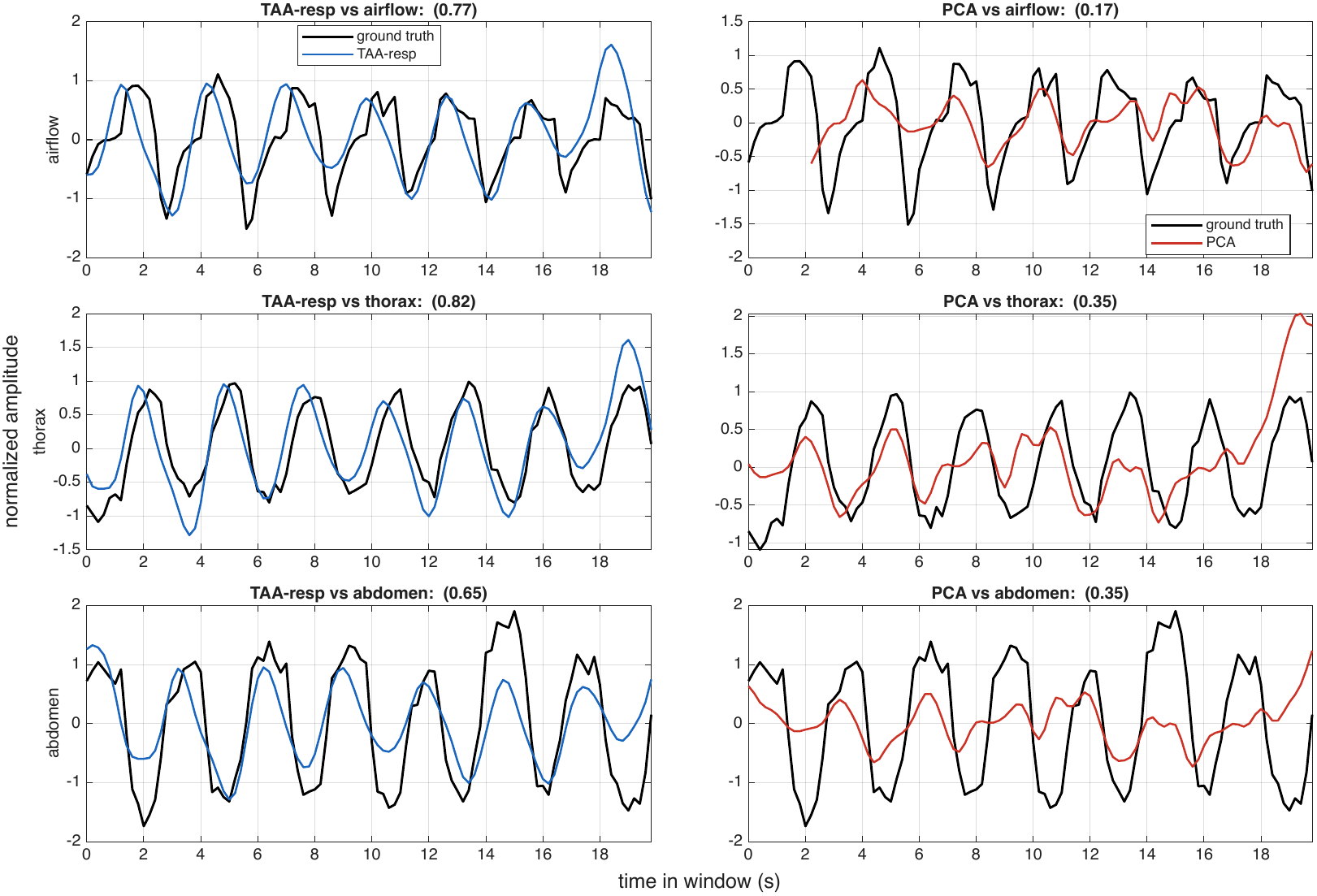}
    \includegraphics[width=\textwidth]{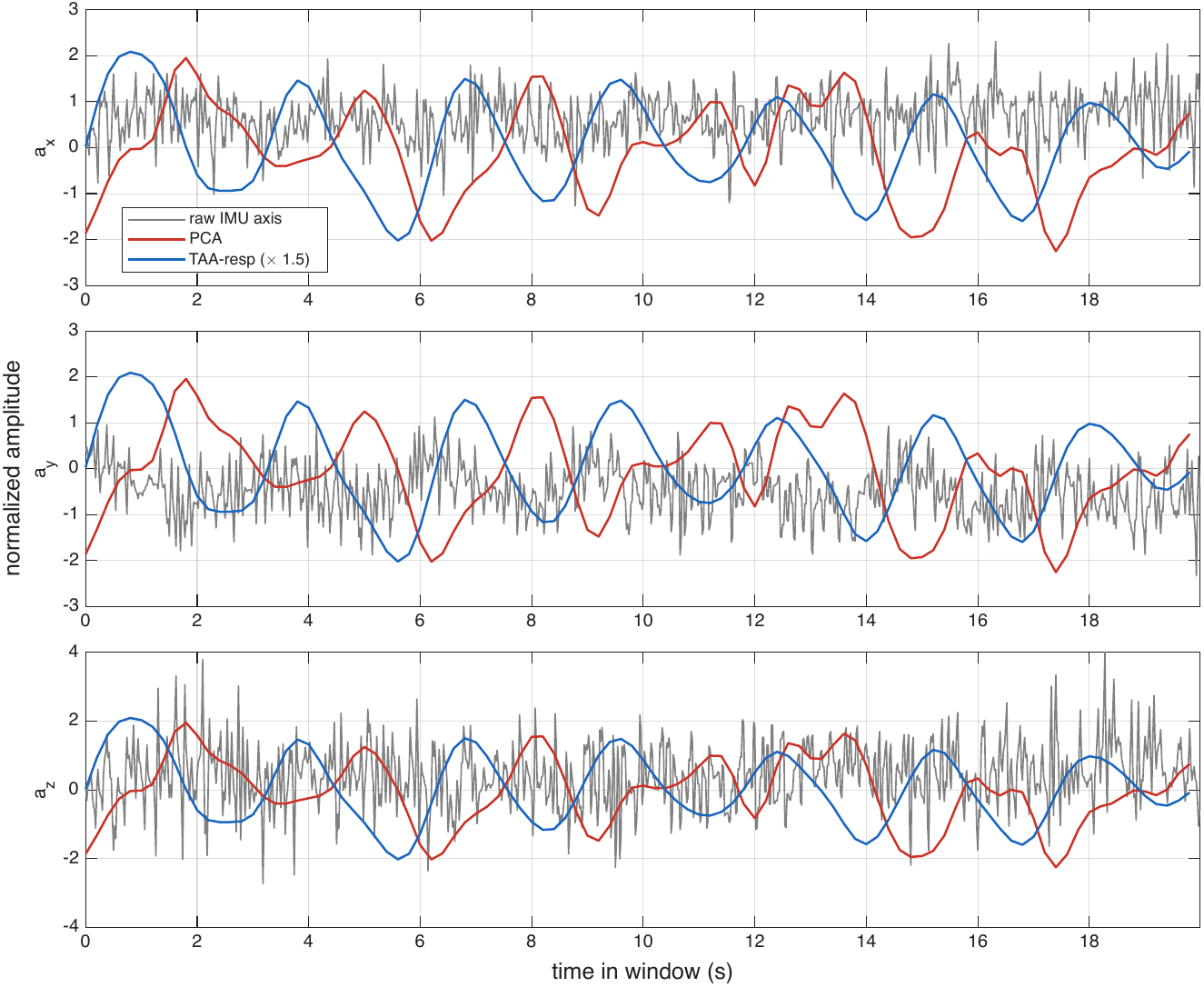}
    \caption{Comparison between respiratory surrogates derived via TAA-resp and PCA-based axes fusion over another 20 second moderate TAA segment superimposed with the ground truth (top) and both surrogates against the raw IMU data (bottom) where each surrogate is demeaned and scaled to unit standard deviation over the displayed interval.}
    \label{fig:PCAvsTAAresp2}
\end{figure}

To further understand the PCA axes-fusion algorithm, we observe that in several moderate segments, such as in Figures \ref{fig:PCAvsTAAresp} and \ref{fig:PCAvsTAAresp2}, the respiratory oscillation is directly visible in the raw accelerometer axes and is recovered by TAA-resp while the PCA surrogate is unable to track it. These are the segments that make up the lower lobe of the PCA RMI distribution and form a separate cluster beneath the identity line in Figure \ref{fig:RMI-violin}. We found that the PCA surrogate either matches TAA-resp closely or fails, and its median falls between the two behaviors without describing either.

\end{document}